\documentclass[10pt,twocolumn,aps,american,pra,floatfix,nofootinbib,superscriptaddress,longbibliography]{revtex4-2}
\usepackage[utf8]{inputenc}
\usepackage[english]{babel}
\usepackage[T1]{fontenc}
\usepackage{amsmath,epsfig,amssymb}
\usepackage{bbm}
\usepackage{bm}
\usepackage{amsthm}
\usepackage{xcolor}
\usepackage[caption=false]{subfig}
\usepackage{multirow}
\usepackage[normalem]{ulem}
\usepackage{braket}
\usepackage{enumitem}
\usepackage{afterpage}
\usepackage{amsfonts,amssymb,dsfont}
\usepackage{graphicx}
\usepackage{tikz}
\usetikzlibrary{quantikz}
\usepackage{dsfont}
\usepackage{soul}
\usepackage[title]{appendix}
\usepackage{hyperref}
\hypersetup{
    colorlinks=true,       % false: boxed links; true: colored links
    linkcolor=red,          % color of internal links
  citecolor=magenta,        % color of links to bibliography
    filecolor=magenta,      % color of file links
    urlcolor=cyan,           % color of external links
    runcolor=cyan
}

\DeclareMathOperator*{\argmin}{arg\,min}

\def\sx{\sigma^x}

\def\sz{\sigma^z}

\def\Tr{\text{Tr}}

\usepackage{thmtools}
\usepackage{thm-restate}

\newcommand{\pr}[1]{\left(#1\right)} % parentheses
\newcommand{\br}[1]{\left[#1\right]} % brackets
\newcommand{\sr}[1]{\left\{#1\right\}} % curly brackets
\newcommand{\abs}[1]{\left|#1\right|} % absolute value
\newcommand{\absq}[1]{\left|#1\right|^2} % absolute value squared
\newcommand{\aabs}[1]{\left\|#1\right\|} % norm
\newcommand{\expect}[1]{\left\langle#1\right\rangle} % expectation
\let\Tr\undefined
\newcommand{\Tr}[1]{\mathrm{Tr}\br{#1}}

\newcommand{\maxsat}[1]{MAX \ensuremath{#1}-SAT}

\newcommand{\groundstate}{\ensuremath{C_\mathrm{min}}}
\newcommand{\maxcoststate}{\ensuremath{C_\mathrm{max}}}

\newcommand{\identity}{\mathds 1}

\newcommand{\reals}{\mathbb R}

\makeatletter
\renewcommand{\maketag@@@}[1]{\hbox{\m@th\normalsize\normalfont#1}}%
\makeatother

\newcommand{\red}[1]{{\color{red}{#1}}}

\begin{document}
\title{Quantum Optimization with Classical Chaos}
\author{Malick A. Gaye}
\email[]{mgaye@ur.rochester.edu}
%\thanks{Present affiliation: University of Rochester}
\thanks{Now at University of Rochester}
\affiliation{William H. Miller III Department of
Physics \& Astronomy, Johns Hopkins University, Baltimore, Maryland 21218, USA}

\author{Omar Shehab}
\email[]{Omar.Shehab@ibm.com}
\thanks{Work completed while at IonQ, Inc.}
\affiliation{IBM Quantum, IBM Thomas J Watson Research Center, Yorktown Heights, NY, USA}

\author{Paraj Titum}
\email[]{Paraj.Titum@jhuapl.edu}
\affiliation{Johns Hopkins Applied Physics Laboratory, Laurel, Maryland, 20723, USA}

\author{Gregory Quiroz}
\email[]{Gregory.Quiroz@jhuapl.edu}
\affiliation{William H. Miller III Department of
Physics \& Astronomy, Johns Hopkins University, Baltimore, Maryland 21218, USA}
\affiliation{Johns Hopkins Applied Physics Laboratory, Laurel, Maryland, 20723, USA}

\begin{abstract}
    The Quantum Approximate Optimization Algorithm (QAOA) is a powerful tool in solving various combinatorial problems such as Maximum Satisfiability and Maximum Cut. Hard computational problems, however, require deep circuits that place high demands on classical variational parameter optimization. Ultimately, this has necessitated investigations into alternative methods for effective QAOA parameterizations. Here, we study a parameterization scheme based on classical chaotic recursive mapping, which enables significant reductions in the scaling of the variational parameter space. 
    Through numerical investigations of hard Maximum Satisfiability problems, we demonstrate that the chaotic mapping can effectively match the performance of standard QAOA when subject to a limited number of classical optimization iterations and short-depth circuits. Insight into this behavior is elucidated through the lens of classical dynamical systems and used to inform hybridized schemes that leverage both standard and chaotic parameterizations. It is shown that these hybridized approaches can boost QAOA performance beyond that of the standard approach alone, especially for deep circuits. Through this study, we
    provide a new perspective that introduces a generalized framework for specifying performant, dynamical-map-based QAOA parameterizations.
\end{abstract}
\maketitle

%-------------------------
%
%  INTRODUCTION
%
%-------------------------
\section{Introduction}
In modern quantum computing, variational quantum algorithms form the cornerstone of practical computational tasks seeking to leverage quantum hardware~\cite{cerezo2021variational}. These algorithms are designed to circumvent challenges associated with limited qubit resources and faulty gate operations in ways not afforded by traditional, fault-tolerant quantum algorithms. The quantum approximate optimization algorithm (QAOA) is a particular variational algorithm that aims to find approximate solutions to combinatorial optimization problems~\cite{farhi2014quantum, blekos2024review}. In addition to proofs of universality~\cite{lloyd2018quantum, morales2020universality}, QAOA potentially offers a pathway to quantum advantage~\cite{farhi2016quantum, crooks2018performance, niu2019optimizing, bravyi2020obstacles, an2022quantum, zhang2022applying, lykov2023sampling, boulebnane2024solving, shaydulin2024evidence}. In recent years, a number of experimental demonstrations have emerged across numerous quantum platforms, illustrating the potential of large-scale realizations~\cite{shaydulin2024evidence, pagano2020quantum, harrigan2021quantum, ebadi2022quantum, nguyen2023quantum, montanez2024towards, sack2024largescale}.

QAOA is characterized by two components: a parametrized quantum circuit and a classical optimization routine. Canonical implementations of QAOA define the quantum circuit as $p$ repetitions of two unitary operators: the cost evolution, which encodes the computational problem and a mixer evolution for spreading quantum information throughout the search space. Each alternating unitary is parametrized by a distinct tuning parameter, typically referred to as a variational parameter. The tuning of the $2p$ parameters is accomplished via a classical optimizer, a wide variety of which have been considered~\cite{pellow2021comparison, fernandez2022study, acampora2023genetic, pellow2024effect}. Ultimately, the objective of the optimizer is to train the quantum circuit to discover approximate solutions to the computational problem encoded within the QAOA. 

A key challenge of practical QAOA implementations is parameter training. The expressive power of the algorithm is dictated in part by the QAOA order $p$. As such, in order to achieve a desired quality of solution for hard computational problems, one may need a large number of QAOA layers. The number of variational parameters scales linearly with the QAOA depth which implies that larger QAOA circuits place increasing demands on the classical optimizer. An overall increase in the number of training parameters can be highly detrimental to the algorithm's performance. The training landscape can quickly become riddled with local minima traps and vanishing gradients, i.e., barren plateaus~\cite{barren_plateaus}. Furthermore, it may drastically increase training time and the usage of valuable quantum resources.

Strategies employed to circumvent this obstacle have sought to alter the quantum or classical subroutine. In the case of the former, the focus has been altering the parametrized quantum operations. For example, adaptively learning the circuit composition~\cite{zhu2022adaptive} or exploiting underlying symmetries in the computational problem~\cite{hadfield2019quantum, shaydulin2021exploiting, Shaydulin2021classical, tsvelikhovskiy2024symmetries, kazi2024analyzing}. While exhibiting utility, adaptive approaches require discovering a QAOA ansatz for each optimization problem; there is no generic structure. Symmetry-exploiting methods, on the other hand, by definition are scoped towards computational problems that exhibit known symmetries. Furthermore, cases exist where symmetry-preserving evolution may present obstacles for finding ground states of certain Hamiltonians~\cite{bravyi2020obstacles}.

Modifications to the classical subroutine have sought to improve variational parameter training and specification. Gradient-free protocols have become the preferred methods for variational training, with comparisons between existing methods and novel approaches highlighting the dependence of protocol performance on problem specifications~\cite{guerreschi2017compare, nannicini2019compare, shaydulin2019multistart, bonet2023compare, hao2024fewshots}. Machine learning based parameter discovery has also become quite popular, seeking to exploit underlying structure in the variational parameters and its relation to the problem instances to identify near-optimized parameters~\cite{alam2020accelerating, khairy2020learning, yao2020policy, moussa2022unsupervised, wauters2020rl, xie2023quantum, patel2024reinforcement}. A number of studies have shown that variational parameters can exhibit concentration that extends across random problem instances~\cite{brandao2018fixedcontrol, basso2022concen, farhi2022quantum, sureshbabu2024parameter, vijendran2025nearoptimalparametertuninglevel1} and measurements~\cite{farhi2014quantum, farhi2022quantum}. As such, in addition to machine learning based training, many studies have conveyed parameter transferability within and across computational problems~\cite{galda2021transfer, shaydulin2023parameter, montanez2024transfer, sakai2024linearly, sureshbabu2024parameter, sud2024qaoa, dehn2025linearramp}. This technique offers potential enhancements in parameter initialization or sufficient performance to circumvent further training.

Parameter reduction has gained significant attention as a method for reducing the dimensionality of the parameter search space. Reducing the $2p$ parameters to a smaller subset of trainable parameters potentially alleviates pressure on the classical optimizer and lessens the number of quantum processor calls required for gradient estimation. A common approach has been inspired by the intimate relationship between quantum annealing and QAOA. Namely, utilizing a linear or nonlinear ramp schedule that is dictated by only a few parameters~\cite{sack2021quantum, kremenetski2021linearramp, kremenetski2023linramp, montanezbarrera2024qaoa, dehn2025linearramp}. The schedules are designed for driving the evolution from a mixer-dominant evolution to a cost-dominant evolution similar to quantum annealing. This methodology has been further extended to functional expansions, which offer parameter reduction and global parameter updates through the training of weight coefficients prepended to basis functions~\cite{apte2025iterativeinterp}. The success of these parameter reduction techniques begs the question of whether there exists more generic frameworks for QAOA schedules---and variational algorithms more generally---that can enhance parameter training while reducing dimensionality.

In this work, we investigate the utility of classical chaos optimization for enhancing QAOA training, an approach we refer to as quantum approximate chaotic optimization algorithms (QACOAs). Well known in the classical community, chaotic optimization seeks to strike a balance between exploration and exploitation by drawing on chaotic maps and conventional global optimization frameworks~\cite{yang2007efficiency, luo2008chaos, zhang2024chaos}. Unlike purely random processes, chaotic maps exhibit deterministic yet pseudo-random behavior, which can prevent premature convergence and improve global optima searching. As a result, chaotic mappings can enhance both convergence rate and probability~\cite{zhang2024chaos, luo2008chaos}, with the extent of improvement determined by the location of the global optimum and the statistical properties of the chaotic map~\cite{yang2007efficiency}.

We utilize classical chaotic maps to parametrize the variational parameters and achieve reductions in the dimensionality of the parameter space. In particular, we introduce pure QACOA which solely draws on chaotic mapping. Using a logistic mapping, we show that pure QACOA can realize parameterizations consisting of only two independent variables, a factor of $p$ fewer parameters to optimize. Furthermore, we introduce novel hybridized QACOA protocols that incorporate aspects of chaotic and standard mappings to accelerate convergence. We argue that this approach broadens the pathway for a wide range of new QAOA parameterizations.

Through numerical and analytical studies, we assess the viability of QACOA. Using the Maximum $K$-Satisfiability (MAX $K$-SAT) problem class~\cite{hansen1990algorithms} as a test case, we show that pure QACOA can yield performance (i.e., quality of solution) greater than or equal to that of the standard QAOA parametrization at low circuit depth. However, large-depth pure QACOA can exhibit trainability challenges due to the inherent chaotic behavior of the parameter maps. Analytical examinations of the Lyapunov exponent---a parameter widely employed to quantify chaos in classical dynamical systems~\cite{strogatz2018nonlinear}---yield important insights into trainability and ergodicity in the QACOA framework. Namely, we find that pure QACOA parameterizations that solely draw on chaotic mappings are subject to exponentially large gradients at large $p$.

We utilize this finding to inform the design of hybridized QACOA circuits. Hybrid approaches are shown to enhance the suppression of divergent gradients and enable significant gains in the approximation ratio over standard QAOA for deep circuits. Our results indicate that QACOA can be an advantageous approach to variational algorithm training when a limited number of classical optimization steps are afforded. 
Overall, our work brings together classical and quantum optimization techniques to address a key challenge in variational quantum algorithm design and execution.

The manuscript is structured as follows. We contextualize the study by introducing the general QAOA/QACOA algorithms and circuit structures in Sec.~\ref{sec:qaoa_for_maxksat}. In Sec.~\ref{sec:parameterization}, we outline the standard QAOA and QACOA parameterization schemes discussed in the work. In Sec.~\ref{sec:numerical_comparison}, we simulate several standard QAOA and QACOAs on random MAX $K$-SAT instances for $N=5,8$ qubits, utilizing these results to choose map hyperparameters and investigate algorithm performance.
In Sec.~\ref{sec:discussion_of_ergodicity}, we describe QACOA's phase space dynamics through the lens of classical dynamical systems, enabling analysis of deep-circuit (ergodic) QACOA performance. In doing so, we introduce QACOA's phase-space and cost landscape characteristic exponents in order to explain its qualitative performance in the ergodic (i.e., deep circuit) limit. Lastly, we conclude with discussion of QACOA hybridization in Sec.~\ref{sec:qacoa_hybridization}, where we convey significant boosts in algorithmic performance for deeper circuits.

%-------------------------
%
%  OVERVIEW
%
%-------------------------
%\section{Overview}

%-------------------------
%
%  QAOA INTRODUCTION
%
%-------------------------
\section{QAOA for MAX $K$-SAT}\label{sec:qaoa_for_maxksat}

%====================
%  QAOA Overview
%====================
\subsection{QAOA}
\label{subsec:qaoa}
QAOA is a variational quantum algorithm designed to find approximate solutions to combinatorial optimization problems. Given an objective function $C(\mathbf{x})$, where $\mathbf{x} = (x_1, x_2, \ldots, x_N)$, $x_i\in\{0,1\}$, QAOA produces candidate solutions by driving the quantum system according to parameterized evolution. Commonly denoted by $\bm{\beta}=(\beta_1,\cdots, \beta_\ell)$ and $\bm{\gamma}=(\gamma_1,\cdots, \gamma_\ell)$, the variational parameters are related to the state of the system resulting from the QAOA evolution via
\begin{equation}
    \ket{\psi^{(p)}(\bm{\beta}, \bm{\gamma})} = U^{(p)}_{\rm QAOA}(\bm{\beta}, \bm{\gamma})\ket{\psi_0}.
\end{equation}
The initial state of the system $\rho_0\equiv\ket{\psi_0}\bra{\psi_0}$ is typically given by the equal superposition state $\ket{\psi_0}=\ket{+}^{\otimes N}$, where $\ket{+}=(\ket{0}+\ket{1})/\sqrt2$. The QAOA evolution can be generally described in accordance with the alternating operator ansatz~\cite{farhi2014quantum} as
\begin{equation}
    U^{(p)}_{\rm QAOA}(\bm{\beta}, \bm{\gamma})=\prod^{p}_{m=1}U_M(g_m(\bm{\beta})) U_C(f_m(\bm{\gamma})).
\end{equation}
In this way, the quantum system is subject to evolution dictated by two unitaries, each parameterized by distinct variational parameter functions. The so-called mixer evolution is given by 
\begin{equation}
    U_M(g_m(\bm{\beta}))=e^{-i \pi g_m(\bm{\beta}) H_M},
\end{equation}
where $g_m(\bm{\beta})$ 
is a real-valued function that, in general, may be dependent upon more than one variational parameter; see Sec.~\ref{sec:parameterization} for the standard case and the alternative mapping considered here. The Hamiltonian generating the mixer evolution is frequently defined as
\begin{equation}
    H_M=\sum^{N}_{i=1} \sx_i,
\label{eq:mixer_hamiltonian}
\end{equation}
with $\sigma^\mu_i$, $\mu=x,y,z$, denoting the Pauli operators for the $i$th qubit. Alternative forms of the mixer evolution have also been considered as a way to enforce constraints on the search space when applicable~\cite{blekos2024review}.

Similarly, the cost evolution 
\begin{equation}
    U_C(f_m(\bm{\gamma}))=e^{-i 2\pi f_m(\bm{\gamma}) H_C}
\end{equation}
is characterized by a real-valued function $f_m(\bm{\gamma})$ and a cost Hamiltonian $H_C$. The objective function $C(\bm{x})$ is encoded within $H_C$ by associating $x_i$ with the $z_i\in\{1,-1\}$ eigenvalues of the $\sz_i$ operators. More concretely, this is accomplished by first transforming the objective function according to the bipolar encoding $x_i=(1-z_i)/2$ and then replacing $z_i$ with $\sz_i$. The resulting Hamiltonian is of Ising form and diagonal in the computational basis, i.e.,
\begin{equation}
    H_C\ket{\bm{x}}=C(\bm{z})\ket{\bm{x}},
    \label{eq:Hc-eig}
\end{equation}
with $\bm{z}=(z_1,\ldots, z_N)$.

QAOA obtains approximate solutions by optimizing the variational parameters with respect to the expectation value
\begin{equation}
    F^{(p)}(\bm{\beta}, \bm{\gamma}) = \braket{\psi^{(p)}(\bm{\beta}, \bm{\gamma})|H_C|\psi^{(p)}(\bm{\beta}, \bm{\gamma})}.\label{eq:cost_function}
\end{equation}
When aiming to maximize $F$, the quality of the solution is typically quantified by the approximation ratio \red{(AR)} 
\begin{equation}
    \epsilon^{(p)} = \frac{F^{(p)}(\bm{\beta}^*, \bm{\gamma}^*)}{C_{\rm max}},
\label{eq:ar}
\end{equation}
where $(\bm{\beta}^*, \bm{\gamma}^*)$ denote optimized variational parameters and $C_{\rm max}=\max_x C(x)$ is the global maximum of the classical problem. In this work, we will define a cost Hamiltonian that leads to a minimization of the expectation value. For this reason, we define a modified AR
\begin{equation}
    \epsilon^{(p)} = \frac{\maxcoststate-F^{(p)}(\bm{\beta}^*, \bm{\gamma}^*)}{\maxcoststate-\groundstate},
    \label{eq:modified-ar}
\end{equation}
where $\groundstate$ the minimum (i.e., ground state of $H_C$). Note that, in this setting, $\maxcoststate$ refers to the worst case ($\epsilon^{(p)}=0$) while $\groundstate$ denotes the best solution ($\epsilon^{(p)}=1$).

%====================
%  QAOA Mapping
%====================
\subsection{QAOA mapping for K-SAT instances}
\subsubsection{K-Satisfiability}\label{k_satisfiability}
Satisfiability (SAT) problems form a class of multivariate problems that are defined in terms of $N$ Boolean variables. The variables are related via a set of $M$ constraints, each forming a special form of a clause. In conjunctive normal form (CNF), a SAT instance $\Omega$ can be written as
\begin{equation}
    \Omega = C_1 \land C_2 \land \cdots \land C_M,
    \label{eq:ksat-instance}
\end{equation}
such that there is a logical ANDing between clauses $\{C_j\}^{M}_{j=1}$. Each clause is a logical ORing of $K$ Boolean variables, i.e.,
\begin{equation}
    C_j = \bigvee^{K}_{l=1} v_{j_l} x_{j_l}.
\end{equation}
The additional variables $v_{jl}$ denote a logical unity ($v_{j_l}=1$) or NOT operation ($v_{j_l}=-1$), the specifications of which are determined by the SAT problem. A clause that evaluates to TRUE (FALSE) is called SAT (UNSAT).

$K$-SAT instances can be employed for decision and optimization problems. In the former, the objective is to determine whether there exists an $\bm{x}$ such that Eq.~(\ref{eq:ksat-instance}) evaluates to TRUE, i.e., it is satisfiable. The optimization version of the Boolean $K$-SAT problem is the MAX $K$-SAT problem, where the objective is to find variable assignments that maximize the number of satisfied clauses. The classical computational complexity of the problem type is strongly related to $K$. For instance, 2-SAT is within the complexity class P and therefore, it can be solved in polynomial time~\cite{Cook1971,GAREY1976237}. However, the task of finding high-quality solutions to MAX 2-SAT is known to be NP-hard, similar to 3-SAT~\cite{GAREY1976237} and MAX 3-SAT~\cite{CRESCENZI199965}. These features make MAX $K$-SAT problems attractive for characterizing algorithmic performance with respect to known computationally difficult problems.

\subsubsection{Problem Instance Complexity}
In this study, we leverage random MAX $K$-SAT to study the efficacy of alternative variational quantum circuit parameterization schemes. We identify SAT problems as being particularly useful because they possess an intrinsic tuning knob to control the complexity of a problem instance. More specifically, the clause density
\begin{equation}
    \alpha = M/N
\end{equation}
is a key parameter in describing various regimes of SAT problem hardness.  
As the number of clauses grows at a fixed number of variables, so does the unlikely scenario for satisfying all clauses. In the thermodynamic limit ($N\rightarrow\infty$), the probability of SAT as a function of $\alpha$ exhibits a phase transition at a critical clause density $\alpha_c$. For example, for $k=2$, $\alpha_c=1$~\cite{Goerdt1996}; this case has been studied extensively along with its finite-size scaling~\cite{Akshay_2020,max2sat_108_qubits}. Similarly, for $k=3$, the critical clause density has been numerically shown to be $\alpha_c\approx 4.26$~\cite{Crawford1996}.

\begin{figure*}[t]
    \centering
    \includegraphics[width=\textwidth]{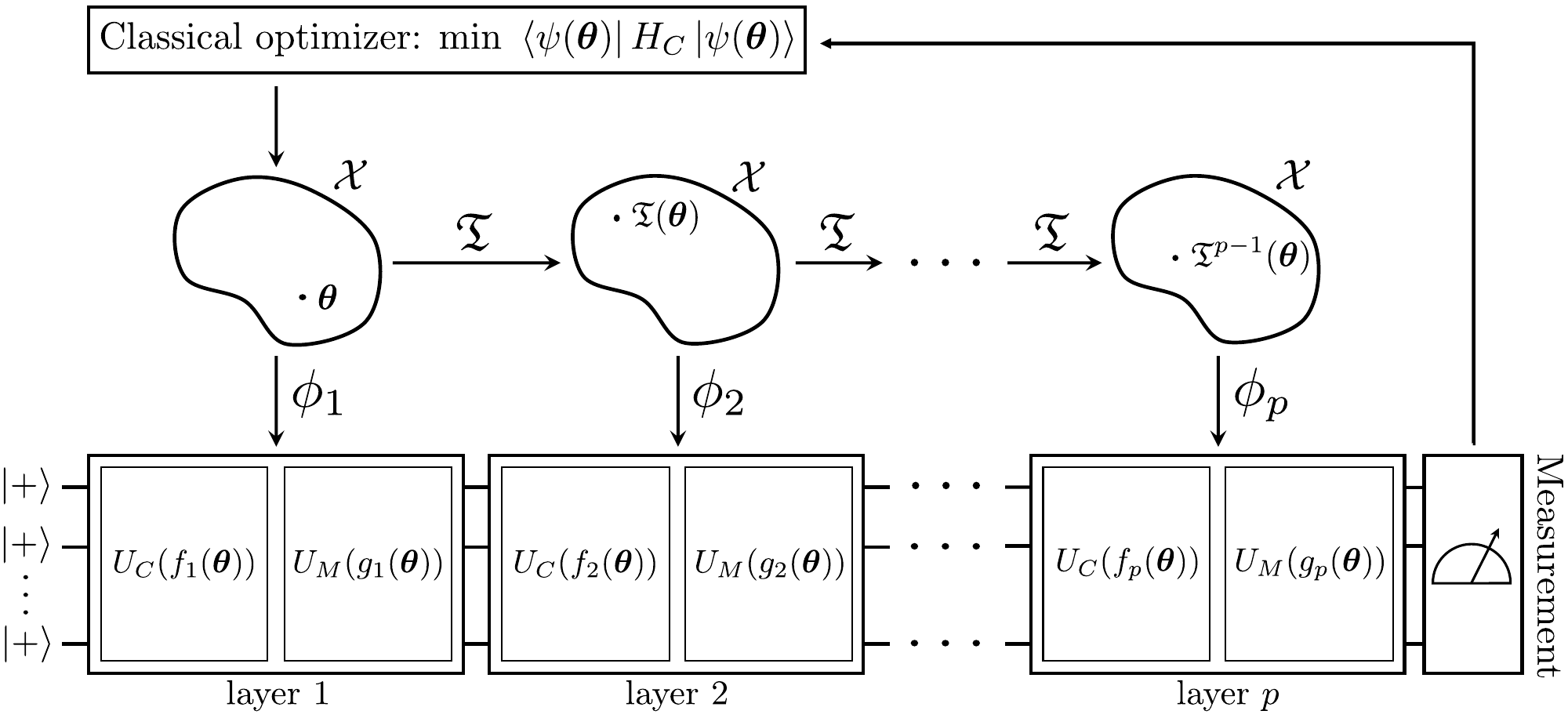}
    \caption{The general procedure for the parameterization $(\mathcal X, \mathfrak T, \Phi=\{\phi_m\}_{m=1}^p)$ of a depth $p$ QACOA/QAOA circuit. Circuit angles are parameterized by trajectories on a phase space $\mathcal X$ of arbitrary dimension transforming under $\mathfrak T$. General `pure' QACOA is given by $\mathfrak T$ ergodic (see Secs.~\ref{subsec:chaotic_parameterization}, \ref{sec:discussion_of_ergodicity}), and standard QAOA (Sec.~\ref{subsec:standard_parameterization}) corresponds to the case where $\mathcal X$ is the unit cube in $2p$ dimensions, $\mathcal T$ is the identity, and $\phi_m$ is a function simply selecting $(\gamma_m,\beta_m)$ from $\mathfrak T^{m-1}(\bm\theta)$. In the version of QACOA implemented here, we take $\mathcal X$ as the unit square, $\mathfrak T=l^c\times l^c$ the decoupled logistic map iterated $c$ times per layer, and $\phi_m$ the identity.}
    \label{fig:diagram}
\end{figure*}

\subsubsection{Construction of $H_C$}
The cost Hamiltonian $H_C$ is constructed by first defining Hamiltonians for each clause. This is accomplished via the variable translation discussed in Sec.~\ref{subsec:qaoa} and by recognizing that the logical OR operation can be represented equivalently as a multiplication. The resulting clause Hamiltonians are given by 
\begin{equation}
H_{C_j}=\prod_{l=1}^k\pr{\frac{\identity+v_{j_l}\sigma_{j_l}^z}{2}},
\end{equation}
where $\identity$ denotes the identity operator. The complete cost Hamiltonian is defined as 
\begin{equation}
    H_C=\sum_{C_j\in \Omega}H_{C_j},
\end{equation}
where the logical AND operation is translated into a summation of clause Hamiltonians.

The cost Hamiltonian is defined such that the ground state is synonymous with the variable configuration leading to a maximum number of clauses being satisfied. As such, the cost Hamiltonian denotes a minimization problem. This can be trivially transformed into a maximization on the modified AR $\epsilon^{(p)}$ [Eq.~\eqref{eq:modified-ar}] by taking $H_C\mapsto -H_C$ in accordance with the typical QAOA objective; we will do so, and moving forward we will refer to the modified AR as simply the AR.

%-------------------------
%
%  PARAMETRIZATION SCHEMES
%
%-------------------------
\section{Parameterization Schemes}
\label{sec:parameterization}

Here, we establish a general framework for describing QAOA parametrization schemes. This framework will inevitably cover both the standard QAOA and QACOA parametrization presented in this work. We define $\bm\theta\in\mathcal X$ as a classical parameter vector in an $n_\theta$-dimensional phase space $\mathcal X$; at times, we may use $\bm\theta$ and $(\bm\beta,\bm\gamma)$ interchangeably. $\mathfrak T:\mathcal X\to\mathcal X$ is a transformation on the phase space, with the iterated transformation $\mathfrak T^i\equiv\mathfrak T\circ\mathfrak T\circ\dots=\mathfrak T(\mathfrak T(\dots))$ giving the evolved parameters $\mathfrak T^{m-1}(\bm\theta)$ at depth (time) $m$. We will primarily consider transformations $\mathfrak T$ that are joint transformations on multiple parameters, denoted by $\mathfrak T=\mathcal T_1\times\mathcal T_2\times\dots$, with $\mathcal T_i$ a primitive transformation on a subset of the parameters, and $\times$ here acting as $(f\times g)(x,y)=(f(x),g(y))$. Preferably, $\mathcal T_i$ is well-studied in classical contexts. Let $\Phi\equiv\{\phi_m\}^{p}_{m=1}$ be a sequence of functions $\phi_m:\mathcal X\to\reals$ taking said evolved parameters to the $m$th layer angles. The $m$th layer angles are then written as
\begin{equation}
    (f_m(\bm\theta),g_m(\bm\theta))\equiv\pr{\phi_m\circ\mathfrak T^{m-1}}(\bm\theta).
\end{equation}
For this reason, we'll refer to $(\mathcal X, \mathfrak T,\Phi)$ as the \emph{parameterization}. We will also define the unit interval as $I\equiv[0,1]$. A diagram of the general QAOA parameterization scheme discussed here is shown in Fig.~\ref{fig:diagram}.

\subsection{Standard parameterization}\label{subsec:standard_parameterization}
In the standard parameterization scheme, the independent parameter vector is $\bm\theta=(\gamma_1,\beta_1,\dots,\gamma_p,\beta_p)\in\mathcal X=I^{2p}$, with parameterization functions $\mathfrak T=\mathrm{id}_{\mathcal X}$ and $\phi_m(\bm\theta) = (\gamma_m,\beta_m)$ (i.e., a projection of $\bm\theta$). As such, our $m$th layer circuit angles are given by
\begin{equation}
    (f_m(\bm\theta), g_m(\bm\theta))=(\gamma_m,\beta_m).
\end{equation}
This mapping results in each cost and mixer unitary possessing a distinct parameter that must be optimized, with $n_\theta=2p$ parameters total.
In QAOA, problem hardness is directly related to the order $p$ required to achieve a desired quality in the solution~\cite{Akshay_2020,Akshay2022}. Thus, for hard ($\alpha>\alpha_c$) optimization problems, large $p$ circuits and correspondingly large parameter vectors are necessary to locate high-quality solutions; this is a notable point of concern for algorithmic training that has been well-studied in recent years~\cite{barren_plateaus}. The problem of dimension reduction motivates our search for a non-naive parameterization method.

\subsection{Chaotic parametrization}\label{subsec:chaotic_parameterization}
We draw on concepts from classical chaotic optimization to define the variational parameters for QACOA circuits. The parametrization scheme entails the use of a chaotic map to recursively encode cost and mixer unitaries. 
We consider maps on a lower-dimensional parameter space than that of standard QAOA ($n_\theta< 2p$), with $\bm\theta=(\theta_1,\theta_2)=(\gamma_1,\beta_1)$. We present schemes for designing general QACOA circuits under arbitrary parameterizations. However, in this work, we primarily consider those of the form $(I^2,\mathfrak T,\{\mathrm{id}_{I^2}\}_m)$, with $\mathfrak T\equiv\mathcal T_1\times\mathcal T_2$ representing a 2D decoupled chaotic map. Under this parameterization, the $m$th layer circuit angles are given by
\begin{eqnarray}
\pr{f_m(\bm\theta),g_m(\bm\theta)}&=&\pr{f_m(\theta_1),g_m(\theta_2)}\nonumber\\
&=&\pr{\mathcal T_1^{m-1}(\theta_1),\mathcal T_2^{m-1}(\theta_2)}.
\end{eqnarray}
As we show below, this approach affords a substantial reduction in the number of required training parameters.

We will refer to the setting described above as a \emph{pure} QACOA parameterization. It is composed of the same transformation throughout the circuit and thus, remains strictly ergodic. This is in contrast to \emph{hybrid} QACOA (see Sec.~\ref{sec:qacoa_hybridization}), where the parametrization changes part of the way through the circuit. Pure QACOAs are advantageous from an analysis point of view due to characteristics such as the almost-everywhere convergence of its characteristic exponents; see Sec.~\ref{sec:qacoa_ergodicity} for further details. Hybrid QACOAs do not rigorously maintain such features.

In this study, we utilize the logistic map $l$ defined by the recursion relation 
\begin{equation}
    l(x_n)=x_{n+1}=rx_n(1-x_n).
    \label{eq:logistic_map}
\end{equation}
to construct the primitive transformation $\mathcal T_i$. Here, $x_1\in[0,1]$ is a free parameter, with $x_{n>1}$ being a function of $x_1$. Typically, the parameter $r\in[0, 4]$; we use $r=4$, giving a maximally chaotic logistic map on $I$~\cite{Whittaker1991,Hirsch2013}. Although the methods presented in this work generalize to arbitrary parameterization functions, we choose the $r=4$ logistic map due to its prominence in chaos theory literature~\cite{Hirsch2013, strogatz2018nonlinear} in order to facilitate the analysis of QACOA's performance in the ergodic (i.e., $c(p-1)\gg1$) limit. 

The ergodic limit is defined with respect to the `map speed' $c$ (a positive integer) corresponding to the number of transformations between circuit layers.
Moreover, we define the primitive transformations as $\mathcal T_1=\mathcal T_2=l^c$, denoting $l^c$ as $c$ compositions of the $r=4$ logistic map $l$. Note that for a given $c$, the $m$th layer will correspond to the initial parameters $(\theta_1,\theta_2)$ ``evolved'' $c(m-1)$ times under $l$. Throughout this study, we choose $c=100$ for reasons covered in Sec.~\ref{qacoa_performance_propagation_factor}.

Under this pure QACOA parameterization $(I^2,l^c\times l^c, \{\mathrm{id}_{I^2}\})$, the resulting $m$th layer circuit angles are given as
\begin{equation}
    \pr{f_m(\bm\theta),g_m(\bm\theta)}=\pr{l^{c(m-1)}(\theta_1),l^{c(m-1)}(\theta_2)}.
    \label{eq:parameterization_scheme}
\end{equation}
By design, $f_m(\bm\theta)$ and $g_m(\bm\theta)$ are dependent on only one parameter each, $\theta_1$ and $\theta_2$, respectively, and the number of free variables is now independent of the circuit depth $p$ ($n_\theta=2$). As we will show in the following section, this alternative parameterization strategy can yield results similar to the standard QAOA encoding at low depth using a factor of $p$ less parameters given limited classical resources.

%-------------------------
%
%  NUMERICAL RESULTS
%
%-------------------------
\section{Numerical Comparison}\label{sec:numerical_comparison}
The efficacy of the chaotic map parametrization is investigated via an empirical study of QAOA versus QACOA. Using the \maxsat{K} problem class as our testbed, we compare ARs and what we will refer to as misassignment rates for each approach as a function of algorithmic hyperparameters and specifications of the problem hardness.

\subsection{Problem specification and optimizer details}\label{problem_specification_and_optimizer_details}
In order to assess the performance of the chaotic parameterization scheme, we generate several \maxsat2, \maxsat3 problems of varying problem hardness (i.e., the clause density $\alpha$) with $N\in\{5,8\}$. Specifically, we have 50 $N=8$ problems and 5 $N=5$ problems. The $N=5$ problems will allow us to probe QACOA performance as a function of the map speed $c$, motivating our selection of $c=100$ in subsequent optimizations. In doing so, we will only showcase the results for a single $N=5$ \maxsat3 problem near the critical clause density, in the interest of focusing on performance comparisons across $c$; the conclusions drawn from such were consistent across other $N=5$ data. For each problem instance $\Omega$, we construct a problem Hamiltonian $H_C$ from which a QAOA or QACOA ansatz of order $p$ is assembled. The mixer Hamiltonian is chosen to be the standard transverse field Hamiltonian described in Eq.~\eqref{eq:mixer_hamiltonian}.

The performance of the ansatz is determined by minimizing the  objective function $F^{(p)}(\bm\theta)$ using a stochastic optimizer. We utilize a constrained first-order Simultaneous Perturbation Stochastic Approximation (SPSA) optimizer~\cite{SPALL1997109,spall_spsa_implementation,SADEGH1997281} to perform the minimization. The optimizer is run for up to $j_{\max}=1000$ iterations, with $j\leq j_{\max}$ the iteration number. First-order SPSA estimates gradients via a finite difference formula, which requires tuning hyperparameters. We follow the typical parameter selection rules in accordance by previous SPSA studies~\cite{spall_spsa_implementation}. For further information regarding the optimizer, see Appendix~\ref{appendix:spsa_optimizer}.

Here, we use the modified AR given by Eq.~\eqref{eq:modified-ar}. For a given problem instance, the AR is $ \epsilon^{(p)}_\mu$, where $\mu\in\{\text{std},\text{chaotic}\}$ denotes the standard QAOA and QACOA parametrization, respectively, and $\bm\theta^*$ the optimum parameters. In the subsequent analysis, we compare each scheme via $\overline{\epsilon}^{(p)}_\mu$, the AR averaged over five SPSA optimization runs. We find this to be a sufficient number of SPSA optimizations to generate averages representative of typical circuit performance.

\subsection{QACOA Performance and the Map Speed}\label{qacoa_performance_propagation_factor}
The QACOA parameterization primarily studied in this work features a single hyperparameter: the map speed $c$, indicating how many times the parameter vector is transformed per circuit layer. Due to the increasing sensitivity of the primitive transformation $l^c$ in $c$~\cite{Hirsch2013}, we expect QACOA performance to be sensitive to the choice of the map speed in the non-ergodic limit.

Here, we study QACOA performance dependence on the map speed, showing results specifically for a random $N=5$ \maxsat3 problem instance near the critical clause density. The relative performances across $c$ in this set of results are consistent with that of other problem instances, so we only showcase one data set for the sake of clarity. Within this comparison, we study the performance of QACOA against that of the standard QAOA, focusing on variations observed with respect to number of optimization iterations and circuit depth $p$.

\begin{figure}[t]
    \centering
    \includegraphics[width=\linewidth]{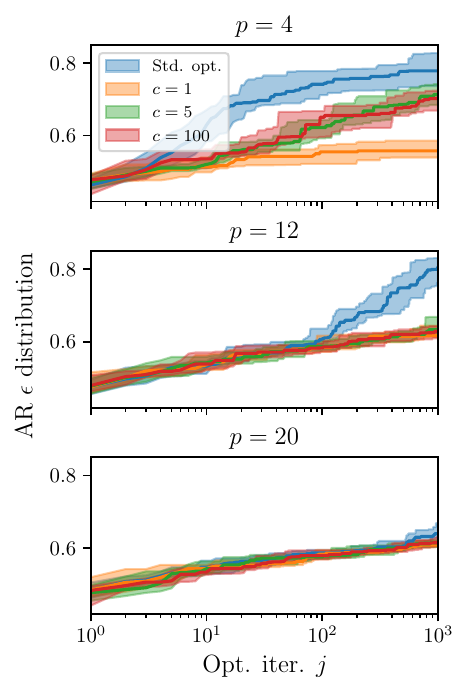}
    \caption{The distribution of ARs $\epsilon^{(p)}_\mu$ achieved by the QACOA algorithm over the optimization iteration $j$, compared to those yielded by standard QAOA. Results are shown for a fixed $N=5$ \maxsat3 problem near $\alpha=4.2$ for circuit depths $p=4,12,20$ and map speeds $c=1,5,100$. The blue curve corresponds to standard QAOA, while the others convey QACOA data. Each curve takes into account 20 optimizations, with the shaded regions representing IQRs and the solid lines are the medians. The comparison indicates that the larger map speeds yield the best performance.}\label{fig:approximation_ratios_over_j_c_fixed_problem}
\end{figure}

We analyze the effect of changing map speed for fixed circuit depth in Fig.~\ref{fig:approximation_ratios_over_j_c_fixed_problem}. The AR for $c=1,5,10$ is tracked as a function of optimization iteration for 20 SPSA optimizations. The resulting AR distributions are used to determine median (solid lines) and interquartile range (IQR; shaded region) for QACOA and the standard QAOA parameterization.

Our findings convey that $c=5, 100$ QACOAs consistently outperform the $c=1$ QACOA, indicating that pure QACOA performance is maximized when $c$ is not small. Moreover, the $c=5,100$ AR curves significantly overlap, so our results are somewhat insensitive to the exact choice of sufficiently large $c$. Based on this evidence, we choose a large value of $c$ that will more quickly bring our control into the ergodic limit (i.e., large $c(p-1)$), which will additionally be helpful for analysis purposes. As a result, moving forward we will primarily use a map speed of $c=100$ in our simulations, unless noted otherwise, to best facilitate discussion of QACOA behavior in the ergodic limit. While the search for an optimum value of $c$ was not exhaustive, this choice of $c$ will reveal much about QACOA's qualitative performance for a large range of $c$ as demonstrated in Sec.~\ref{sec:discussion_of_ergodicity}.

\begin{figure}
    \centering
    \includegraphics[width=\linewidth]{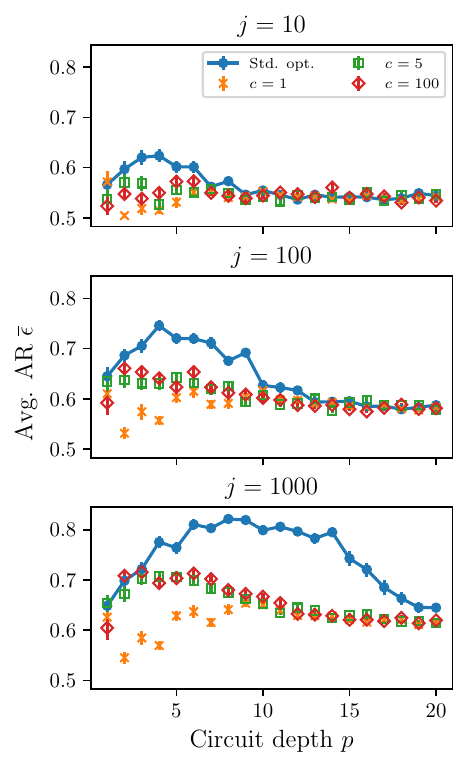}
    \caption{Pure QACOA performance results after various optimization iterations as a function of the circuit depth $p$ and map speed $c$. Results are shown for a fixed $N=5$ \maxsat3 problem at $\alpha=4.2$. Each data point is averaged over 20 optimizations, and the error bars represent a $68\%$ confidence interval estimate for the mean $\overline\epsilon$. QACOA performance improves with the optimization iteration $j$ most significantly at small $p$. However, at large $p$, we see evidence of a trainability deficit as the mean AR results improve slowly with $j$. This behavior is seemingly insensitive to the choice of the map speed $c$ at large depths.} 
\label{fig:performance_saturation}
\end{figure}

Notably, whether the QACOA achieves optimal performance seems to have a stronger dependence on the circuit depth than standard QAOA. In Fig.~\ref{fig:approximation_ratios_over_j_c_fixed_problem}, we find that increasing $p$ eventually leads to performance convergence between QACOA and QAOA for fixed $j$. This is likely an indication of a failure in the optimizer rather than a unique feature of the QACOA algorithm. However, an examination of QACOA performance as a function of circuit depth conveys a more subtle behavior not captured in the fixed $p$ analysis.

In Fig.~\ref{fig:performance_saturation}, we show the QACOA performance as a function of the circuit depth for a fixed number of optimization iterations. We focus on the average AR calculated over the 20 SPSA optimizations, with error bars denoted 68\% confidence intervals. This is accompanied by convergence in the performances across choices of the map speed $c$ at depth $p\gg1$. Specifically, this can be seen in the $j_{\max}=1000$ analysis, where QACOA is rapidly outperformed by QAOA beyond a critical circuit depth. This is not necessarily due to a change in the functionality of the optimizer, but rather an intrinsic numerical challenge inherited in our use of the chaotic map. In the following subsections, we will introduce another figure of merit and alternate problem specifications in order to test how consistently these effects arise over a number of QACOA optimizations. We will later show their analytical origin in Sec.~\ref{sec:discussion_of_ergodicity} and how to hybridize QACOAs to combat what we will refer to as \emph{trainability deficits} in Sec.~\ref{sec:qacoa_hybridization}.

\subsection{Hamming distance and state distribution analysis}\label{subsec:hamming_distance}
A QACOA output state $\ket{\psi(\bm\theta)}$ will be a mixture of candidate (bitstring) solutions to the problem $\Omega$, each of which may be associated with some distance from the best solution states. We may quantify distance between a QACOA state and the set of target states by averaging the minimum number of bitflips on a  state $\ket u$ required to reach any solution state, with weights $\absq{\braket{u|\psi(\bm\theta)}}$. That is, we are interested in counting the average number of variable misassignments in candidate solutions generated by QACOA circuits. For this reason, we consider a figure of merit $h$ based on the Hamming distance $d_h(u_1,u_2)$, defined as the number of (elementwise) mismatches between bitstrings $u_1,u_2$.

In order to define $h$, let $U\equiv\{0,1\}^N$ be the set of all $N$-variable bitstrings and the solution bitstrings given in $S(\Omega)=\{\argmin_{u\in U}\braket{u|H_C|u}\}\subseteq U\}$. We first define the \textit{minimum} Hamming distance between an arbitrary bitstring $u$ and the solution states $S(\Omega)$ as
\begin{equation}
    \tilde d_h(u,\Omega)\equiv\min_{s\in S(\Omega)}d_h(u,s).
\end{equation}
$\tilde d_h$ is interpreted as the number of misassignments in $u$ with respect to the `nearest' solution (in the Hamming sense). We only consider the distance from the nearest solution because given $u$, not all solution states $s\in S(\Omega)$ will have the same Hamming distance from $u$ (e.g., $u=000$, $S(\Omega)=\{001,011\}$). We use it to define the quantity $h(\bm\theta,\Omega)$ as
\begin{equation}
    h(\bm\theta,\Omega)\equiv\sum_{u\in U}\absq{\braket{u|\psi(\bm\theta)}}\tilde d_h(u,\Omega).
\end{equation}
$h(\bm\theta,\Omega)$ is thus the average number of variable misassignments given by the QACOA state $\ket{\psi(\bm\theta)}$.

Note that $U$ may be partitioned into disjoint subsets $U_d(\Omega)$ ($d=0,1,\dots,N$) with $U_d(\Omega)\equiv\{u\in U|\tilde d_h(u,\Omega)=d\}$ for any problem specification $\Omega$. This partitioning results in the following equivalent definition for $h(\bm\theta,\Omega)$.
\begin{equation}
    h(\bm\theta,\Omega)=\sum_{d=0}^Nd\sum_{u\in U_d(\Omega)}\absq{\braket{u|\psi(\bm\theta)}}
\end{equation}
The second sum over $u\in U_d(\Omega)$ is the probability that $\ket{\psi(\bm\theta)}$ is measured in the subspace of states with $d$ incorrect bits from the solution states $S(\Omega)=U_0(\Omega)$. If $\ket{\psi(\bm\theta)}=\sum_{s\in S(\Omega)}c_s\ket s$ lies entirely within the solution space, then we will always measure $h(\bm\theta,\Omega)=0$ incorrect assignments. Likewise, if the state is entirely in the subspace corresponding to $U_N(\Omega)$ then $h(\bm\theta,\Omega)=N$ is maximized. Thus, moving forward, we will consider the normalized quantity $h(\bm\theta,\Omega)/N\in[0,1]$, the \emph{misassignment rate}. 

\begin{figure}[t]
    \centering
    \includegraphics[width=\linewidth]{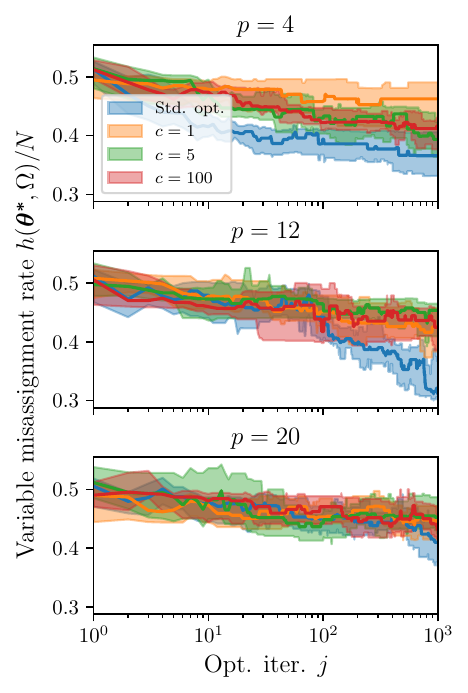}
        \caption{The misassignment rates $h(\bm\theta^*,\Omega)/N$ corresponding to the data shown in Fig.~\ref{fig:approximation_ratios_over_j_c_fixed_problem}. As was done in said figure, we have the misassignment rates for a fixed $N=5$ \maxsat3 problem under standard QAOA and $c=1,5,100$ QACOAs for three circuit depths $p=4,12,20$. This figure supports the conclusions drawn from Fig.~\ref{fig:approximation_ratios_over_j_c_fixed_problem}: $c=5,100$ QACOAs can be said to outperform $c=1$ at short depth, but performance at high depth is more or less indistinguishable in our results.}
    \label{fig:misassignment_rate_pure}
\end{figure}

In Fig.~\ref{fig:misassignment_rate_pure}, we show the misassignment rate as a function of the circuit depth and optimization iteration for the same data set as the previous subsection. We again see QACOA trains best in the short-depth limit. In particular, for $p=12$, we find that QACOA and QAOA possess similar misassignment rates for small $j$, with the two methods beginning to diverge as the number of optimization iterations increases. In contrast, in the ergodic limit, we see evidence for diminishing improvements to QACOA output states that holds regardless the number of optimization iterations. As we demonstrate below, this behavior is not confined to a specific hard instance of \maxsat3, but is instead a general characteristic of pure QACOA. Fortunately, it can be mitigated by modifying QACOA---a strategy we will present in Sec.~\ref{sec:qacoa_hybridization}. 

\subsection{QACOA performance as a function of problem specification} 
\label{subsec:qacoa_performance_function_of_circuit_depth}

\begin{figure}[t]
    \centering
    \includegraphics[width=\linewidth]{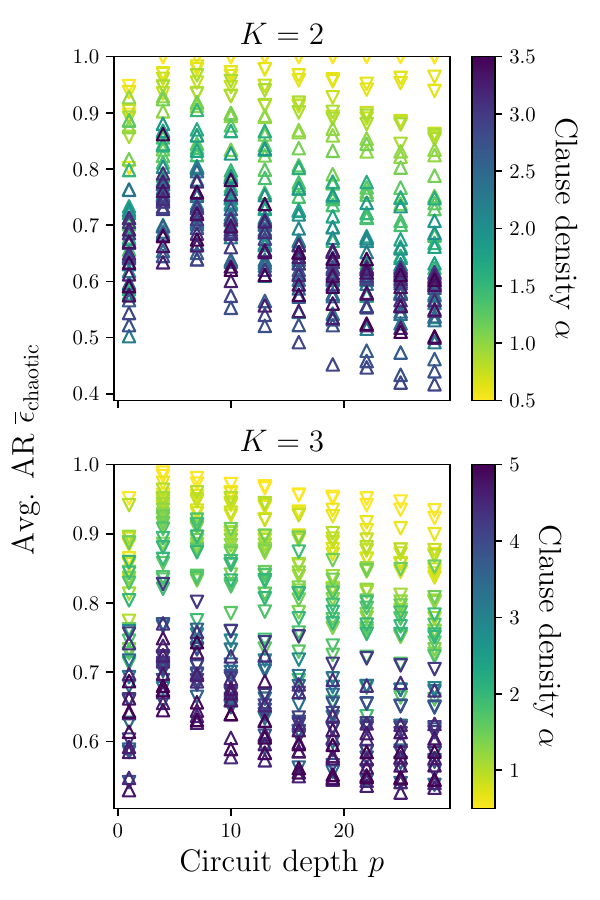}
    \caption{The average ARs achieved under chaotic parameterization as a function of the circuit depth $p$, for various \maxsat 2 (top) and \maxsat 3 (bottom) problems, colored by the clause density $\alpha$, for $N=8$ qubits after 1000 optimization iterations. Down-triangle ($\triangledown$) markers indicate a problem for which $\alpha<\alpha_c$, whereas up-triangle ($\triangle$) markers indicate $\alpha\geq\alpha_c$.}
    \label{fig:approximation_ratio_results_chaotic}
\end{figure}

\begin{figure*}[t]
    \centering
    \includegraphics[width=\linewidth]{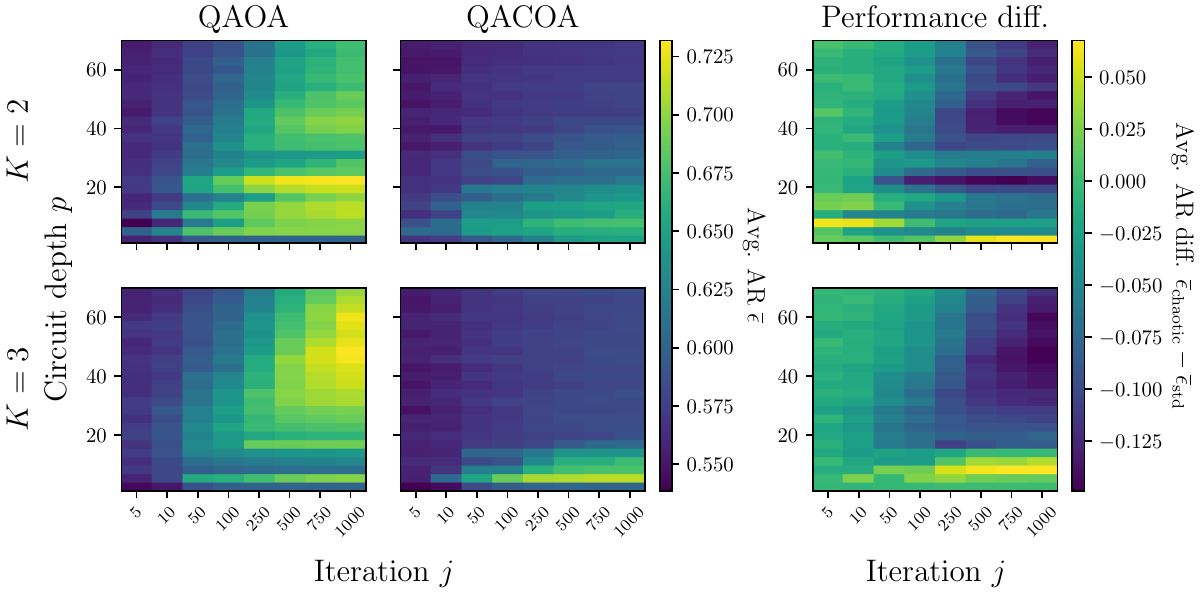}
    \caption{Averaged ARs $\overline\epsilon^{(p)}_\mu$ during optimization associated with a sample \maxsat 2 ($\alpha=1.125$) and \maxsat 3 ($\alpha=4.125$) problem, both with size $N=8$. For each figure, the horizontal axis corresponds to the number of optimization iterations $j$ and the vertical axis corresponds to the number of QAOA layers $p$. The first row corresponds to the \maxsat 2 problem, and the second row to the \maxsat 3 problem. The first column shows the average AR $\overline{\epsilon}_\mathrm{std}$ for optimized QAOA circuits under standard parameterization (Sec.~\ref{subsec:standard_parameterization}). The second column shows the same result for circuits under chaotic parameterization (Sec.~\ref{subsec:chaotic_parameterization}), $\overline{\epsilon}^{(p)}_\mathrm{chaotic}$. The third column shows the difference in performance between the chaotic and standard parameterization schemes, $\overline\epsilon^{(p)}_\mathrm{chaotic}-\overline\epsilon^{(p)}_\mathrm{std}$.}
    \label{fig:optimization_grid}
\end{figure*}

Here, we investigate the performance of QACOA for a broader set of problem instances. We focus on ensembles of 50 $N=8$ variable \maxsat2, \maxsat3 problems with varying problem hardness $\alpha$. Previous studies have noted the dependence of QAOA performance on SAT problem hardness~\cite{Akshay_2020}. Our goal is to determine whether such dependence exists for QACOA. Furthermore, we wish to assess how this dependence relates to relative performance improvements between QAOA and QACOA. 
In Fig.~\ref{fig:approximation_ratio_results_chaotic}, we compare the average AR to the number of QACOA layers over a range of clause densities for both \maxsat2 and \maxsat3. The average is computed over 5 SPSA training sessions and 50 problem instances for an $N=8$ qubit system. The map speed is chosen in accordance with the results of the previous section, i.e., $c=100$.

The comparison indicates a peak at small circuit depth, consistent with the results of Sec.~\ref{qacoa_performance_propagation_factor}.
This feature of QACOA's performance suggests that its use should be restricted for circuits with depth $p<\tilde p$, where $\tilde p$ is a ``critical depth''. This depth will generally have dependence on the choice of chaotic map $\mathfrak T$, and consequently, the map speed $c$ we use to define it. For the $c=100$ data shown in Fig.~\ref{fig:approximation_ratio_results_chaotic}, we identify $\tilde p\sim 10$.

Note that QACOA's performance across circuit depth $p$ is distinct from that of typical QAOA. For QAOA, it is known that increasing $p$---in particular, past a critical QAOA circuit depth $p^*$ determined by $\alpha,N$---is conducive to generating improved approximate solutions to \maxsat K problems~\cite{Akshay_2020,Akshay2022,max2sat_108_qubits}. However, for pure QACOA, it seems we must target depths $p<\tilde p$ determined by a finite $j_{\max}$. In the small $p$ regime, we see that QACOA can match standard QAOA performance in the limit of a small number of optimization iterations. We also note that QACOA is subject to a substantial algorithmic performance drop as the MAX $K$-SAT problem hardness $\alpha$ increases, as occurs with standard QAOA; see Appendix~\ref{appendix:qacoa_performance_function_of_problem_hardness} for further details.

We explore this behavior further and its impact on performance in Fig.~\ref{fig:optimization_grid}. The average AR is compared against circuit depth and training iterations for \maxsat2 and \maxsat3. Each algorithm is subject to the same problem instances (100 in total), as well as the same optimizer specification. We show the results for two randomly selected \maxsat2 and \maxsat3 problems, with hardness $\alpha=1.125$ and $\alpha=4.125$, respectively (near $\alpha_c$). These results highlight stark qualitative differences in QACOA and QAOA behavior, which will be important to understand in order to design improved QACOAs. 

Both algorithms improve in performance as a function of training iteration, however, exhibit distinct behavior for increasing $p$. QACOA can perform similarly to QAOA for small $p$, but trails in the $p>\tilde p$ regime. This feature is further evidenced by the relative difference plot between $\overline\epsilon_\mathrm{chaotic}$ and $\overline\epsilon_\mathrm{std}$. 
These results indicate that pure QACOA can be a viable alternative to standard QAOA at short depths, especially for classically hard problems with finite $j_{\max}$ afforded, and with the added bonus of a markedly reduced search space size. 

Although short-depth QACOA has shown some promise in the context of limited classical budgets, it has become glaringly apparent that adjustments must be made to pure QACOA in order to boost performance \textit{beyond} that of standard QAOA for deeper circuits, where the issue of large parameter spaces is increasingly relevant. In order to justify these adjustments, we must first turn focus to analysis of QACOA's phase space dynamics, as this will elucidate the properties of our classical map that deteriorate algorithmic performance.

\section{Discussion of Ergodicity}\label{sec:discussion_of_ergodicity}
In the previous section, we observed that pure QACOA's performance consistently deteriorates at sufficiently large circuit depth, for a finite number of optimization iterations. In order to diagnose and amend this trainability deficit, we will seek to understand QACOA's phase space dynamics using tools of classical chaos theory. We will start with a rigorous treatment of pure QACOA for which certain quantities, namely its Lyapunov exponents, can be shown to converge to an ergodic value almost everywhere. These results are then used to characterize the rate at which the set of ``useful'' optimizer perturbation vectors shrink, before they become effectively inaccessible in the ergodic limit $c(p-1)\gg1$. Our findings are generally insensitive to the problem instance and the parameter vector $\bm\theta$, conveying the broad applicability of this classical approach to the analysis of QACOA trainability.

\subsection{Dynamical description of general pure QACOA}\label{subsec:dynamical_description_pure_qacoa}
We may study QACOA using the theory of measure-preserving dynamical systems~\cite{Kechris1995,Kallenberg2021,Einsiedler2011} as follows. With the independent parameter vectors $\bm\theta$ and the total number of independent parameters $n_\theta$, let $\mu(B)\equiv\int_Bd^{n_\theta}\bm\theta\,\rho(\bm\theta)$ denote a Borel probability measure on $\mathfrak B(\mathcal X)$, with probability density $\rho(\bm\theta)$. Here, $\mathfrak B(\mathcal X)$ is the smallest $\sigma$-algebra containing subintervals of the sample (phase) space $\mathcal X\equiv X_1\times X_2\times \dots\times X_{n_\theta}\subseteq\reals^{n_\theta}$. We define a measurable transformation $\mathfrak T:\mathcal X\to\mathcal X$, where it must satisfy:
\begin{itemize}
    \item[(1)] $\mu(B) = \mu(\mathfrak T^{-1}B)\phantom i\forall\phantom iB\in\mathfrak B(\mathcal X)$ ($\mu$-preserving),
    \item[(2)] $\mathfrak T^{-1}B=B\in\mathfrak B(\mathcal X)\implies\mu(B)\in\{0,1\}$ (ergodic)
\end{itemize}
by definition. Requiring $\mathfrak T$ to have these properties restricts our choice of the density $\rho:\mathcal X\to\reals$ defining the measure $\mu$. We avoid placing explicit restrictions on the maps $\phi_m$ in the interest of not over-complicating subsequent analysis, and take these as more or less trivially-defined objects for now. The most general version of what we refer to as pure QACOA may be described using the now-complete system $(\mathcal X, \mathfrak B(\mathcal X), \mu, \mathfrak T)$. 

QACOA variants with respect to the version we simulate in this work [see Eq.~(\ref{eq:parameterization_scheme})] may be constructed through the tools offered in this section. One might choose to look at implementations of QACOA using completely-decoupled transformations of the independent variables, i.e., $\mathfrak T = \mathcal T_1\times\dots\times\mathcal T_{n_\theta}$ as we do in this work. In the completely-decoupled case, the measure might take the form $\mu=\mu_1\times\dots\times\mu_{n_\theta}$; if the invariant density $\rho_i:X_i\to\reals$ is known for an ergodic and $\mu_i$-preserving primitive transformation $\mathcal T_i:X_i\to X_i$, the joint density predicting QACOA statistics is constructed straightforwardly, as we will show in Appendix~\ref{appendix:2d_logistic_map}. 

One may also consider partially-coupled and totally-coupled versions of QACOA, e.g., $\mathfrak T = \mathcal T_1\times\mathcal T_{2,3}\times \mathcal T_4\times\dots$ with $\mathcal T_{i}:X_{i}\to X_{i}$ (density $\rho_i$), $\mathcal T_{i,j}:X_i\times X_j\to X_i\times X_j$ (density $\rho_{i,j}$). The primitive transformations $\mathcal T_i$ are chosen to be chaotic maps for which the densities are known. This formulation of QACOA is highly modular and offers significant potential for future work. For example, optimizing QACOA transformations $\mathfrak T$ and dimension $n_\theta$ for particular problem classes/sizes, and exploiting QACOA asymptotics to cheaply estimate trajectory-averaged QACOA statistics. In particular, we will show in Sec.~\ref{sec:trainability_decay} that the results derived in this framework can explain the systematic failure of the optimizer in certain limits, which will provide a lead into how QACOA may be hybridized to circumvent this issue.

\subsection{Ergodic result for QACOA statistics}\label{sec:qacoa_ergodicity}
An essential set of results for pure QACOA as described by $(\mathcal X, \mathfrak B(\mathcal X), \mu, \mathfrak T)$ are the \textit{ergodic theorems} for measurable functions $f$. They state that given ergodic transformations $\mathfrak T$, \textit{temporal} averages of $f$ converge to \textit{spatial} averages almost everywhere~\cite{Birkhoff1931,Moore2015}. That is,
\begin{align}
    \lim_{n\to\infty}\frac1n\sum_{i=1}^nf\circ\mathfrak T^{i-1}=\int_{\mathcal X}d\mu\, f=\expect{f}_{\mathcal X}\label{eq:ergodic_theorem}
\end{align}
where $\mathfrak T^{i-1}$ is the iterated transformation discussed in Sec.~\ref{sec:parameterization} and $\expect{\cdot}_{\mathcal X}$ is the phase space (trajectory) average. For the $r=4$ logistic map, the invariant density $\rho$ defining the measure $\mu$ is given by $\rho(x,y)=\rho_l(x)\rho_l(y)$ with $\rho_l(x)=1/\pi\sqrt{x(1-x)}$; for more information, see Appendix~\ref{appendix:2d_logistic_map}. This result can help explain the eventual saturation of averaged QACOA observed for the AR and misassignment rate results in previous sections. Furthermore, it will assist in assessing saturations found for $\mathcal X$'s `characteristic exponents' which we will define and discuss in Sec.~\ref{subsec:lyapunov}. Importantly, the result of Eq.~(\ref{eq:ergodic_theorem}) will enable us to assess the saturation in terms of constants dependent only on $H_C$, $N$, $j_{\max}$, and the transformation itself.

\subsection{Characteristic exponent analysis}\label{subsec:lyapunov}
We may now use ideas from the previous subsections to diagnose pure QACOA's trainability deficit in the ergodic (deep-circuit) limit. This is accomplished by characterizing circuit gradients via their Lyapunov exponents (LEs) $\lambda$, commonly studied in the context of chaotic dynamical systems~\cite{Eckmann1985,Abarbanel1991,Eckhardt1993}. The LEs are of interest to us because their almost-everywhere convergence to a positive constant for pure QACOAs (as formulated in Sec.~\ref{subsec:dynamical_description_pure_qacoa}) straightforwardly predicts QACOA's limiting performance in the ergodic limit, as we will show in Sec.~\ref{sec:trainability_decay}. 

We begin by introducing the exponents associated with generic functions in normed vector spaces~(Sec.~\ref{subsec:context_and_definitions_lyapunov}), and then specify the LEs in phase space~(Sec.~\ref{subsec:phase_space_LEs}) and on the cost landscape~(Sec.~\ref{subsec:cost_landscape_LEs}). A correspondence between these results will allow us to confidently treat pure QACOA's trainability deficit as an approximately classical effect, for which the decay rate is straightforwardly given by the ergodic phase space LE (Sec.~\ref{sec:trainability_decay}). Knowledge of this decay rate will prove to be useful in improving QACOA performance via hybridization (Sec.~\ref{sec:qacoa_hybridization}).

\subsubsection{Context and definitions}\label{subsec:context_and_definitions_lyapunov}
In order to facilitate a comprehensive discussion of QACOA's LEs, we will first define and discuss the LEs associated with some generic function $A^{(p)}$, with $p$ labeling the discrete time. We will later choose $A^{(p)}(\bm\theta)=\mathfrak T^{p-1}(\bm\theta)$ (the parameter vector) to characterize the phase space LEs and $A^{(p)}(\bm\theta)=F^{(p)}(\bm\theta) $ (the cost function after $p$ circuit layers) to characterize the cost landscape LEs.

Let $\bm\theta\in\mathcal X$ be the free parameter vector with $n_\theta$ elements. We will use $\bm{\hat e}_{\theta_i}$ to describe unit magnitude basis (row) vectors for $\mathcal X$. $\bm\theta=\sum_{i=1}^{n_\theta}\theta_i\bm{\hat e}_{\theta_i}$ is used to compute the $m$-th layer circuit angles $g_m(\bm\theta), f_m(\bm\theta)$. The differential of some generic well-behaved function $A^{(p)}:\mathcal X\to Y$, with $Y$, a normed vector space and depth (discrete time) $p$, is given by $\delta A^{(p)}(\bm\theta)=A^{(p)}(\bm\theta+\bm{\delta\theta})-A^{(p)}(\bm\theta)$. In the case that $\bm{\delta\theta}$ is small, $\delta A^{(p)}(\bm\theta)$ is approximated using a first-order Taylor expansion in the free parameters:
\begin{equation}
\delta A^{(p)}(\bm\theta) \approx \sum_{i=1}^{n_\theta}\delta\theta_i\partial_{\theta_i}A^{(p)}(\bm\theta)=\bm{\delta\theta}\,\mathcal{D}\left(A^{(p)}(\bm\theta)\right)
\label{eq:cost_function_differential}
\end{equation}
with $\mathcal{D}$ the total derivative.

We wish to assess Eq.~\eqref{eq:cost_function_differential} as a function of increasing $p$. $\aabs{\delta A^{(p)}(\bm\theta)}$ quantifies some notion of distance on the landscape at time $p$ generated by an infinitesimal separation between two trajectories, $\bm\theta$ and $\bm\theta+\bm{\delta\theta}$. As we consider only real-scalar (cost function) and real-vector (parameter vectors) valued $A^{(p)}$ in this work, we choose the norm $\|\cdot\|$ as the Euclidean distance. However, alternate choices for $\|\cdot\|$ may be made if $A^{(p)}$ is tensor-valued or complex, for example; we do not consider such cases in this work. 

We may characterize $\delta A^{(p)}(\bm\theta)$'s growth rate via $\lambda^{(p)}(\bm\theta)$ defined by
\begin{align}
    e^{\lambda^{(p)}(\bm\theta)(p-1)}&=\frac{\aabs{\delta A^{(p)}(\bm\theta)}}{\aabs{\delta A^{(1)}(\bm\theta)}}\\
     \implies\lambda^{(p)}(\bm\theta)&=\frac1{p-1}\ln\frac{\aabs{\delta A^{(p)}(\bm\theta)}}{\aabs{\delta A^{(1)}(\bm\theta)}}\label{eq:lyapunov_exponent}
\end{align}
That is, $e^{\lambda^{(p)}(\bm\theta)}$ is something of a geometric mean of the growth rate of the distance, over all past timesteps $m=1,\dots,p-1$. $\lambda^{(p)}(\bm\theta)$ is referred to as the \textit{local Lyapunov exponent} (LLE)~\cite{Abarbanel1991,Eckhardt1993}, defined for a finite time $p>1$. It gives the local average rate of separation between two QAOA/QACOA trajectories initialized infinitely close in phase space (i.e., by $\bm{\delta\theta}$). Positive LLEs will indicate instabilities on the cost landscape~\cite{Abarbanel1991} (typically associated with chaotic behavior), and negative LLEs commonly indicate attractive fixed points (i.e., stability)~\cite{Eckmann1985}.

An additional restriction often imposed on the definition of the LLE in Eq.~(\ref{eq:lyapunov_exponent}) is $\aabs{\delta A^{(1)}(\bm\theta)}\to0$. This is an enforcement of the linear approximation~[Eq.~(\ref{eq:cost_function_differential})], and we will also do the same when computing LEs numerically in the following subsections by considering the \textit{spectrum} of LEs. The LE spectrum $\{\lambda_i^{(p)}(\bm\theta)\}$ is found by choosing the perturbation to be in only one parameter, i.e., $\bm{\delta\theta}=\delta\theta_i\bm{\hat e}_{\theta_i}$. As such, Eq.~\eqref{eq:cost_function_differential} reduces to $\delta A^{(p)}(\bm\theta)=\delta\theta_i\partial_{\theta_i}A^{(p)}(\bm\theta)$. $\delta\theta_i$ cancels out in Eq.~\eqref{eq:lyapunov_exponent}, giving an expression for $\lambda_i^{(p)}(\bm\theta)$ which is computed exactly.

We may use the properties of QACOA transformations $\mathfrak T$ to show that certain kinds of LLEs approach an ergodic value almost everywhere. Consider the case where the trajectory differential is of the form $\delta A^{(p)}(\bm\theta)=a\prod_{m=1}^{p-1} (b\circ\mathfrak T^{m-1})(\bm\theta)$, with $a$ some constant (in particular, in the next sections we will choose $b$ to be a derivative). In the case that $\aabs{\prod_{m=1}^{p-1} (b\circ\mathfrak T^{m-1})(\bm\theta)}=\prod_{m=1}^{p-1}\aabs{(b\circ\mathfrak T^{m-1})(\bm\theta)}$ (e.g., scalars, certain types of operators), we may write the LLE in the large $p$ limit as
    \begin{align}
        \lambda^{(p\gg1)}(\bm\theta)&=\frac1{p-1}\sum_{m=1}^{p-1}\ln\aabs{\pr{b\circ\mathfrak T^{m-1}}(\bm\theta)}\\
        &\to\int_{\mathcal X}d\mu(\bm\theta)\,\ln\aabs{b(\bm\theta)}
        \label{eq:lyapunov_lle_ergodic}
    \end{align}
almost everywhere, by ergodicity of $\mathfrak T$. Note that this result presumes that $b$ meets conditions required to apply Eq.~(\ref{eq:ergodic_theorem}). Eq.~(\ref{eq:lyapunov_lle_ergodic}) is referred to as the \textit{global Lyapunov exponent} (GLE), as it is a trajectory-independent quantity. We will take the convergence of LLEs to GLEs to be loosely indicative of the transition to the ergodic regime, $\lambda^{(\tilde p)}(\bm\theta)\stackrel{\mathrm{def}}\approx\int_{\mathcal X}d\mu(\bm\theta)\,\ln\aabs{b(\bm\theta)}$. Although, we emphasize that the dependence of $\tilde p$ on both $j_{\max}$ and the chaotic map are not easily unentangled.

\subsubsection{Phase space LEs}\label{subsec:phase_space_LEs}
\begin{figure}
    \centering
    \includegraphics[width=\linewidth]{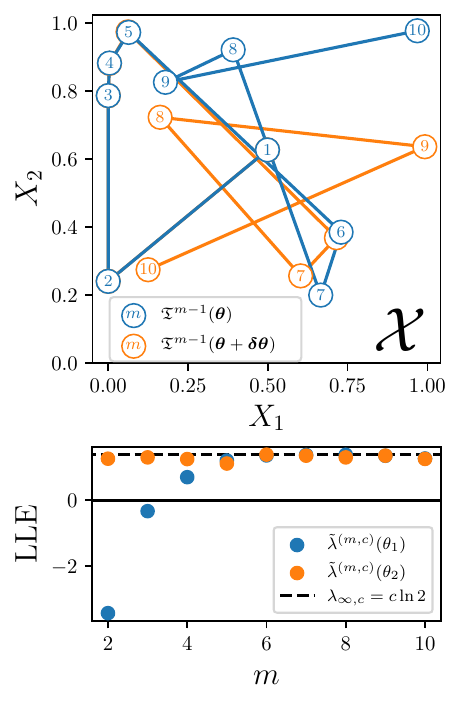}
    \caption{Top: The unperturbed (perturbed) QACOA angles $\mathfrak T^{m-1}(\bm\theta)$ ($\mathfrak T^{m-1}(\bm\theta+\bm{\delta\theta})$) under the $c=2$ chaotic parameterization scheme $\mathfrak T = l^{c}\times l^{c}$ as a function of `time' $m=1,2,\dots,10$ (labeling the markers). Note that both unperturbed and perturbed angles overlap for $m=1-5$. Bottom: the LE spectrum over $m$, $\lambda_i^{(m)}(\bm\theta)=\tilde\lambda^{(m,c)}(\theta_i)$. While $\theta_2\approx 0.6268$ is drawn randomly from $U(0,1)$, $\theta_1=1/2-10^{-3}\approx1/2$ is deliberately chosen here such that $\tilde\lambda^{(m,c)}(\theta_1)$ changes sign as $m$ increases, indicating that an instability is produced at $\theta_1$ between the third and fourth QACOA layers. The LEs saturate at $\lambda_{\infty,c}=c\ln\br2$, the GLE for the pure QACOA primitive transformation $\mathcal T_i=l^c$.}
    \label{fig:orbit_plot}
\end{figure}

In order to assess the behavior of functions parameterized by outputs of chaotic maps, we must start by considering chaos in the phase space $\mathcal X$. By rigorously showing pointwise convergence in the phase space LEs, we build the foundation for addressing QACOA trainability in the subsequent sections. We showcase only the main result and include further detail in Appendix~\ref{appendix:phase_space_le_expressions}.

We will discuss the result for the primary parameterization of interest, $(I^2,l^c\times l^c,\{\mathrm{id}_{I^2}\})$. Let $A^{(p)}(\bm\theta)=\mathfrak T^{p-1}(\bm\theta)=(f_p(\bm\theta),g_p(\bm\theta))$, the normalized QACOA angles generated for layer $p$. As such, in this section, we consider the rate that closely-initialized trajectories diverge in phase space. By choosing $\bm{\delta\theta}=\delta\theta_i\bm{\hat e_{\theta_i}}$, we have the LE spectrum $\lambda_i^{(p)}(\bm\theta)$ for $i=1,2$, corresponding to the cost unitary and mixer unitary parameter, respectively. As the transformations on $\theta_1,\theta_2$ are the same, their LLEs will have the same functional form. As such, we can write $\lambda_i^{(p)}(\bm\theta)=\tilde\lambda^{(p,c)}(\theta_i)$, with $\tilde\lambda^{(p,c)}$ given by
\begin{equation}
\tilde\lambda^{(p,c)}(\theta)=\frac1{p-1}\sum_{i=1}^{c(p-1)}\ln\abs{r(1-2l^{i-1}(\theta)}
\end{equation}
with $r=4$. By Eq.~\eqref{eq:ergodic_theorem}, in the ergodic limit [$c(p-1)\gg1$],
\begin{equation}
\tilde\lambda^{(p,c)}(\theta)\to \lambda_{\infty,c}\equiv c\ln 2,
\end{equation}
Thus, $\lambda_i^{(p)}(\bm\theta)=\tilde\lambda^{(p,c)}(\theta_i)$ converges to the GLE $\lambda_{\infty,c}=c\ln 2$ almost everywhere on $\mathcal X$.

In Fig.~\ref{fig:orbit_plot}, the convergence of the LLE is assessed numerically for the parameterization $(I^2,l^c\times l^c,\{\mathrm{id}_{I^2}\})$. These results consider a trajectory $\mathfrak T^{m-1}(\bm\theta)$ initialized at $\theta_1\approx1/2$ (i.e., near a singularity of $\tilde\lambda^{(p,c)}(\theta_1)$), alongside a perturbed trajectory $\mathfrak T^{m-1}(\bm\theta+\bm{\delta\theta})$. Although these trajectories are initially close, they diverge to arbitrary distances on $\mathcal X$ with sufficient $m$---this divergence begins immediately along $X_2$, and after $m=3$ along $X_1$ in the example shown in the top panel of Fig.~\ref{fig:orbit_plot}. The rate of this separation is characterized by the corresponding LLEs $\tilde\lambda^{(m,c)}(\theta_i)$, which we show converge to the GLE $\lambda_{\infty,c}$ in the bottom panel of Fig.~\ref{fig:orbit_plot}. As the GLE is positive, perturbations in the QACOA parameterization vector will grow very quickly under $\mathfrak T$, almost everywhere on $\mathcal X$ after sufficient time $m$. We will show in the next section that this effect is straightforwardly apparent in the cost landscape LLE results, and that it may impact QACOA trainability.

\subsubsection{Cost landscape LEs}
\label{subsec:cost_landscape_LEs}
In this section, we turn our attention to the LEs associated with the cost function, i.e., the average energy. Although we will not have expressions for cost landscape GLEs, we can leverage the analytical results of the previous section in conjunction with numerical studies of LEs on the cost landscape to gain insight into QACOA behavior. We may quantify instability on the cost landscape by choosing the generic function as the cost function $A^{(p)}(\bm\theta)=F^{(p)}(\bm\theta)$ associated with a problem instance $\Omega$. Consequently, it will be useful to express Eq.~(\ref{eq:cost_function_differential}) as
\begin{align}
    \delta F^{(p)}(\bm\theta)&=\sum_{m=1}^p\br{\delta f_m(\bm\theta)\partial_{f_m(\bm\theta)}+\delta g_m(\bm\theta)\partial_{g_m(\bm\theta)}}F^{(p)}(\bm\theta),
\end{align}
with derivatives given by
\begin{align}
    (\delta f_m(\bm\theta),\delta g_m(\bm\theta)) &= \bm{\delta\theta}\,\mathcal D\pr{(\phi_m\circ\mathfrak T^{m-1})(\bm\theta)}\label{eq:angle_differential},\\
    \partial_{f_m(\bm\theta)} F^{(p)}(\bm\theta)&=2\pi i\Tr{\tilde H_{C,p:m}\br{\rho_{m-1:1},H_C}}\label{eq:cost_gamma_gradient},\\
    \partial_{g_m(\bm\theta)} F^{(p)}(\bm\theta)&=\pi i \Tr{\tilde{H}_{C,p:m+1}\br{\rho_{m:1},H_M}}.
    \label{eq:cost_beta_gradient}
\end{align}
Above, we have $Q_{j:k}=\prod_{m=k}^jU_M(g_m(\bm\theta))U_C(f_m(\bm\theta))$, a windowed QACOA unitary.  $\tilde H_{C,j:k}=Q^\dag_{j:k}H_CQ_{j:k}$ and $\rho_{j:k}=Q_{j:k}\rho_0Q^\dag_{j:k}$ denote the cost Hamiltonian and initial state rotated by $Q_{j:k}$, respectively. For QACOA implemented in this work, we have $n_\theta=2$ and $\phi_m=\mathrm{id}_{I^2}$, so Eq.~(\ref{eq:angle_differential}) is given by $\bm{\delta\theta}\,\mathcal{D}\mathfrak T^{p-1}\pr{\bm\theta}$.

\begin{figure}
    \centering
    \includegraphics[width=\linewidth]{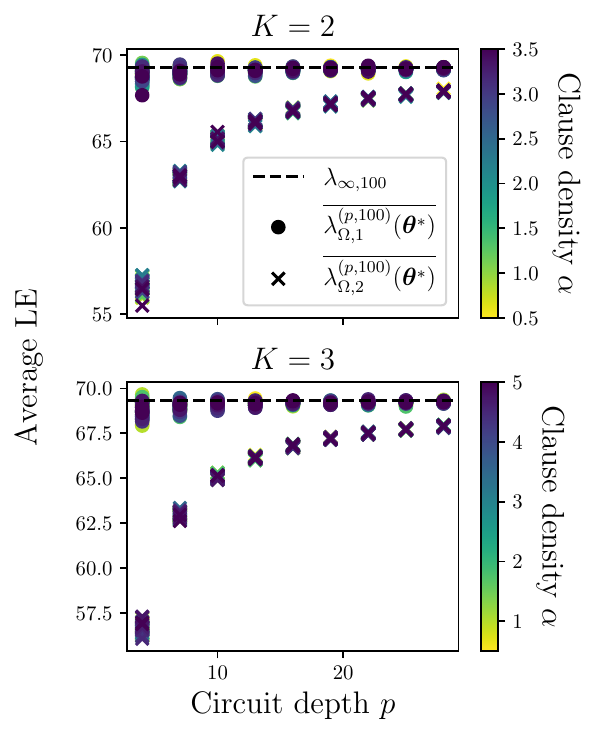}
     \caption{The LLE spectrum (at the optimum $\bm\theta^*$ found by the optimizer) associated with the cost function (Eqs.~\eqref{eq:cost_lle_spectrum_1} and \eqref{eq:cost_lle_spectrum_2}) for \maxsat2 and \maxsat3, averaged over five independent optimizations (similar to the AR in Sec.~\ref{sec:numerical_comparison}). The LEs appear to be insensitive to the problem hardness $\alpha$, and approach $\lambda_{\infty,100}=100\ln2$, the phase space GLE under the $c=100$ chaotic parameterization.}
    \label{fig:lyapunov_exponent_results}
\end{figure}

We will again consider the spectrum of exponents by choosing $\bm{\delta\theta}$ in a single direction at a time, as we did in the last subsection. Let $\lambda_{\Omega,i}^{(p,c)}(\bm\theta)$ be the exponent at $\bm\theta$ associated with the single-parameter perturbation $\bm{\delta\theta}=\delta\theta_i\bm{\hat e}_{\theta_i}$ for the problem instance $\Omega$; recall that $i=1,2$ correspond to the first layer angles $\gamma_1,\beta_1$, respectively. The superscripts $(c, p)$ indicate the map speed and circuit depth. Then, we have
\begin{align}
    \lambda_{\Omega,1}^{(p,c)}(\bm\theta) &= \frac1{p-1}\ln\frac{\abs{\sum_{m=1}^ph_{m,c}(\theta_1)\partial_{f_m(\bm\theta)} F^{(p)}(\bm\theta)}}{\abs{h_{1,c}(\theta_1)\partial_{f_1(\bm\theta)} F^{(1)}(\bm\theta)}}\label{eq:cost_lle_spectrum_1},\\
    \lambda_{\Omega,2}^{(p,c)}(\bm\theta) &= \frac1{p-1}\ln\frac{\abs{\sum_{m=1}^ph_{m,c}(\theta_2)\partial_{g_m(\bm\theta)} F^{(p)}(\bm\theta)}}{\abs{h_{1,c}(\theta_2)\partial_{g_1(\bm\theta)} F^{(1)}(\bm\theta)}}\label{eq:cost_lle_spectrum_2},
\end{align}
with the iterated logistic map derivatives given as
\begin{eqnarray}
    h_{m,c}(x) & \equiv&\partial_{x}l^{c(m-1)}(x)=\prod_{i=1}^{c(m-1)}r(1-2l^{i-1}(x)),\quad\quad\\
    \partial_xh_{m,c}(x)&=&h_{m,c}(x)\sum_{i=1}^{c(m-1)}\frac{l^{i-1}(x)\tilde h_{i-1}(x)}{l^{i-1}(x)-1/2}.
\end{eqnarray}
In the $c(p-1)\gg1$ limit, convergence of the cost landscape LLEs to an ergodic value $\lambda_{\Omega,i}^{(p,c)}(\bm\theta)\to \lambda_{\infty,c}=c\ln2$ is evidenced in our numerical LLE results for the $N=8$ problem instances discussed in Sec.~\ref{sec:numerical_comparison} and shown in Fig.~\ref{fig:lyapunov_exponent_results}. This observation indicates that the ``stretching'' of optimizer perturbation vectors is dominated by the phase space dynamics, rather than the details of the cost function. In the next section, we will exploit this relative insensitivity to $\Omega$ to justify considering solely the effect of the classical phase space dynamics on QACOA trainability. 

\section{Trainability Deficits}\label{sec:trainability_decay}
In Secs.~\ref{subsec:phase_space_LEs} and \ref{subsec:cost_landscape_LEs}, we have rigorously shown that phase space LLEs converge almost everywhere to a positive constant $\lambda_{\infty,c}$, and that the cost landscape LLEs also indicate convergence to $\approx \lambda_{\infty,c}$. As the cost landscape LEs are intimately tied to the cost function's gradients, we can characterize pure QACOA's diminishing trainability in terms of $\lambda_{\infty,c}$. We have empirically demonstrated global convergence of phase space and cost landscape LEs to $\lambda_{\infty,c}$ under the pure QACOA, and thus, there is an opportunity to derive general insights about QACOA trainability from this observation.

First, we introduce some relevant context. We assume that due to some minimum precision $\bm\epsilon_\theta\equiv\epsilon_{\theta,i}\bm{\hat e}_{\theta_i}$ in elements of a nonzero parameter differential $\bm{\delta\theta}=\delta\theta_i\bm{\hat e}_{\theta_i}$ with $0\leq\epsilon_{\theta,i}\leq\abs{\delta\theta_i}\ll1$ and at least one $\epsilon_{\theta,i}>0$, our optimizer produces a small distance on the cost landscape $\delta\mathcal F^{(p)}(\bm\theta|\bm\epsilon_\theta)\equiv\abs{\delta F^{(p)}(\bm\theta|\bm\epsilon_\theta)}$. We introduce the notion of machine precision here to emphasize that the classical optimizer cannot truly represent a continuum of parameters; this will have concrete implications in Sec.~\ref{subsec:nonlinearizability}. 

In order for practical optimization methods to succeed, we must be able to achieve (1) ${\delta\mathcal F^{(p)}(\bm\theta|\bm\epsilon_\theta)\ll1}$ so that the training protocol can make small steps towards extrema $\bm\theta^*$ without overshooting, and (2) $\delta\mathcal F^{(p)}(\bm\theta^*|\bm\epsilon_\theta)\approx0$ such that $\bm\theta^*$ may be detected as an extremum. Crucially, for many gradient-based optimizers (including SPSA), $\delta\mathcal F^{(p)}(\bm\theta|\bm\epsilon_\theta)$ generated by a small parameter differential $\bm{\delta\theta}$ must be approximately linear for the optimizer to succeed. In the following subsections, we discuss how QACOA parameterization may impact an optimizer's ability to satisfy these conditions.

\subsection{Vanishing linearizable perturbations}\label{subsec:nonlinearizability}

We are now prepared to address a limitation of chaotic optimization at large depths. Positive cost landscape LEs indicate that at large $c(p-1)$, the cost function at a perturbed trajectory $\bm\theta+\bm{\delta\theta}$ will typically be nonlinearizable about the unperturbed trajectory $\bm\theta$ when the precision is not arbitrarily small---as is the case in numerical simulations with a finite machine precision. As SPSA's gradient estimation procedure relies on nonlinearities close to $\bm\theta$ being small, these gradient estimates will see significant error as $c(p-1)$ increases, effectively making the landscape untrainable. In this subsection, we aim to address the question of how large perturbation vector elements must be before nonlinearities in the cost function become significant, and thus, how quickly the landscape becomes untrainable under the pure QACOA parameterization $(I^2,l^c\times l^c, \{\mathrm{id}_{I^2}\})$.

We assume that the optimizer is not at a local extrema in the cost landscape 
after the first circuit layer, i.e., ${\delta\mathcal F^{(1)}(\bm\theta|\bm\epsilon_\theta)>0}$. We can consider $\delta\mathcal F^{(p)}(\bm\theta|\bm\epsilon_\theta)$ ``too large'' when the error of the linear approximation is appreciable:  
\begin{equation}
    \aabs{\bm{\delta\theta} \mathcal{D}F^{(p)}(\bm\theta)}\sim \frac12\aabs{\bm{\delta\theta}\br{\mathcal{D}^2F^{(p)}(\bm\theta)}\bm{\delta\theta}^\top}.
    \label{eq:linear_approximation_breakdown}
\end{equation}
That is, we say that nonlinearities near $\bm\theta$ are significant when nonlinear contributions to the Taylor series of the cost function become significant. Eq.~\eqref{eq:linear_approximation_breakdown} then offers a simple criterion by which perturbations $\bm{\delta\theta}$ can be categorized as ``useful'' to the optimizer (i.e., small enough) or not useful (i.e., generating nonlinear $\delta\mathcal F^{(p)}(\bm\theta|\bm\epsilon_\theta)$, resulting in a poor gradient estimate). Although certain numerical strategies, such as taking two-sided gradient estimates, can mitigate dependence of the estimate's error on second derivatives, the condition above will broadly capture the effect we are aiming to characterize, given our choice of chaotic map.

With regards to the parameterization $(I^2,l^c\times l^c,\{\mathrm{id}_{I^2}\})$, it is simpler to express this condition in phase space. This is because our numerical results~(Fig.~\ref{fig:lyapunov_exponent_results}) suggest that variations in the cost landscape LEs, proportional to $\ln\aabs{\bm{\delta\theta} \mathcal DF^{(p)}(\bm\theta)}$, can be almost entirely attributed to large phase space LEs which tend to $\lambda_{\infty,c}$. That is, we utilize the fact that $\lambda^{(p,c)}_{\Omega,i}(\bm\theta)\to\lambda_{\infty,c}$ to relax Eq.~\eqref{eq:linear_approximation_breakdown} to a phase-space analogue,
\begin{equation}
\abs{\delta\theta_ih_{p,c}(\theta_i)}\sim\abs{(1/2)\delta\theta_i^2 \partial_{\theta_i} h_{p,c}(\theta_i)}.
\label{eq:linearizable_parameter_differential_condition}
\end{equation}
So, parameter differentials $\bm{\delta\theta}$ that generate nonlinear differentials $\delta\mathcal F^{(p)}(\bm\theta|\bm\epsilon_\theta)$ on the cost landscape have some $\abs{\delta\theta_i}\gtrsim\eta_{p,c}(\theta_i)$, where we define
\begin{equation}
    \eta_{p,c}(\theta_i)\equiv\abs{\frac{2h_{p,c}(\theta_i)}{\partial_{\theta_i}h_{p,c}(\theta_i)}}
    \label{eq:linearizable_parameter_differential_upper_bound}
\end{equation}
for $r=4$. We take $\eta_{p,c}(\theta_i)$ as the $\delta\theta_i$ that satisfies Eq.~\eqref{eq:linearizable_parameter_differential_condition}. As such, it may be interpreted as a loose upper bound for the elements $\delta\theta_i$ of linearizable parameter differentials. If some $\abs{\delta\theta_i}\gtrsim\eta_{p,c}(\theta_i)$, we would expect to generate nonlinear contributions to $\delta\mathcal F^{(p)}(\bm\theta|\bm\epsilon_\theta)$, negatively impacting the quality of our gradient estimates. This effect is studied in the appendix~\eqref{appendix:nonlinear_control_errors}.

\begin{figure}
    \centering
    \includegraphics[width=\linewidth]{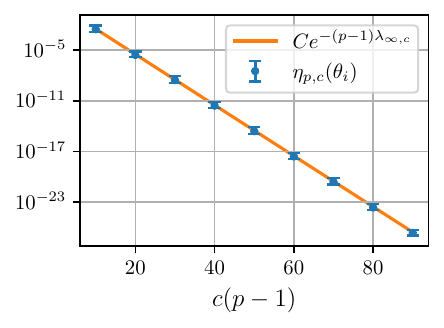}
    \caption{The rough lower bound $\eta_{p,c=10}(\theta_i)$ for non-linearizable perturbation vector elements $\abs{\delta\theta_i}$, aggregated over $10^4$ samples of $\theta_i$ drawn from a standard uniform distribution. The data shown is the median value, with the error bars indicating the IQR. The exponential decay of $\eta_{c, p}(\theta_i)$ in $p$ suggests that useful perturbation vectors are exponentially small in the circuit depth. The constant $C$ is dependent on how strictly one enforces the linearization condition $\abs{\delta\theta_ih_{p,c}(\theta_i)}\gtrsim\abs{(1/2)\delta\theta_i^2\partial_{\theta_i}h_{p,c}(\theta_i)}$ (and thus is dependent on superfluous details of the simulation). This is a purely classical effect.}
    \label{fig:eta_scaling}
\end{figure}

\begin{figure}
    \centering
    \includegraphics[width=\linewidth]{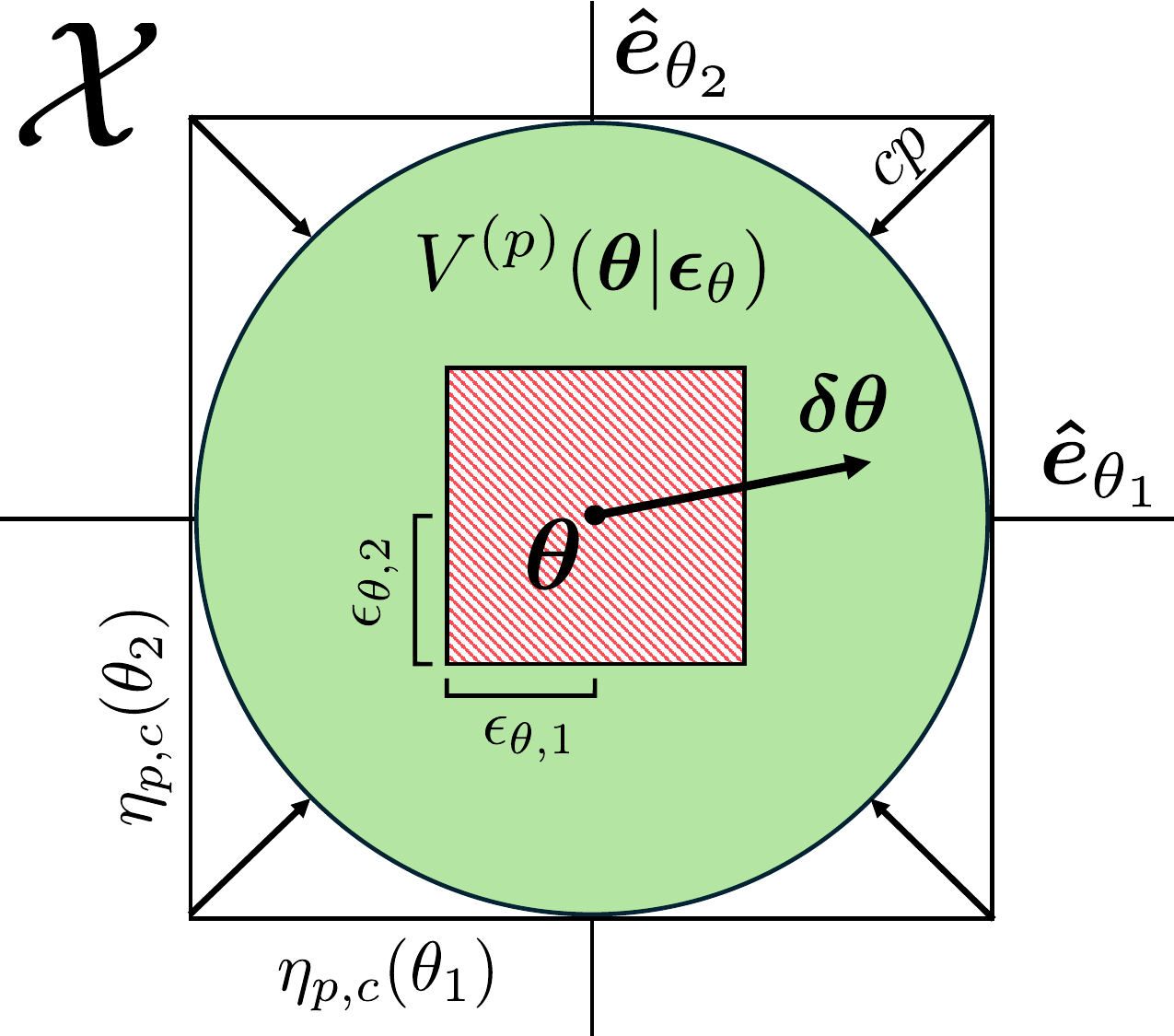}
    \caption{An illustration of the trainability deficit described in Sec.~\ref{subsec:nonlinearizability}, for $n_\theta=2$ parameters. In this figure, we are in the neighborhood of the parameter vector $\bm\theta$. The red region represents a ``forbidden zone'' which requires smaller perturbations $\bm{\delta\theta}$ than allowed by the precision $\epsilon_{\theta,i}$. The set $V^{(p)}(\bm\theta|\bm\epsilon_{\theta})$  (denoted by the green shaded region) represents a superset of the useful perturbations. The shape of its boundary is context-dependent, but its size is taken as $\eta_{p,c}(\theta_i)\sim e^{-(p-1)\lambda_{\infty,c}}$. This means $V^{(p)}(\bm\theta|\bm\epsilon_{\theta})$ shrinks exponentially fast as we transition to the ergodic limit, $c(p-1)\gg1$. It can even shrink into the forbidden zone, making useful perturbations $\bm{\delta\theta}$ virtually inaccessible.}
    \label{fig:trainability_decay}
\end{figure}

We empirically observe that $\expect{\eta_{p,c}(\theta_i)}\sim e^{-(p-1)\lambda_{\infty,c}}$ in Fig.~\ref{fig:eta_scaling}, indicating that pure QACOA's useful perturbation vectors (i.e., those conducive to linearization) live inside a contracting ball of radius $\sqrt{\sum_i\eta_{p,c}(\theta_i)^2}\sim e^{-(p-1)\lambda_{\infty,c}}$; this is an example of a precision loss effect induced by use of the classical chaotic map $l^c$. 
Additionally, recall that $\abs{\delta\theta_i}$ is lower bounded by the precision $\epsilon_{\theta,i}\leq\abs{\delta\theta_i}$, so effectively, the useful perturbation vectors $\bm{\delta\theta}$ are ``squeezed'' as $\epsilon_{\theta,i}\leq\abs{\delta\theta_i}\leq\eta_{p,c}(\theta_i)$. If we define $V^{(p)}(\bm\theta|\bm\epsilon_\theta)\equiv\sr{\bm{\delta\theta}\in\mathcal X|\epsilon_{\theta,i}\leq \abs{\delta\theta_i}< \eta_{p,c}(\theta_i)}$ as a superset of the useful perturbation vectors, then $\eta_{p,c}(\theta_i)$ decreasing gives $V^{(p+1)}(\bm\theta|\bm\epsilon_\theta)\subseteq V^{(p)}(\bm\theta|\bm\epsilon_\theta)$ by construction. Thus, the results of Fig.~\ref{fig:eta_scaling} indicate that the measure of $V^{(p)}(\bm\theta|\bm\epsilon_\theta)$ is typically non-increasing in $p$. That is, the useful perturbations $\bm{\delta\theta}$ are effectively squeezed into an exponentially shrinking region in $\mathcal X$. An illustration of this effect is shown in Fig.~\ref{fig:trainability_decay}. 

A consequence of this squeezing effect is that, in the ergodic limit where $\eta_{p,c}(\theta_i)\leq\epsilon_{\theta,i}$, the useful perturbation vectors become inaccessible to the classical optimizer. As a result, one cannot expect distinct results from those generated using random parameterization strategies. In particular, for the parameterization of interest, at large $c(p-1)$, parameters will eventually be drawn pseudorandomly according to an arcsine distribution; see Appendix~\ref{appendix:2d_logistic_map} for further elaboration. This effect may be mitigated by intelligent choice of the hyperparameters, a procedure which is built into many modern optimizers (e.g., SPSA); however, ultimately, the exponential scaling precision requirement will serve as a challenge for deep circuits. In particular, $F^{(p)}(\bm\theta|\bm\epsilon_\theta)$ will become Lipschitz discontinuous almost everywhere for large $p$ due to exponentially large differentials, explaining failure of the SPSA optimizer at large depths~\cite{maryak2001}. 
Note that the results of this section do not apply for a negligibly small set of points---notably those for which certain derivatives of the logistic map vanish (where LLEs and $\eta$ might diverge). However, we are concerned with broad QACOA performance and thus are not concerned with characterizing these sets.

\begin{figure*}
    \centering
    \includegraphics[width=\linewidth]{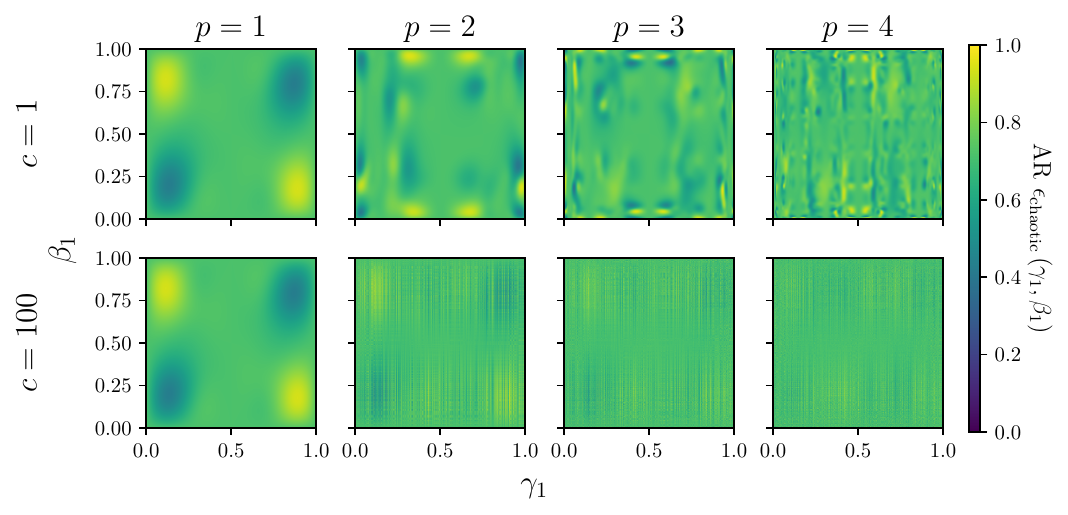}
    \caption{The mixing of the cost landscape associated with a random $N=8$ MAX 3-SAT problem of hardness $\alpha=1.125$, shown for map speeds $c=1,c=100$. In the $c(p-1)\gg1$ limit, global extrema become increasingly localized, and pure QACOAs can be expected to perform similarly in our simulations with finite $j_{\max}$.}
    \label{fig:overmixing}
\end{figure*}

\subsection{Resolution of extrema}

The squeezing of $V^{(p)}(\bm\theta|\bm\epsilon_\theta)$ is associated with the stretching of its elements as $\mathcal X$ is iteratively transformed. We may get a clearer picture of the stretching of these differentials produced on the landscape at time $p$ by considering the distribution of $\delta\mathcal F^{(p)}(\bm\theta|\bm\epsilon_\theta)$ generated by random parameter differentials $\bm{\delta\theta}\in V^{(p)}(\bm\theta|\bm\epsilon_\theta)$. Note that in this subsection we will have all differentials implicitly conditioned on the precision $\bm\epsilon_\theta$ for the sake of legibility. We will also explicitly note dependence on $\bm{\delta\theta}$ when appropriate.

We define the cumulative density function of the cost function differentials $\delta\mathcal F^{(p)}(\bm\theta,\bm{\delta\theta})$ as
\begin{equation}
\varphi^{(p,c)}(\Delta|\bm\theta)\equiv \Pr\br{\delta\mathcal F^{(p)}(\bm\theta,\bm{\delta\theta})<\Delta}.
\end{equation}
A distribution as $\varphi^{(p,c)}(\Delta|\bm\theta)\ll1$ might indicate high curvature near $\bm\theta$ and so estimates of the gradient computed near $\bm\theta$ by the SPSA optimizer may suffer from large error. This effect will impact QACOA performance at large depth, as is apparent in Fig.~\ref{fig:approximation_ratio_results_chaotic}.

In order to demonstrate how this differential distribution evolves for the parameterization $(I^2,l^c\times l^c,\{\mathrm{id}_{I^2}\})$, let us consider a small differential $\delta\mathcal F^{(p)}(\bm\theta,\bm{\delta\theta})$ generated by the perturbation $\bm{\delta\theta}=\sum_{i=1,2}\delta\theta_i\bm{\hat e}_{\theta_i}\in V^{(p)}(\bm\theta)$:
\begin{align}
    \delta\mathcal F^{(p)}(\bm\theta,\bm{\delta\theta})
    &\approx \abs{\sum_{i=1,2}s_{i}\delta\mathcal F^{(p)}(\bm\theta,\bm{\delta\theta}_i)},
\end{align}
where $s_{i}\equiv\mathrm{sgn}\br{\delta F^{(p)}(\bm\theta,\bm{\delta\theta}_i)}$ is the sign of the partial cost function differential, and thus has a fixed magnitude. We may then use the definition for the cost function LLEs, $\delta\mathcal F^{(p)}(\bm\theta,\bm{\delta\theta}_i)=\delta\mathcal F^{(1)}(\bm\theta,\bm{\delta\theta}_i)e^{(p-1)\lambda_{\Omega,i}^{(p,c)}(\bm\theta)}$, to write
\begin{align}
    \varphi^{(p,c)}(\Delta|\bm\theta)&\approx\Pr\br{\abs{\sum_{i=1,2}s_{i}\delta\mathcal F^{(1)}(\bm\theta,\bm{\delta\theta}_i)e^{(p-1)\lambda_{\Omega,i}^{(p,c)}(\bm\theta)}}<\Delta}
\end{align}
where $\delta\mathcal F^{(1)}(\bm\theta,\bm{\delta\theta}_i)$ is independent of $p$ and $c$. Our results for the parameterization $(I^2,l^c\times l^c,\{\mathrm{id}_{I^2}\})$ additionally indicate that $\lambda_{\Omega,i}^{(p,c)}(\bm\theta)\sim\lambda_{\infty,c}$ in the ergodic limit. So, for large $p, c$ we have
\begin{align}
    \varphi^{(p,c)}(\Delta|\bm\theta)\sim\Pr\br{\abs{\sum_{i=1,2}s_{i}\delta\mathcal F^{(1)}(\bm\theta,\bm{\delta\theta}_i)}<\Delta e^{-(p-1)\lambda_{\infty,c}}}.
    \label{eq:trainability_decay}
\end{align}
We see that $\varphi^{(p,c)}(\Delta|\bm\theta)\to0$ almost everywhere for nonzero $\Delta$ in the ergodic limit $c(p-1)\gg1$, since the left hand side of the inequality is in the set
\begin{equation}
    \sr{\abs{\delta\mathcal F^{(1)}(\bm\theta,\bm{\delta\theta}_1)-\delta\mathcal F^{(1)}(\bm\theta,\bm{\delta\theta}_2)},\sum_{i=1,2}\delta\mathcal F^{(1)}(\bm\theta,\bm{\delta\theta}_i)}
\label{eq:triangle_inequality_bounds}\end{equation}
by the triangle inequality and the reverse triangle inequality. Equation~\eqref{eq:triangle_inequality_bounds} does not depend on $p$, whereas $\Delta e^{-(p-1)\lambda_{\infty,c}}\to0$ exponentially decays.

The result of Eq.~\eqref{eq:trainability_decay} may be interpreted as such: almost everywhere on $\mathcal X$, our capability of resolving $\abs{\delta F^{(p)}(\bm\theta,\bm{\delta\theta})}\ll1$ (and thus extrema) is lost in the ergodic limit $c(p-1)\gg1$.  That is, in the ergodic limit, the optimizer loses the ability to create reliably small, linear differentials on the cost landscape. This implies that we can neither compute high-fidelity gradient estimates, nor can we reliably detect extrema $\delta\mathcal F^{(p)}(\bm\theta,\bm{\delta\theta}_i)\ll1$. These are the root causes of the trainability deficit that impacts pure QACOA performance at high depth.

We further complement our analytical findings with numerical results in Fig.~\ref{fig:overmixing}. Here, we show that the trainability deficit is attributed to a mixing of the cost landscape by action of the transformation $\mathfrak T$, for $\mathfrak T=l^c\times l^c$. We focus specifically on an $N=8$ instance of \maxsat3. In the $p=1$ panels, macroscopic valleys are well-defined on the landscape, but, however, they narrow as $c(p-1)$ increases due to the action of the chaotic map---that is, rapid expansion of differentials with significant nonlinear contributions. Eventually, global extrema can locally narrow to width $\lesssim\bm\epsilon_\theta$ on $\mathcal X$ (smaller than the precision), after which they can no longer be reliably detected. This is illustrated in Fig.~\ref{fig:overmixing} by the significant localization of extrema as $c(p-1)$ increases.

Note that the QACOA result shown in Eq.~\eqref{eq:trainability_decay} contrasts with that of standard QAOA, in the sense that deep-circuit QAOA trainability suffers from barren plateaus (exponentially \emph{diminishing} gradients)~\cite{barren_plateaus,blekos2024review}, whereas QACOA trainability suffers from exponentially \textit{large} gradients. For this reason, we expect alternative forms of QACOA to more aptly balance trainability deficits, while still offer the advantage of a reduction in the number of training parameters. Below, we briefly discuss two such examples, but note that this is a rich space for future exploration.

\begin{figure}[t]
    \centering
    \includegraphics[width=\linewidth]{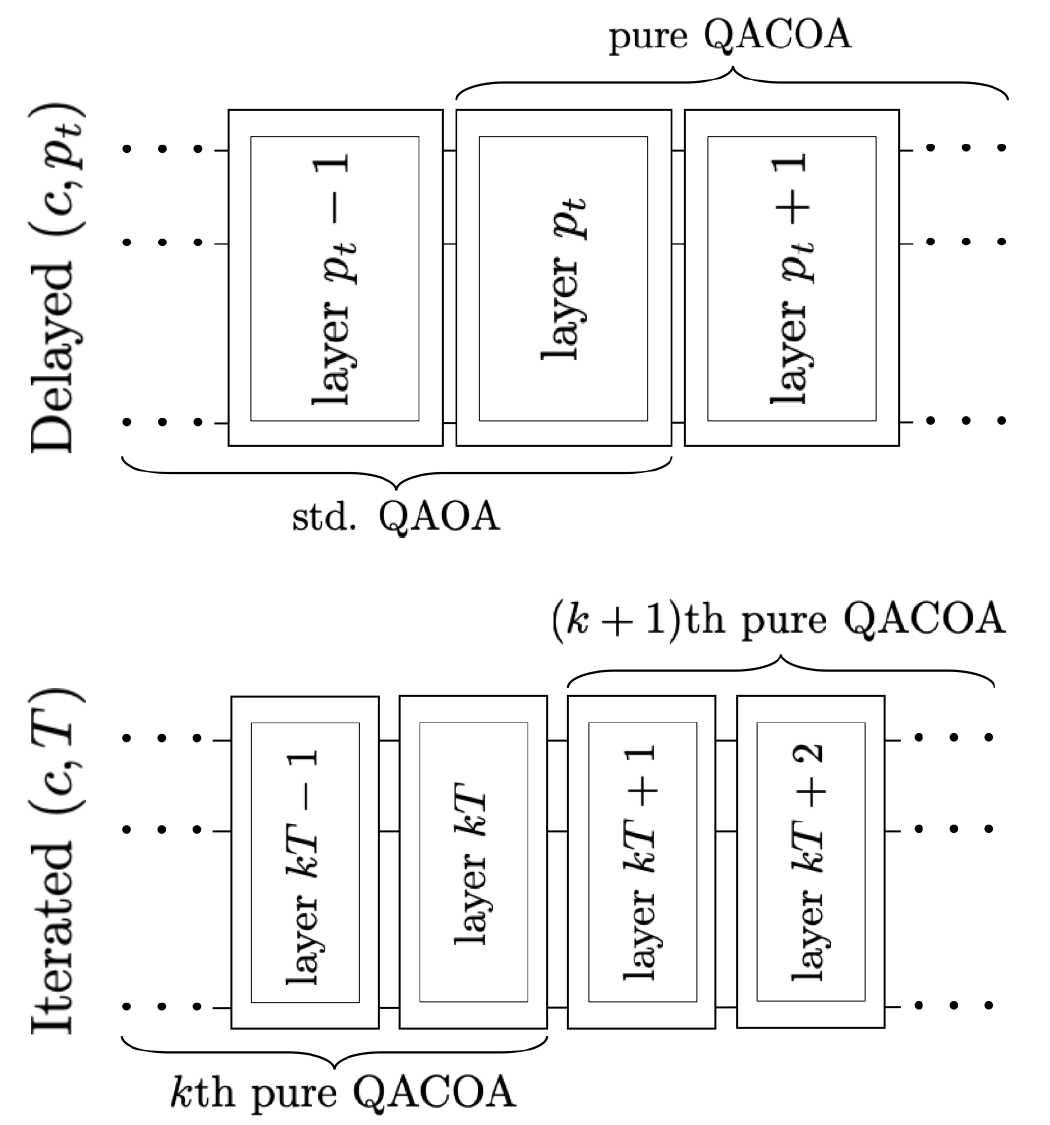}
    \caption{Depictions of the hybrid QACOA structures discussed in Sec.~\ref{sec:qacoa_hybridization}. Each pure QACOA referenced in the figure is described by the $(I^2,l^c\times l^c,\{\mathrm{id}_{I^2}\})$ parameterization. For the delayed hybrid variant, the first $p_t$ layers are equivalent to that of a standard QAOA circuit, before switching to a pure QACOA-like segment with the iterated logistic map $l^c$ as our primitive transformation on the layer $p_t$'s parameters. This circuit will have $2p_t=\mathcal{O}(1)$ parameters total. For the iterated hybrid variant, one concatenates independent pure QACOAs of depth $T$; this circuit has $2(\lfloor(p-1)/T\rfloor+1)\leq 2p$ parameters, with equality at $T=1$ (which recovers std. QAOA).}
    \label{fig:hybrid_diagram}
\end{figure}

\section{QACOA hybridization}\label{sec:qacoa_hybridization}
The reduction in QACOA performance due to exponentially large gradients begs the question of how QACOA may be mended to perform well for deep circuits. We present a simple means by which \emph{pure} QACOAs (as implemented in this work) may be turned into \emph{hybrid} QACOAs that (1) leverage aspects of standard and chaotic parameterizations and (2) can outperform pure variants for deeper circuits. We will discuss two hybrid QACOAs variants constructed by concatenating standard and pure QACOAs as introduced in Sec.~\ref{sec:parameterization}. Illustrations of the hybrid QACOA circuit structures are shown in Fig.~\ref{fig:hybrid_diagram}.
\begin{figure}
    \centering
    \includegraphics[width=\linewidth]{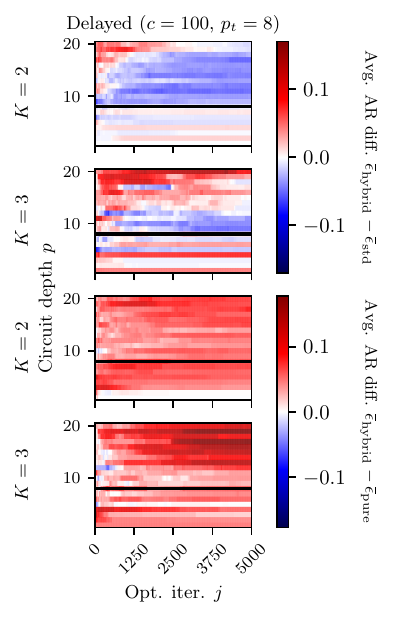}
    \caption{Density plots showing the performance boost $\overline\epsilon_\mathrm{hybrid}-\overline\epsilon_{\mathrm{std},\mathrm{pure}}$ gained from the use of the delayed hybrid variant, with respect to results from standard QAOA (first two rows) and pure QACOA implementations (last two rows). A single \maxsat2 and \maxsat3 problem is selected near the respective critical clause density, and is optimized at depth $p\leq20$ for up to $j_{\max}=5000$ iterations to compute ARs $\overline\epsilon_\mathrm{hybrid},\overline\epsilon_\mathrm{std},\overline\epsilon_\mathrm{pure}$. The black line denotes $p_t=8$, past which the number of variational parameters is fixed at $2p_t=16$. This hybrid can outperform pure QACOA and standard QAOA at a low number of optimization iterations $j$; this fact can be exploited for maximal performance by using the hybrid to ``jumpstart'' optimization at a large depth, before transitioning to a ``seeded'' standard QAOA optimization at large $j$.}
    \label{fig:delayed_hybrid_results}
\end{figure}

We refer to the first hybrid as the ``delayed'' variant. The delayed hybrid takes the phase space as that of standard QAOA at a truncated depth $p_t\leq p$, which limits the total number of independent parameters to $n_\theta=2p_t=\mathcal{O}(1)$. In the case that $p_t$ is chosen to be $p_t\geq p$, the entire circuit will be equivalent to a standard QAOA. Otherwise, the first $p_t$ circuit layers have their parameters generated as standard QAOA, while the rest of the circuit has its parameters generated as pure QACOA with angles generated by the $p_t$th layer angles via application of $l^c$. 
Formally, the parameterization is given by $(\mathcal X, \mathfrak T, \Phi)=(I^{2p_t}, \mathrm{id}_{I^{2p_t}}, \{\phi_m\})$, with $\bm\theta=(\gamma_1,\beta_1,\dots,\gamma_{p_t},\beta_{p_t})$ and $\phi_m(\bm\theta) = (l^{c\pr{\max(m,p_t)-p_t}}(\gamma_{\min(m,p_t)}), l^{c(\max(m,p_t)-p_t)}(\beta_{\min(m,p_t)}))$. 

\begin{figure}
    \centering
    \includegraphics[width=\linewidth]{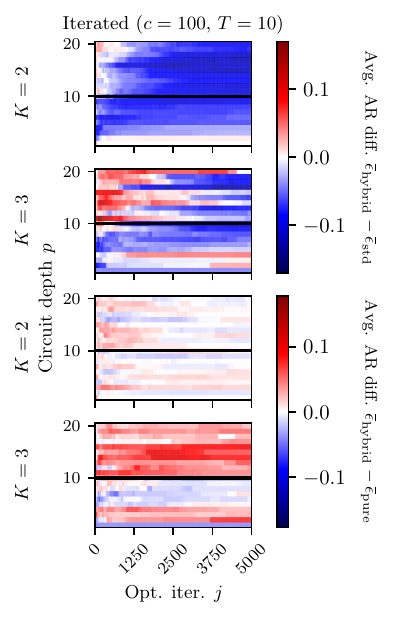}
    \caption{Density plots showing the iterated hybrid performance with respect to standard QAOA and pure QACOA results, using the same \maxsat K problems and $\overline\epsilon_\mathrm{std}$, $\overline\epsilon_\mathrm{pure}$ results as in Fig.~\ref{fig:delayed_hybrid_results}. The black line denotes $T=10$, the depth at which a new pure QACOA subcircuit is introduced. Although this iterated hybrid's performance does not measure up to that of the delayed hybrid, it still may be used to boost solution quality at large circuit depth, which is especially notable in the $K=3$ results.}
    \label{fig:iterated_hybrid_results}
\end{figure}
The second hybrid that we examine is referred to as the ``iterated'' variant. Here, we concatenate pure QACOAs of depth $T$, gaining two parameters per $T$ circuit layers. This gives $n_\theta=2(\lfloor(p-1)/T\rfloor+1)$ total parameters. Although this hybrid's $n_\theta$ scales linearly with $p$ (just as standard QAOA does), choosing $T\geq2$ results in an overall reduction in the number of parameters, and thus may be advantageous for deep circuits. The formal parameterization is given by $(\mathcal X, \mathfrak T, \Phi)=(I^{n_\theta},\mathrm{id}_{I^{n_\theta}},\{\phi_m\})$ with $\bm\theta=(\gamma_1,\beta_1,\dots,\gamma_{i_{p,T}},\beta_{i_{p,T}})$ and $\phi_m(\bm\theta)=(\tilde l^{m,c,T}(\gamma_{i_{m,T}}),\tilde l^{m,c,T}(\beta_{i_{m,T}}))$ where $\tilde l^{m,c,T}\equiv l^{c((m-1)\,\mathrm{mod}\,T)}$, $i_{m,T}\equiv\lfloor(m-1)/T\rfloor+1$.

We use the delayed and iterated hybrid QACOAs to optimize two randomly generated $N=8$ MAX 2-SAT and MAX 3-SAT problem instances of hardness $\alpha=1.125, 4.125$ respectively, for up to circuit depth $p=20$. We choose $p_t=8$ for the delayed variant ($n_\theta\leq16$) and $T=10$ for the iterated hybrid ($n_\theta\leq4$ for $p\leq20$). We also perform a standard QAOA optimization as a control for the same problems. The optimizer parameters remain the same as in Sec.~\ref{sec:numerical_comparison}, and our averages correspond to five independent optimizations. We choose a map speed $c=100$ where relevant. 

In Figs.~\ref{fig:delayed_hybrid_results} and \ref{fig:iterated_hybrid_results}, the gain in the average AR (with respect to standard QAOA) is shown as a function of the circuit depth and the optimization iteration, with black lines indicating $p_t$ and multiples of $T$. We find that these hybrid QACOAs are subject to a significant performance boost compared to pure QACOA [Fig.~\ref{fig:approximation_ratios_over_j_c_fixed_problem}].
Importantly, we find that, for a limited number of training iterations and larger circuit depth, hybrid QACOA can outperform standard QAOA. This is particularly evident in the $K=3$ results for $j\lesssim10^3$ optimization iterations. As the trainability of QAOA decreases in $p$, hybrid QACOA performance remains fairly consistent for the depths tested.

Notably, there is a reduction in the delayed QACOA hybrid performance at intermediate depth $p$ greater than but close to $p_t$, and in the iterated hybrid performance where $(m-1)\,\mathrm{mod}\,T$ is close to $T$ (with $T$ sufficiently large). Thus, while the hybrid QACOAs studied here convey performance improvements over pure QACOA, alternative hybridizations must be considered to consistently outperform standard QAOA. It will likely require a careful balancing between standard and chaotic mappings to maintain high ARs as circuit depth increases.

For completeness, we include a direct comparison between pure QACOA and the hybrid approaches in Figs.~\ref{fig:delayed_hybrid_results} and \ref{fig:iterated_hybrid_results} as well. We find that the iterated hybrid and pure QACOA perform similarly at $p\leq T$ (as expected by construction), whereas we see some improvement with respect to pure QACOA beyond $T$; however, there exists some dependence on the problem structure. Utilizing the delayed hybrid QACOA variant results in more favorable ARs for all $p$ values. Unlike the comparison to standard QAOA, we do not observe a strong dependent on the number of optimization iterations. Nevertheless, the hybrid approaches can indeed offer higher performance than pure QACOA and standard QAOA.

\section{Conclusion}
In this study, we have introduced a general framework for incorporating classical chaotic optimization into QAOA; referred to as QACOA. This framework enables a broad class of functional parameterizations for QAOA that can drastically reduce the number of required training parameters. Crucially, we show that such reductions can be achieved while still maintaining performance on-par with canonical QAOA implementations.

Our study highlights two particular instantiations of QACOA. The first, referred to as pure chaotic QACOA, enables a factor of $p$ reduction in QAOA training parameters. However, the substantial reduction is complemented by challenges in ansatz trainability at large circuit depths. We provide a rigorous analysis of these results via the theory of measure-preserving dynamical systems. This enables a classical description of QACOA's phase space dynamics that affords analytical insights into the stability of QACOA training as a function of circuit depth. 

While pure chaotic QACOA is shown to offer expansive searching of the cost landscape, it suffers from exponentially increasing gradients beyond a critical circuit depth. Notably, this is a contrasting phenomenon from that observed in standard QAOA, where barren plateaus (i.e., vanishing gradients) emerge. Nevertheless, we show that it is possible to circumvent this challenge by introducing hybrid approaches that leverages both the standard and chaotic maps. In addition to offering improved trainability, this modification yields enhanced performance that is comparable to---in some instances better than---standard QAOA under fixed classical resources.

In this study, we focus on phase-space transformations that are a product of decoupled transformations that incorporate well-studied classical chaotic maps. However, there is rich potential for future work in alternate parameterizations and hybridized circuit structures within our framework. One may even consider alternative QAOA ansatze based on our results. For example, one can envision leveraging the quick convergence of hybrid QACOAs to quickly search for ``good'' initial parameters for pure QAOAs. As such, we believe the methods introduced in this manuscript can illuminate novel methods of control for variational quantum algorithms that address algorithmic performance concerns associated with the standard QAOA paradigm.

\section{Acknowledgments}
This work was supported in part by the U.S. Department of Energy, Office of Science, Office of Advanced Scientific Computing Research under Award Number DE-SC0024163.

\appendix

\section{SPSA optimizer}\label{appendix:spsa_optimizer}
In this work, we utilize an implementation of the SPSA optimizer as described and contextualized in Refs.~\cite{SPALL1997109,spall_spsa_implementation}. The update to the parameter vector $\bm\theta_j$ at the $j$th optimization iteration is given by $-a_j\hat{\mathcal D}\pr{F^{(p)}(\bm\theta_j)}$, where $\hat{\mathcal D}$ denotes a numerical estimate and $a_j\equiv a/(A+j+1)^\alpha$ is a \emph{gain coefficient}. The numerical gradient estimate is computed by perturbing the cost function $F^{(p)}(\bm\theta_j)$ by $\bm{\delta\theta}_j=c_j\sum_{i=1,2}s_i\hat{\bm e}_{\theta_i}$, where $s_i=\pm1$ is randomly chosen with equal probability and $c_j=c/(j+1)^\gamma$ is another gain coefficient. 

To ensure the optimizer works reasonably well, we must choose a reasonable set of hyperparameters $a, A, \alpha, c_0, \gamma$. We use the recommended values $\alpha=0.602$ and $
\gamma=0.101$. Accordingly, we choose $A = j_{\mathrm{max}}/100$, following the guideline that it should not exceed 10\% of the maximum number of iterations~\cite{spall_spsa_implementation}. The parameter $a$ must then be chosen such that the magnitude of the smallest update in the variational parameters early in the optimization is ``reasonable" given the ranges of said parameters. Since our parameters are normalized angles in $[0, 1]$, we target this early small update to be as $\Delta\theta_\mathrm{min}\sim0.01$; note that this does not prevent finer tuning of the angles as optimization progresses. We let $a\equiv\Delta\theta_\mathrm{min}\times(A+1)^\alpha/\tilde g$, with $\tilde g\equiv e^{\overline{\log\abs{\hat{\mathcal D}(F^{(p)}(\bm\theta_1))_{\ell}}}}$; the overline here denotes an average over the components (labeled by $\ell$) of the gradient estimate $\hat{\mathcal D}(F^{(p)}(\bm\theta_1))$. $\tilde g$ characterizes the average scale of the cost function gradients at $\bm\theta_1$ by taking the geometric mean of its nonzero elements. Defining $a$ as such will give early updates to the angles on the order of $0.01$, enabling SPSA to make larger yet reasonable updates to the parameters $\bm\theta_j$ early in optimization ($j$ small), while more precise tuning occurs as the algorithm progresses. Lastly, for the hyperparameter $c_0$, we choose $c_0=0.1$ in order to promote an initially aggressive probing of the cost landscape at small $j$.

Throughout the work, we keep these SPSA hyperparameters fixed, in the interest of avoiding tailoring the optimizer towards certain parameterization schemes. However, the results of Sec.~\ref{subsec:nonlinearizability} suggest that the gain coefficients should have an $\mathcal O(e^{-\lambda_{\infty,c}(p-1)})$ scaling to produce high-fidelity gradient estimates at arbitrary depth for the pure QACOA presented in this study. As the scaling of $c_j$ over $j$ follows a power law, we have $\eta_{c,p}(\theta_i)<c_{j}$ almost everywhere for all $j=1,\dots,j_{\max}$, for sufficiently large $p$. With $p$ fixed, SPSA-trainability (i.e., the $c_{j}\lesssim\expect{\eta_{c,p}(\theta_i)}$ regime) can be recovered by increasing $j$. If we use the result $\expect{\eta_{c,p}(\theta_i)}=Ce^{-\lambda_{\infty,c}(p-1)}$ for the pure QACOA parameterization, the SPSA-trainable regime is given by $j\gtrsim\mathcal O(c_0^{1/\gamma}e^{(p-1)\lambda_{\infty,c}/\gamma})$. This insight serves as a possible explanation for pure QACOA's performance deficits at large depths, which is especially evident in Fig.~\ref{fig:performance_saturation}: in these results, we have not reached the SPSA-trainable regime for deep circuits. The critical depth $\tilde p$, as identified in Sec.~\ref{subsec:qacoa_performance_function_of_circuit_depth}, is thus an indication of the transition out of the SPSA-trainable regime at fixed $j$. Adjusting the SPSA gain sequences accordingly is a straightforward means of tailoring the optimizer for additional QACOA-robustness.

\begin{figure}[t]
    \centering
    \includegraphics[width=\linewidth]{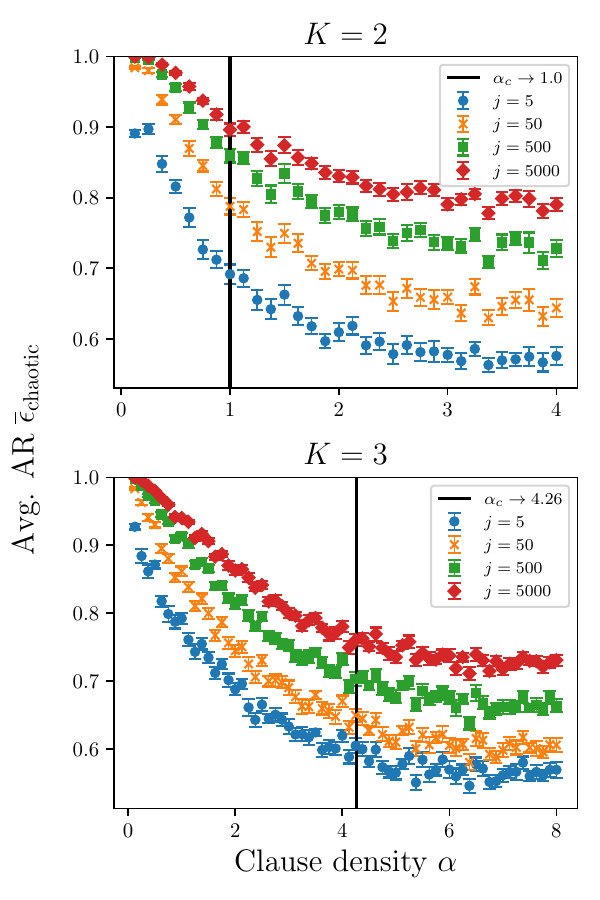}
    \caption{Mean pure QACOA solution qualities as a function of the classical problem hardness $\alpha$, shown for a few choices of $j\leq j_{\max}=5000$, the optimization iteration. Error bars denote the standard error of the mean. $25$ $N=8$ variable MAX $K$-SAT are generated per $\alpha$ up to a maximum value dependent on $K$. QACOA's performance over the clause density aligns with that of standard QAOA as reported in other works~\cite{Akshay_2020}.}
    \label{fig:qacoa_performance_over_clause_density}
\end{figure}

\section{QACOA performance as a function of problem hardness}\label{appendix:qacoa_performance_function_of_problem_hardness}
In \maxsat K problems, the typical solution quality achieved by QAOA has previously been shown to be strongly dependent on the hardness (i.e., clause density $\alpha$~\cite{Akshay_2020}). In this section, we investigate QACOA's performance as a function of $\alpha$ for a fixed depth ($p=8$), under the pure parameterization $(I^2,l^c\times l^c,\{\mathrm{id}_{I^2}\})$. In order to do so, we randomly generate $25$ random $N=8$ MAX $K$-SAT problems per $\alpha$ for $\alpha=1/N,2/N,\dots$, up to $\alpha=4$ for $K=2$ and $\alpha=8$ for $K=3$. We minimize the energy of the cost Hamiltonian using the same SPSA parameters described in previous sections, with $j_{\max}=5000$ optimization iterations; the results are shown in Fig.~\ref{fig:qacoa_performance_over_clause_density}. We find that QACOA's performance broadly decreases as the classical problem hardness increases, a result that aligns with the conclusions drawn from prior works.

\section{Nonlinear control errors}\label{appendix:nonlinear_control_errors}
Many gradient-based optimizers rely on the assumption that the cost function is locally linear, $F^{(p)}(\bm\theta+\bm{\delta\theta})\approx F^{(p)}(\bm\theta)+\bm{\delta\theta}\mathcal DF^{(p)}(\bm\theta)$, to make high-fidelity estimates of the gradient. As such, extreme nonlinearities near $\bm\theta$ may degrade the optimizer's gradient estimates, negatively impacting the optimization as a whole. In Sec.~\ref{sec:trainability_decay}, we showed that the chaotic parameterization can lead to significant local nonlinearities, most notably in the ergodic limit.

Here, we study the characteristics of these nonlinearities---which we refer to as nonlinear control noise---accumulated by perturbing a trajectory $\bm\theta$ in the chaotic parameterization $(I^2,l^c\times l^c,\{\mathrm{id}_{I^2}\})$. Specifically, we will investigate their stationarity. We note that stationary control errors have been previously studied in the context of QAOA~\cite{quiroz2025qaoa}. Here, we show that control errors generated by nonlinearities lead to unique, nonstationary error profiles.

We make a similar argument as in Sec.~\ref{sec:trainability_decay} to restrict our analysis to errors in the phase space $\mathcal X$, due to the large impact of the parameterization on the cost function's gradients. That is to say that nonlinearities in $F^{(p)}(\bm\theta)$ are dominated by those generated by the control $f_m(\bm\theta),g_m(\bm\theta)$ for the parameterization of interest. Since $g_m$ is defined similarly to $f_m$, we focus on the latter, as the results of this section will be similar. The exact expression for the $m$th layer angles at a perturbed trajectory $\bm\theta+\bm{\delta\theta}$ is given by $(f_m(\bm\theta+\bm{\delta\theta}),g_m(\bm\theta+\bm{\delta\theta}))=(\phi_m\circ\mathfrak T^{m-1})(\bm\theta+\bm{\delta\theta})$, Taylor expanded as
\begin{align}
(f_m(\bm\theta+\bm{\delta\theta}),g_m(\bm\theta+\bm{\delta\theta}))&=(\phi_m\circ\mathfrak T^{m-1})(\bm\theta)\nonumber\\
&\phantom=+\bm{\delta\theta}\,\mathcal D\pr{(\phi_m\circ\mathfrak T^{m-1})(\bm\theta)}\nonumber\\
&\phantom=+\xi_{m}(\bm\theta,\bm{\delta\theta}).%\\
\end{align}
$\bm{\delta\theta}\,\mathcal D\pr{(\phi_m\circ\mathfrak T^{m-1})(\bm\theta)}+\xi_{m}(\bm\theta,\bm{\delta\theta})$ is an exact differential, with $\bm{\delta\theta}\,\mathcal D\pr{(\phi_m\circ\mathfrak T^{m-1})(\bm\theta)}$ being linear in the perturbation $\bm{\delta\theta}$, and $\xi_{m}(\bm\theta,\bm{\delta\theta})$ denoting a sum of nonlinear terms. As a large $\xi_{m}(\bm\theta,\bm{\delta\theta})$ generally coincides with significant nonlinear terms in an exact expansion of $F^{(p)}(\bm\theta+\bm{\delta\theta})$, we take $\xi_{m}(\bm\theta,\bm{\delta\theta})$ as a kind of control noise only affecting the optimizer.

\begin{figure}[t]
    \centering
    \includegraphics[width=\linewidth]{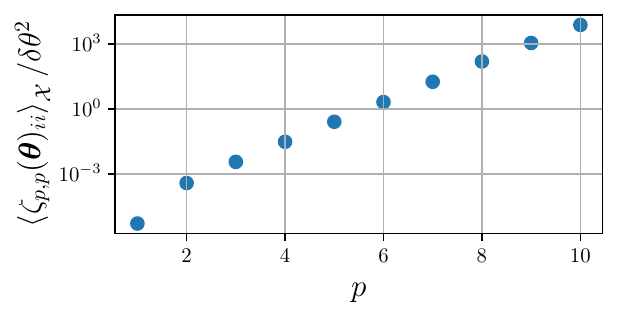}
    \caption{The average second moment of the noise on control $i=1$, $\zeta_{p,p}(\bm\theta)_{ii}$, over the phase space $\mathcal X$. For pure QACOA ($\mathfrak T=l^c\times l^c$ with $c=1$), the $i=2$ results will be identical. We fix and normalize by $\delta\theta=10^{-18}$ a perturbation scale factor that SPSA might use. Perturbation vector elements generating the control noise are drawn as $\delta\theta_i\in\{\pm\delta\theta\}$ with equal probability. $\bm\theta$ is randomly sampled $10^4$ times to estimate the average $\expect{\zeta_{p,p}(\bm\theta)_{ii}}_{\mathcal X}$. We observe that $\expect{\zeta_{p,p}(\bm\theta)_{ii}}_{\mathcal X}$ grows exponentially in $p$, the circuit depth. At large $p$, one sees $\zeta_{p,p}(\bm\theta)_{ii}\gg\delta\theta^2$ (i.e., the control noise becomes dominant).}
    \label{fig:average_control_noise_second_moment}
\end{figure}

We define the correlation function for the noise as
\begin{equation}
\zeta_{m_1,m_2}(\bm{\theta})\equiv\expect{\xi_{m_1}(\bm\theta,\bm{\delta\theta})^\top\xi_{m_2}(\bm\theta,\bm{\delta\theta})}_{\delta\mathcal X},
\label{eq:autocorrelation}\end{equation}
where $\expect\cdot_{\delta\mathcal X}$ denotes a classical average over $\delta\mathcal X(\bm\theta)$, the perturbations $\bm{\delta\theta}$ produced by the optimizer at $\bm\theta$ in a single iteration (assuming edges of $\mathcal X$ are ignored, i.e., that $\bm\theta+\bm{\delta\theta}\in\mathcal X$). As suggested in Ref.~\cite{spall_spsa_implementation}, our implementation of SPSA draws $\delta\theta_i\in\{\pm\delta\theta\}$ from a Bernoulli distribution with equal probability of $1/2$, so Eq.~\eqref{eq:autocorrelation} 
can be computed exactly. $\zeta_{m_1,m_2}(\bm\theta)$ is a $2\times 2$ matrix with elements $\zeta_{m_1,m_2}(\bm\theta)_{ij}$ interpreted as the cross-correlation (autocorrelation) function for the control noise when $i\neq j$ ($i=j$); the $i=j$, $m_1=m_2=p$ case gives a second moment of the $\gamma$/$\beta$ control noise at depth $p$. In Fig.~\ref{fig:average_control_noise_second_moment}, we showcase estimates for the mean of the second moment $\zeta_{p,p}(\bm{\theta})_{ii}$ over the phase space $\mathcal X$, under the $l^c\times l^c$ control ($c=1$). These results indicate that our control noise $\xi_p(\bm\theta,\bm{\delta\theta})$ is nonstationary since $\zeta_{m_1,m_2}(\bm\theta)$ cannot be written in terms of the lag $m_2-m_1$; its variance is exponentially growing with $p$. These exponentially large errors are a direct cause of the trainability deficit described in Sec.~\ref{sec:trainability_decay}; hence, the correspondence between the exponentially large $\zeta_{p,p}(\bm{\theta})_{ii}$ and the exponentially small $V^{(p)}(\bm\theta|\bm{\epsilon}_\theta)$ described in Sec.~\ref{subsec:nonlinearizability}.

\section{Decoupled logistic map system}\label{appendix:2d_logistic_map}
\subsection{Dynamical system description}
Now we show how the dynamical system representation can be utilized to examine pure QACOA. The description outlined here is effective for building a ``simple'' QACOA, where transformations are decoupled and invariant densities are already known. The version of QACOA that we implement has just two independent parameters, $x=\gamma_1,y=\beta_1$. For the sake of notational convenience, we work with these parameters defined on the unit interval; hence, our definitions for the cost and mixer unitaries include factors of $2\pi$ and $\pi$, respectively, to conform with typical QAOA angle ranges~\cite{farhi2014quantum}. Under these specifications, our sample space is then $\mathcal X=I^2$. Now, let us define the transformation $\mathfrak T:\mathcal X\to\mathcal X$ as $\mathfrak T\equiv \mathcal T_1\times\mathcal T_2=l^c\times l^c$ where $l:I\to I$ is the $r=4$ logistic map.

\subsection{Invariant density derivation}
We will now outline the derivation for the invariant density, which defines the measure $\mu$ associated with the parameterization $(I^2,l\times l, \{\mathrm{id}_{I^2}\})$. We have let $c=1$ here for simplicity but note that the results hold for general $c$. We solve for the density $\rho:\mathcal X\to\reals$, such that $\mathfrak T=l\times l$ is $\mu$-preserving, by first computing the preimage of $B=(x_1,x_2)\cup(y_1,y_2)\stackrel{\mathrm{def}}\subset \mathcal X$:
\begin{equation}
\mathfrak T^{-1}B = k(x_1,x_2)\times k(y_1,y_2)\subset\mathcal X,
\end{equation}
with
\begin{align}
    k(z_1,z_2)&\equiv l_1(z_1,z_2)\cup l_2(z_1,z_2),\\
    l_1(z_1,z_2)&\equiv(h_-(z_1), h_-(z_2)),\\
    l_2(z_1,z_2)&\equiv(h_+(z_2), h_+(z_1)),\\
    h_\pm(z) &\equiv\frac{1\pm\sqrt{1-z}}{2}=\mathcal T_i^{-1}(z).
\end{align}
Here, $l_i(z_1,z_2)$ are subintervals of $I$ and $h_\pm(z_i)$ are their endpoints. The expressions above come from solving $z=\mathcal T_i(h_\pm(z))$ for $h_\pm(z)$. Note that $0\leq h_-(z)\leq1/2\leq h_+(z)\leq1$ for any $z\in I$, so $l_1, l_2$ are disjoint. 
The measure of $B$'s preimage is then given by
\begin{align}
    \mu(\mathfrak T^{-1}B)&=\mu(k(x_1,x_2)\times k(y_1,y_2))\nonumber\\
    &=\sum_{i,j\in\{1,2\}}\mu(l_i(x_1,x_2)\times l_j(y_1,y_2))
    \label{eq:logistic_equation_probability}
\end{align}
since $l_1(z_1,z_2)\cap l_2(z_1,z_2)=\emptyset$ for any choice of $z_i\in I$. 

Although we previously defined $B$ as an open set on $\mathcal X$, the same result follows for closed/semi-closed sets since points and line segments have measure zero. For $\mathfrak T$ to be $\mu$-preserving, we require that Eq.~(\ref{eq:logistic_equation_probability}) is equal to $\mu(B)$, and thus,
%\small
\begin{align}
   \pr{\int_{x_1}^{x_2}\int_{y_1}^{y_2}-\sum_{i,j\in\{1,2\}}\int_{l_i(x_1,x_2)}\int_{l_j(y_1,y_2)}}dxdy\,\rho(x,y)=0.
    \label{eq:preserved_measure_def}
\end{align}
Applying $\partial_{x_i}\partial_{y_j}$ (for any choice of $i,j\in\{1,2\}$) to Eq.~(\ref{eq:preserved_measure_def}) yields the following equation for candidate densities $\rho:\mathcal X\to\reals$:
\begin{equation}
\rho(x,y)=\br{\frac{1}{4\sqrt{1-x}}}\br{\frac{1}{4\sqrt{1-y}}}\sum_{i,j\in\{+,-\}}\rho(h_i(x),h_j(y)).
\label{eq:invariant_density_requirement_a}
\end{equation}

By construction of the transformation $\mathfrak T=l\times l$, we have $x, y$ decoupled, and consequently, we seek solutions of the form $\rho(x, y)=\rho_l(x)\rho_l(y)$ (product states, with $\rho_l:I\to\reals$). Using this fact in Eq.~(\ref{eq:invariant_density_requirement_a}) gives the recursive equation
\begin{align}
\rho_l(z)&=\frac{\rho_l(h_+(z))+\rho_l(h_-(z))}{4\sqrt{1-z}}\\
&=\frac{dh_-}{dz}\rho_l(h_-(z))-\frac{dh_+}{dz}\rho_l(h_+(z)).
\label{eq:invariant_density_requirement_b}
\end{align}
The single-parameter invariant density $\rho_l$ \textit{must} satisfy this equation. Using $h_\pm(z)$ previously specified, we have the solution
\begin{align}
    \rho_l(z)&=\frac1{\pi\sqrt{z(1-z)}}\\
    \rho(x,y)&=\rho_l(x)\rho_l(y)
\end{align}
with support on $I,I^2$ respectively. The factor $1/\pi$ ensures normalization ($\mu$ is a probability measure). $\rho_l(z)$ is known as the invariant density for the $r=4$ logistic map~\cite{Falk1984}.

\section{Phase space LE expressions}\label{appendix:phase_space_le_expressions}
In this section, we discuss the LEs associated with the parameterization $(I^2,l\times l,\{\mathrm{id}_{I^2}\})$. We choose $A^{(p)}(\bm\theta)=\mathfrak T^{p-1}\pr{\bm\theta}$, the evolved parameter vector at depth $p$. The derivative of $A$ is then given by, for $p>1$,
\begin{equation}
    \partial_{\theta_i}A^{(p)}(\bm\theta)=\bm{\hat e}_{\theta_i}\mathcal{D}\mathfrak T^{p-1}\pr{\bm\theta}\label{eq:phase_space_gradient}
\end{equation}
where
\begin{equation}
    \mathcal{D}\mathfrak T^{p-1}(\bm\theta)=\mathcal{D}\mathfrak T^{p-1}\pr{\mathfrak T^{p-2}\pr{\bm\theta}}\circ\dots\circ \mathcal{D}\mathfrak T^{1}\pr{\mathfrak T^{0}\pr{\bm\theta}}
    \label{eq:decomposed_transformation_derivative}
\end{equation}
and $\mathcal{D}\mathfrak T^{0}(\bm\theta)$ is the identity operator. $\mathcal{D}\mathfrak T^{p-1}\pr{\bm\theta}$ is block-diagonal with shape $n_\theta\times n_\theta$, with each individual block corresponding to a set of primitive transformations $\mathcal T_i$ decoupled from the rest; its elements are given by $(\mathcal{D}\mathfrak T^{p-1}\pr{\bm\theta})_{ij}=\partial_{\theta_i}\mathfrak T^{p-1}\pr{\bm\theta}_j$. This gives the differential of the transformed parameter vector as
\begin{equation}
\delta \mathfrak T^{p-1}\pr{\bm\theta}\approx \bm{\delta\theta}\,\mathcal{D}\mathfrak T^{p-1}\pr{\bm\theta}.
\label{eq:parameter_vector_differential}
\end{equation}
This result assumes that $\bm{\delta\theta}\in V^{(p)}(\bm\theta|\bm{\epsilon}_\theta)$ in making the linear approximation.
The corresponding phase space LLE, denoted by $\lambda_\theta^{(p)}$, for general QACOA may be computed as given in Eq.~(\ref{eq:lyapunov_exponent}):
\begin{equation}
    \lambda^{(p)}_\theta(\bm\theta)=\frac1{p-1}\ln\frac{\aabs{\bm{\delta\theta}\mathcal{D}\mathfrak T^{p-1}\pr{\bm\theta}}}{\aabs{\bm{\delta\theta}}}
\end{equation}
For the version of QACOA implemented in this work [Eq.~(\ref{eq:parameterization_scheme})], we use $n_\theta=2$ with $\mathfrak T=l^c\times l^c$, so $\mathcal D\mathfrak T^{p-1}\pr{\bm\theta}$ is a diagonal $2\times 2$ matrix given by
\begin{equation}
\begin{split}
    \mathcal D\mathfrak T^{p-1}\pr{\bm\theta}&=\sum_{i=1,2}\bm{\hat e}_{\theta_i}^\top\bm{\hat e}_{\theta_i}\prod_{j=1}^{c(p-1)}r\pr{1-2l^{j-1}(\theta_i)},
\end{split}
\end{equation}
with $r=4$. Since we use a decoupled transformation of our two independent parameters, we consider here the Lyapunov exponents associated with perturbations along a fixed axis. These are computed by choosing $\bm{\delta\theta}=\delta\theta_i\bm{\hat e}_{\theta_i}$, which yields the spectrum of exponents $\{\lambda_i^{(p)}(\bm\theta)\}_{i}$ for $i=1,2$. In our case, the form for these exponents are similar. As such, we will define a new function $\tilde\lambda^{(p,c)}(\theta)$ below as a stand-in for $\lambda_i^{(p)}(\bm\theta)$. If we let $\lambda_1^{(p)}(\bm\theta)=\tilde\lambda^{(p,c)}(\theta_1)$ and $\lambda_2^{(p)}(\bm\theta)=\tilde\lambda^{(p,c)}(\theta_2)$, we can compute its large $p$ limit as
\begin{align}
    \tilde\lambda^{(p,c)}(\theta) &\equiv \frac1{p-1}\sum_{i=1}^{c(p-1)}\ln\abs{4(1-2l^{i-1}(\theta))}\label{eq:phase_space_lle}\\
    &\stackrel{p\gg 1}{\to} c\iint_{I^2}d\theta_1d\theta_2\,\ln\abs{4(1-2\theta_1)}\rho_l(\theta_1)\rho_l(\theta_2)\nonumber\\
    &=c\ln2\equiv\lambda_{\infty,c}.
    \label{eq:phase_space_gle}
\end{align}
Above, we have used the result of Eq.~(\ref{eq:lyapunov_lle_ergodic}) in the large $p$ limit, and computed the ergodic value using details of our probability space given in Sec.~\ref{appendix:2d_logistic_map}. This shows how we can characterize QACOA's typical long-time behavior in terms of how it evolves in phase space by using the results of Sec.~\ref{sec:qacoa_ergodicity}. $\lambda_{\infty,1}=\ln2$ is the known GLE for the $r=4$ logistic map~\cite{Eckmann1985}, and thus the factor $c$ may be viewed as ``fast-forwarding'' the rate at which our gradients grow (hence the `map speed' name).
In Fig.~\ref{fig:orbit_plot}, we plot unperturbed $\mathfrak T^{m-1}\pr{\bm\theta}$ and perturbed $\mathfrak T^{m-1}\pr{\bm{\theta}+\bm{\delta\theta}}$ trajectories at a random angle $\bm\theta\in\mathcal X$ along with the finite-time LLEs, indicating that they do in fact converge to $\lambda_{\infty,c}$. The result of Eq.~\ref{eq:phase_space_gle} is crucial to our treatment of QACOA trainability in Sec.~\ref{sec:trainability_decay}.

%\clearpage
%\bibliographystyle{apsrev4-2}
%\bibliography{refs}

\begin{thebibliography}{88}%
\makeatletter
\providecommand \@ifxundefined [1]{%
 \@ifx{#1\undefined}
}%
\providecommand \@ifnum [1]{%
 \ifnum #1\expandafter \@firstoftwo
 \else \expandafter \@secondoftwo
 \fi
}%
\providecommand \@ifx [1]{%
 \ifx #1\expandafter \@firstoftwo
 \else \expandafter \@secondoftwo
 \fi
}%
\providecommand \natexlab [1]{#1}%
\providecommand \enquote  [1]{``#1''}%
\providecommand \bibnamefont  [1]{#1}%
\providecommand \bibfnamefont [1]{#1}%
\providecommand \citenamefont [1]{#1}%
\providecommand \href@noop [0]{\@secondoftwo}%
\providecommand \href [0]{\begingroup \@sanitize@url \@href}%
\providecommand \@href[1]{\@@startlink{#1}\@@href}%
\providecommand \@@href[1]{\endgroup#1\@@endlink}%
\providecommand \@sanitize@url [0]{\catcode `\\12\catcode `\$12\catcode
  `\&12\catcode `\#12\catcode `\^12\catcode `\_12\catcode `\%12\relax}%
\providecommand \@@startlink[1]{}%
\providecommand \@@endlink[0]{}%
\providecommand \url  [0]{\begingroup\@sanitize@url \@url }%
\providecommand \@url [1]{\endgroup\@href {#1}{\urlprefix }}%
\providecommand \urlprefix  [0]{URL }%
\providecommand \Eprint [0]{\href }%
\providecommand \doibase [0]{https://doi.org/}%
\providecommand \selectlanguage [0]{\@gobble}%
\providecommand \bibinfo  [0]{\@secondoftwo}%
\providecommand \bibfield  [0]{\@secondoftwo}%
\providecommand \translation [1]{[#1]}%
\providecommand \BibitemOpen [0]{}%
\providecommand \bibitemStop [0]{}%
\providecommand \bibitemNoStop [0]{.\EOS\space}%
\providecommand \EOS [0]{\spacefactor3000\relax}%
\providecommand \BibitemShut  [1]{\csname bibitem#1\endcsname}%
\let\auto@bib@innerbib\@empty
%</preamble>
\bibitem [{\citenamefont {Cerezo}\ \emph {et~al.}(2021)\citenamefont {Cerezo},
  \citenamefont {Arrasmith}, \citenamefont {Babbush}, \citenamefont {Benjamin},
  \citenamefont {Endo}, \citenamefont {Fujii}, \citenamefont {McClean},
  \citenamefont {Mitarai}, \citenamefont {Yuan}, \citenamefont {Cincio} \emph
  {et~al.}}]{cerezo2021variational}%
  \BibitemOpen
  \bibfield  {author} {\bibinfo {author} {\bibfnamefont {M.}~\bibnamefont
  {Cerezo}}, \bibinfo {author} {\bibfnamefont {A.}~\bibnamefont {Arrasmith}},
  \bibinfo {author} {\bibfnamefont {R.}~\bibnamefont {Babbush}}, \bibinfo
  {author} {\bibfnamefont {S.~C.}\ \bibnamefont {Benjamin}}, \bibinfo {author}
  {\bibfnamefont {S.}~\bibnamefont {Endo}}, \bibinfo {author} {\bibfnamefont
  {K.}~\bibnamefont {Fujii}}, \bibinfo {author} {\bibfnamefont {J.~R.}\
  \bibnamefont {McClean}}, \bibinfo {author} {\bibfnamefont {K.}~\bibnamefont
  {Mitarai}}, \bibinfo {author} {\bibfnamefont {X.}~\bibnamefont {Yuan}},
  \bibinfo {author} {\bibfnamefont {L.}~\bibnamefont {Cincio}}, \emph
  {et~al.},\ }\href
  {https://doi.org/https://doi.org/10.1038/s42254-021-00348-9} {\bibfield
  {journal} {\bibinfo  {journal} {Nature Reviews Physics}\ }\textbf {\bibinfo
  {volume} {3}},\ \bibinfo {pages} {625} (\bibinfo {year} {2021})}\BibitemShut
  {NoStop}%
\bibitem [{\citenamefont {Farhi}\ \emph {et~al.}(2014)\citenamefont {Farhi},
  \citenamefont {Goldstone},\ and\ \citenamefont {Gutmann}}]{farhi2014quantum}%
  \BibitemOpen
  \bibfield  {author} {\bibinfo {author} {\bibfnamefont {E.}~\bibnamefont
  {Farhi}}, \bibinfo {author} {\bibfnamefont {J.}~\bibnamefont {Goldstone}},\
  and\ \bibinfo {author} {\bibfnamefont {S.}~\bibnamefont {Gutmann}},\
  }\href@noop {} {\bibinfo {title} {A quantum approximate optimization
  algorithm}} (\bibinfo {year} {2014}),\ \Eprint
  {https://arxiv.org/abs/1411.4028} {arXiv:1411.4028 [quant-ph]} \BibitemShut
  {NoStop}%
\bibitem [{\citenamefont {Blekos}\ \emph {et~al.}(2024)\citenamefont {Blekos},
  \citenamefont {Brand}, \citenamefont {Ceschini}, \citenamefont {Chou},
  \citenamefont {Li}, \citenamefont {Pandya},\ and\ \citenamefont
  {Summer}}]{blekos2024review}%
  \BibitemOpen
  \bibfield  {author} {\bibinfo {author} {\bibfnamefont {K.}~\bibnamefont
  {Blekos}}, \bibinfo {author} {\bibfnamefont {D.}~\bibnamefont {Brand}},
  \bibinfo {author} {\bibfnamefont {A.}~\bibnamefont {Ceschini}}, \bibinfo
  {author} {\bibfnamefont {C.-H.}\ \bibnamefont {Chou}}, \bibinfo {author}
  {\bibfnamefont {R.-H.}\ \bibnamefont {Li}}, \bibinfo {author} {\bibfnamefont
  {K.}~\bibnamefont {Pandya}},\ and\ \bibinfo {author} {\bibfnamefont
  {A.}~\bibnamefont {Summer}},\ }\href
  {https://doi.org/https://doi.org/10.1016/j.physrep.2024.03.002} {\bibfield
  {journal} {\bibinfo  {journal} {Physics Reports}\ }\textbf {\bibinfo {volume}
  {1068}},\ \bibinfo {pages} {1} (\bibinfo {year} {2024})}\BibitemShut
  {NoStop}%
\bibitem [{\citenamefont {Lloyd}(2018)}]{lloyd2018quantum}%
  \BibitemOpen
  \bibfield  {author} {\bibinfo {author} {\bibfnamefont {S.}~\bibnamefont
  {Lloyd}},\ }\href@noop {} {\bibinfo {title} {Quantum approximate optimization
  is computationally universal}} (\bibinfo {year} {2018}),\ \Eprint
  {https://arxiv.org/abs/1812.11075} {arXiv:1812.11075 [quant-ph]} \BibitemShut
  {NoStop}%
\bibitem [{\citenamefont {Morales}\ \emph {et~al.}(2020)\citenamefont
  {Morales}, \citenamefont {Biamonte},\ and\ \citenamefont
  {Zimbor{\'a}s}}]{morales2020universality}%
  \BibitemOpen
  \bibfield  {author} {\bibinfo {author} {\bibfnamefont {M.~E.}\ \bibnamefont
  {Morales}}, \bibinfo {author} {\bibfnamefont {J.~D.}\ \bibnamefont
  {Biamonte}},\ and\ \bibinfo {author} {\bibfnamefont {Z.}~\bibnamefont
  {Zimbor{\'a}s}},\ }\href
  {https://doi.org/https://doi.org/10.1007/s11128-020-02748-9} {\bibfield
  {journal} {\bibinfo  {journal} {Quantum Information Processing}\ }\textbf
  {\bibinfo {volume} {19}},\ \bibinfo {pages} {1} (\bibinfo {year}
  {2020})}\BibitemShut {NoStop}%
\bibitem [{\citenamefont {Farhi}\ and\ \citenamefont
  {Harrow}(2016)}]{farhi2016quantum}%
  \BibitemOpen
  \bibfield  {author} {\bibinfo {author} {\bibfnamefont {E.}~\bibnamefont
  {Farhi}}\ and\ \bibinfo {author} {\bibfnamefont {A.~W.}\ \bibnamefont
  {Harrow}},\ }\href@noop {} {\bibinfo {title} {Quantum supremacy through the
  quantum approximate optimization algorithm}} (\bibinfo {year} {2016}),\
  \Eprint {https://arxiv.org/abs/1602.07674} {arXiv:1602.07674 [quant-ph]}
  \BibitemShut {NoStop}%
\bibitem [{\citenamefont {Crooks}(2018)}]{crooks2018performance}%
  \BibitemOpen
  \bibfield  {author} {\bibinfo {author} {\bibfnamefont {G.~E.}\ \bibnamefont
  {Crooks}},\ }\href@noop {} {\bibinfo {title} {Performance of the quantum
  approximate optimization algorithm on the maximum cut problem}} (\bibinfo
  {year} {2018}),\ \Eprint {https://arxiv.org/abs/1811.08419} {arXiv:1811.08419
  [quant-ph]} \BibitemShut {NoStop}%
\bibitem [{\citenamefont {Niu}\ \emph {et~al.}(2019)\citenamefont {Niu},
  \citenamefont {Lu},\ and\ \citenamefont {Chuang}}]{niu2019optimizing}%
  \BibitemOpen
  \bibfield  {author} {\bibinfo {author} {\bibfnamefont {M.~Y.}\ \bibnamefont
  {Niu}}, \bibinfo {author} {\bibfnamefont {S.}~\bibnamefont {Lu}},\ and\
  \bibinfo {author} {\bibfnamefont {I.~L.}\ \bibnamefont {Chuang}},\ }\href
  {https://doi.org/https://doi.org/10.48550/arXiv.1905.12134} {\bibinfo {title}
  {Optimizing qaoa: Success probability and runtime dependence on circuit
  depth}} (\bibinfo {year} {2019})\BibitemShut {NoStop}%
\bibitem [{\citenamefont {Bravyi}\ \emph {et~al.}(2020)\citenamefont {Bravyi},
  \citenamefont {Kliesch}, \citenamefont {Koenig},\ and\ \citenamefont
  {Tang}}]{bravyi2020obstacles}%
  \BibitemOpen
  \bibfield  {author} {\bibinfo {author} {\bibfnamefont {S.}~\bibnamefont
  {Bravyi}}, \bibinfo {author} {\bibfnamefont {A.}~\bibnamefont {Kliesch}},
  \bibinfo {author} {\bibfnamefont {R.}~\bibnamefont {Koenig}},\ and\ \bibinfo
  {author} {\bibfnamefont {E.}~\bibnamefont {Tang}},\ }\href
  {https://doi.org/10.1103/PhysRevLett.125.260505} {\bibfield  {journal}
  {\bibinfo  {journal} {Phys. Rev. Lett.}\ }\textbf {\bibinfo {volume} {125}},\
  \bibinfo {pages} {260505} (\bibinfo {year} {2020})}\BibitemShut {NoStop}%
\bibitem [{\citenamefont {An}\ and\ \citenamefont {Lin}(2022)}]{an2022quantum}%
  \BibitemOpen
  \bibfield  {author} {\bibinfo {author} {\bibfnamefont {D.}~\bibnamefont
  {An}}\ and\ \bibinfo {author} {\bibfnamefont {L.}~\bibnamefont {Lin}},\
  }\href {https://doi.org/https://doi.org/10.1145/3498331} {\bibfield
  {journal} {\bibinfo  {journal} {ACM Transactions on Quantum Computing}\
  }\textbf {\bibinfo {volume} {3}},\ \bibinfo {pages} {1} (\bibinfo {year}
  {2022})}\BibitemShut {NoStop}%
\bibitem [{\citenamefont {Zhang}\ \emph {et~al.}(2022)\citenamefont {Zhang},
  \citenamefont {Mu}, \citenamefont {Liu}, \citenamefont {Wang}, \citenamefont
  {Zhang}, \citenamefont {Li}, \citenamefont {Wu}, \citenamefont {Zhao},\ and\
  \citenamefont {Dong}}]{zhang2022applying}%
  \BibitemOpen
  \bibfield  {author} {\bibinfo {author} {\bibfnamefont {Y.}~\bibnamefont
  {Zhang}}, \bibinfo {author} {\bibfnamefont {X.}~\bibnamefont {Mu}}, \bibinfo
  {author} {\bibfnamefont {X.-W.}\ \bibnamefont {Liu}}, \bibinfo {author}
  {\bibfnamefont {X.}~\bibnamefont {Wang}}, \bibinfo {author} {\bibfnamefont
  {X.}~\bibnamefont {Zhang}}, \bibinfo {author} {\bibfnamefont
  {K.}~\bibnamefont {Li}}, \bibinfo {author} {\bibfnamefont {T.}~\bibnamefont
  {Wu}}, \bibinfo {author} {\bibfnamefont {D.}~\bibnamefont {Zhao}},\ and\
  \bibinfo {author} {\bibfnamefont {C.}~\bibnamefont {Dong}},\ }\href
  {https://doi.org/https://doi.org/10.1016/j.asoc.2022.108554} {\bibfield
  {journal} {\bibinfo  {journal} {Applied Soft Computing}\ }\textbf {\bibinfo
  {volume} {118}},\ \bibinfo {pages} {108554} (\bibinfo {year}
  {2022})}\BibitemShut {NoStop}%
\bibitem [{\citenamefont {Lykov}\ \emph {et~al.}(2023)\citenamefont {Lykov},
  \citenamefont {Wurtz}, \citenamefont {Poole}, \citenamefont {Saffman},
  \citenamefont {Noel},\ and\ \citenamefont {Alexeev}}]{lykov2023sampling}%
  \BibitemOpen
  \bibfield  {author} {\bibinfo {author} {\bibfnamefont {D.}~\bibnamefont
  {Lykov}}, \bibinfo {author} {\bibfnamefont {J.}~\bibnamefont {Wurtz}},
  \bibinfo {author} {\bibfnamefont {C.}~\bibnamefont {Poole}}, \bibinfo
  {author} {\bibfnamefont {M.}~\bibnamefont {Saffman}}, \bibinfo {author}
  {\bibfnamefont {T.}~\bibnamefont {Noel}},\ and\ \bibinfo {author}
  {\bibfnamefont {Y.}~\bibnamefont {Alexeev}},\ }\href
  {https://doi.org/https://doi.org/10.1038/s41534-023-00718-4} {\bibfield
  {journal} {\bibinfo  {journal} {npj Quantum Information}\ }\textbf {\bibinfo
  {volume} {9}},\ \bibinfo {pages} {73} (\bibinfo {year} {2023})}\BibitemShut
  {NoStop}%
\bibitem [{\citenamefont {Boulebnane}\ and\ \citenamefont
  {Montanaro}(2024)}]{boulebnane2024solving}%
  \BibitemOpen
  \bibfield  {author} {\bibinfo {author} {\bibfnamefont {S.}~\bibnamefont
  {Boulebnane}}\ and\ \bibinfo {author} {\bibfnamefont {A.}~\bibnamefont
  {Montanaro}},\ }\href {https://doi.org/10.1103/PRXQuantum.5.030348}
  {\bibfield  {journal} {\bibinfo  {journal} {PRX Quantum}\ }\textbf {\bibinfo
  {volume} {5}},\ \bibinfo {pages} {030348} (\bibinfo {year}
  {2024})}\BibitemShut {NoStop}%
\bibitem [{\citenamefont {Shaydulin}\ \emph {et~al.}(2024)\citenamefont
  {Shaydulin}, \citenamefont {Li}, \citenamefont {Chakrabarti}, \citenamefont
  {DeCross}, \citenamefont {Herman}, \citenamefont {Kumar}, \citenamefont
  {Larson}, \citenamefont {Lykov}, \citenamefont {Minssen}, \citenamefont {Sun}
  \emph {et~al.}}]{shaydulin2024evidence}%
  \BibitemOpen
  \bibfield  {author} {\bibinfo {author} {\bibfnamefont {R.}~\bibnamefont
  {Shaydulin}}, \bibinfo {author} {\bibfnamefont {C.}~\bibnamefont {Li}},
  \bibinfo {author} {\bibfnamefont {S.}~\bibnamefont {Chakrabarti}}, \bibinfo
  {author} {\bibfnamefont {M.}~\bibnamefont {DeCross}}, \bibinfo {author}
  {\bibfnamefont {D.}~\bibnamefont {Herman}}, \bibinfo {author} {\bibfnamefont
  {N.}~\bibnamefont {Kumar}}, \bibinfo {author} {\bibfnamefont
  {J.}~\bibnamefont {Larson}}, \bibinfo {author} {\bibfnamefont
  {D.}~\bibnamefont {Lykov}}, \bibinfo {author} {\bibfnamefont
  {P.}~\bibnamefont {Minssen}}, \bibinfo {author} {\bibfnamefont
  {Y.}~\bibnamefont {Sun}}, \emph {et~al.},\ }\href
  {https://doi.org/10.1126/sciadv.adm6761} {\bibfield  {journal} {\bibinfo
  {journal} {Science Advances}\ }\textbf {\bibinfo {volume} {10}},\ \bibinfo
  {pages} {eadm6761} (\bibinfo {year} {2024})}\BibitemShut {NoStop}%
\bibitem [{\citenamefont {Pagano}\ \emph {et~al.}(2020)\citenamefont {Pagano},
  \citenamefont {Bapat}, \citenamefont {Becker}, \citenamefont {Collins},
  \citenamefont {De}, \citenamefont {Hess}, \citenamefont {Kaplan},
  \citenamefont {Kyprianidis}, \citenamefont {Tan}, \citenamefont {Baldwin}
  \emph {et~al.}}]{pagano2020quantum}%
  \BibitemOpen
  \bibfield  {author} {\bibinfo {author} {\bibfnamefont {G.}~\bibnamefont
  {Pagano}}, \bibinfo {author} {\bibfnamefont {A.}~\bibnamefont {Bapat}},
  \bibinfo {author} {\bibfnamefont {P.}~\bibnamefont {Becker}}, \bibinfo
  {author} {\bibfnamefont {K.~S.}\ \bibnamefont {Collins}}, \bibinfo {author}
  {\bibfnamefont {A.}~\bibnamefont {De}}, \bibinfo {author} {\bibfnamefont
  {P.~W.}\ \bibnamefont {Hess}}, \bibinfo {author} {\bibfnamefont {H.~B.}\
  \bibnamefont {Kaplan}}, \bibinfo {author} {\bibfnamefont {A.}~\bibnamefont
  {Kyprianidis}}, \bibinfo {author} {\bibfnamefont {W.~L.}\ \bibnamefont
  {Tan}}, \bibinfo {author} {\bibfnamefont {C.}~\bibnamefont {Baldwin}}, \emph
  {et~al.},\ }\href {https://doi.org/https://doi.org/10.1073/pnas.200637311}
  {\bibfield  {journal} {\bibinfo  {journal} {Proceedings of the National
  Academy of Sciences}\ }\textbf {\bibinfo {volume} {117}},\ \bibinfo {pages}
  {25396} (\bibinfo {year} {2020})}\BibitemShut {NoStop}%
\bibitem [{\citenamefont {Harrigan}\ \emph {et~al.}(2021)\citenamefont
  {Harrigan}, \citenamefont {Sung}, \citenamefont {Neeley}, \citenamefont
  {Satzinger}, \citenamefont {Arute}, \citenamefont {Arya}, \citenamefont
  {Atalaya}, \citenamefont {Bardin}, \citenamefont {Barends}, \citenamefont
  {Boixo} \emph {et~al.}}]{harrigan2021quantum}%
  \BibitemOpen
  \bibfield  {author} {\bibinfo {author} {\bibfnamefont {M.~P.}\ \bibnamefont
  {Harrigan}}, \bibinfo {author} {\bibfnamefont {K.~J.}\ \bibnamefont {Sung}},
  \bibinfo {author} {\bibfnamefont {M.}~\bibnamefont {Neeley}}, \bibinfo
  {author} {\bibfnamefont {K.~J.}\ \bibnamefont {Satzinger}}, \bibinfo {author}
  {\bibfnamefont {F.}~\bibnamefont {Arute}}, \bibinfo {author} {\bibfnamefont
  {K.}~\bibnamefont {Arya}}, \bibinfo {author} {\bibfnamefont {J.}~\bibnamefont
  {Atalaya}}, \bibinfo {author} {\bibfnamefont {J.~C.}\ \bibnamefont {Bardin}},
  \bibinfo {author} {\bibfnamefont {R.}~\bibnamefont {Barends}}, \bibinfo
  {author} {\bibfnamefont {S.}~\bibnamefont {Boixo}}, \emph {et~al.},\ }\href
  {https://doi.org/https://doi.org/10.1038/s41567-020-01105-y} {\bibfield
  {journal} {\bibinfo  {journal} {Nature Physics}\ }\textbf {\bibinfo {volume}
  {17}},\ \bibinfo {pages} {332} (\bibinfo {year} {2021})}\BibitemShut
  {NoStop}%
\bibitem [{\citenamefont {Ebadi}\ \emph {et~al.}(2022)\citenamefont {Ebadi},
  \citenamefont {Keesling}, \citenamefont {Cain}, \citenamefont {Wang},
  \citenamefont {Levine}, \citenamefont {Bluvstein}, \citenamefont {Semeghini},
  \citenamefont {Omran}, \citenamefont {Liu}, \citenamefont {Samajdar} \emph
  {et~al.}}]{ebadi2022quantum}%
  \BibitemOpen
  \bibfield  {author} {\bibinfo {author} {\bibfnamefont {S.}~\bibnamefont
  {Ebadi}}, \bibinfo {author} {\bibfnamefont {A.}~\bibnamefont {Keesling}},
  \bibinfo {author} {\bibfnamefont {M.}~\bibnamefont {Cain}}, \bibinfo {author}
  {\bibfnamefont {T.~T.}\ \bibnamefont {Wang}}, \bibinfo {author}
  {\bibfnamefont {H.}~\bibnamefont {Levine}}, \bibinfo {author} {\bibfnamefont
  {D.}~\bibnamefont {Bluvstein}}, \bibinfo {author} {\bibfnamefont
  {G.}~\bibnamefont {Semeghini}}, \bibinfo {author} {\bibfnamefont
  {A.}~\bibnamefont {Omran}}, \bibinfo {author} {\bibfnamefont {J.-G.}\
  \bibnamefont {Liu}}, \bibinfo {author} {\bibfnamefont {R.}~\bibnamefont
  {Samajdar}}, \emph {et~al.},\ }\href
  {https://doi.org/10.1126/science.abo6587} {\bibfield  {journal} {\bibinfo
  {journal} {Science}\ }\textbf {\bibinfo {volume} {376}},\ \bibinfo {pages}
  {1209} (\bibinfo {year} {2022})}\BibitemShut {NoStop}%
\bibitem [{\citenamefont {Nguyen}\ \emph {et~al.}(2023)\citenamefont {Nguyen},
  \citenamefont {Liu}, \citenamefont {Wurtz}, \citenamefont {Lukin},
  \citenamefont {Wang},\ and\ \citenamefont {Pichler}}]{nguyen2023quantum}%
  \BibitemOpen
  \bibfield  {author} {\bibinfo {author} {\bibfnamefont {M.-T.}\ \bibnamefont
  {Nguyen}}, \bibinfo {author} {\bibfnamefont {J.-G.}\ \bibnamefont {Liu}},
  \bibinfo {author} {\bibfnamefont {J.}~\bibnamefont {Wurtz}}, \bibinfo
  {author} {\bibfnamefont {M.~D.}\ \bibnamefont {Lukin}}, \bibinfo {author}
  {\bibfnamefont {S.-T.}\ \bibnamefont {Wang}},\ and\ \bibinfo {author}
  {\bibfnamefont {H.}~\bibnamefont {Pichler}},\ }\href
  {https://doi.org/10.1103/PRXQuantum.4.010316} {\bibfield  {journal} {\bibinfo
   {journal} {PRX Quantum}\ }\textbf {\bibinfo {volume} {4}},\ \bibinfo {pages}
  {010316} (\bibinfo {year} {2023})}\BibitemShut {NoStop}%
\bibitem [{\citenamefont {Montanez-Barrera}\ and\ \citenamefont
  {Michielsen}(2024{\natexlab{a}})}]{montanez2024towards}%
  \BibitemOpen
  \bibfield  {author} {\bibinfo {author} {\bibfnamefont {J.}~\bibnamefont
  {Montanez-Barrera}}\ and\ \bibinfo {author} {\bibfnamefont {K.}~\bibnamefont
  {Michielsen}},\ }\href
  {https://doi.org/https://doi.org/10.1038/s41534-025-01082-1} {\bibinfo
  {title} {Towards a universal qaoa protocol: Evidence of quantum advantage in
  solving combinatorial optimization problems}} (\bibinfo {year}
  {2024}{\natexlab{a}})\BibitemShut {NoStop}%
\bibitem [{\citenamefont {Sack}\ and\ \citenamefont
  {Egger}(2024)}]{sack2024largescale}%
  \BibitemOpen
  \bibfield  {author} {\bibinfo {author} {\bibfnamefont {S.~H.}\ \bibnamefont
  {Sack}}\ and\ \bibinfo {author} {\bibfnamefont {D.~J.}\ \bibnamefont
  {Egger}},\ }\href {https://doi.org/10.1103/PhysRevResearch.6.013223}
  {\bibfield  {journal} {\bibinfo  {journal} {Phys. Rev. Res.}\ }\textbf
  {\bibinfo {volume} {6}},\ \bibinfo {pages} {013223} (\bibinfo {year}
  {2024})}\BibitemShut {NoStop}%
\bibitem [{\citenamefont {Pellow-Jarman}\ \emph {et~al.}(2021)\citenamefont
  {Pellow-Jarman}, \citenamefont {Sinayskiy}, \citenamefont {Pillay},\ and\
  \citenamefont {Petruccione}}]{pellow2021comparison}%
  \BibitemOpen
  \bibfield  {author} {\bibinfo {author} {\bibfnamefont {A.}~\bibnamefont
  {Pellow-Jarman}}, \bibinfo {author} {\bibfnamefont {I.}~\bibnamefont
  {Sinayskiy}}, \bibinfo {author} {\bibfnamefont {A.}~\bibnamefont {Pillay}},\
  and\ \bibinfo {author} {\bibfnamefont {F.}~\bibnamefont {Petruccione}},\
  }\href {https://doi.org/https://doi.org/10.1007/s11128-021-03140-x}
  {\bibfield  {journal} {\bibinfo  {journal} {Quantum Information Processing}\
  }\textbf {\bibinfo {volume} {20}},\ \bibinfo {pages} {202} (\bibinfo {year}
  {2021})}\BibitemShut {NoStop}%
\bibitem [{\citenamefont {Fern{\'a}ndez-Pend{\'a}s}\ \emph
  {et~al.}(2022)\citenamefont {Fern{\'a}ndez-Pend{\'a}s}, \citenamefont
  {Combarro}, \citenamefont {Vallecorsa}, \citenamefont {Ranilla},\ and\
  \citenamefont {R{\'u}a}}]{fernandez2022study}%
  \BibitemOpen
  \bibfield  {author} {\bibinfo {author} {\bibfnamefont {M.}~\bibnamefont
  {Fern{\'a}ndez-Pend{\'a}s}}, \bibinfo {author} {\bibfnamefont {E.~F.}\
  \bibnamefont {Combarro}}, \bibinfo {author} {\bibfnamefont {S.}~\bibnamefont
  {Vallecorsa}}, \bibinfo {author} {\bibfnamefont {J.}~\bibnamefont
  {Ranilla}},\ and\ \bibinfo {author} {\bibfnamefont {I.~F.}\ \bibnamefont
  {R{\'u}a}},\ }\href
  {https://doi.org/https://doi.org/10.1016/j.cam.2021.113388} {\bibfield
  {journal} {\bibinfo  {journal} {Journal of Computational and Applied
  Mathematics}\ }\textbf {\bibinfo {volume} {404}},\ \bibinfo {pages} {113388}
  (\bibinfo {year} {2022})}\BibitemShut {NoStop}%
\bibitem [{\citenamefont {Acampora}\ \emph {et~al.}(2023)\citenamefont
  {Acampora}, \citenamefont {Chiatto},\ and\ \citenamefont
  {Vitiello}}]{acampora2023genetic}%
  \BibitemOpen
  \bibfield  {author} {\bibinfo {author} {\bibfnamefont {G.}~\bibnamefont
  {Acampora}}, \bibinfo {author} {\bibfnamefont {A.}~\bibnamefont {Chiatto}},\
  and\ \bibinfo {author} {\bibfnamefont {A.}~\bibnamefont {Vitiello}},\ }\href
  {https://doi.org/https://doi.org/10.1016/j.asoc.2023.110296} {\bibfield
  {journal} {\bibinfo  {journal} {Applied Soft Computing}\ }\textbf {\bibinfo
  {volume} {142}},\ \bibinfo {pages} {110296} (\bibinfo {year}
  {2023})}\BibitemShut {NoStop}%
\bibitem [{\citenamefont {Pellow-Jarman}\ \emph {et~al.}(2024)\citenamefont
  {Pellow-Jarman}, \citenamefont {McFarthing}, \citenamefont {Sinayskiy},
  \citenamefont {Park}, \citenamefont {Pillay},\ and\ \citenamefont
  {Petruccione}}]{pellow2024effect}%
  \BibitemOpen
  \bibfield  {author} {\bibinfo {author} {\bibfnamefont {A.}~\bibnamefont
  {Pellow-Jarman}}, \bibinfo {author} {\bibfnamefont {S.}~\bibnamefont
  {McFarthing}}, \bibinfo {author} {\bibfnamefont {I.}~\bibnamefont
  {Sinayskiy}}, \bibinfo {author} {\bibfnamefont {D.~K.}\ \bibnamefont {Park}},
  \bibinfo {author} {\bibfnamefont {A.}~\bibnamefont {Pillay}},\ and\ \bibinfo
  {author} {\bibfnamefont {F.}~\bibnamefont {Petruccione}},\ }\href
  {https://doi.org/https://doi.org/10.1038/s41598-024-66625-6} {\bibfield
  {journal} {\bibinfo  {journal} {Scientific Reports}\ }\textbf {\bibinfo
  {volume} {14}},\ \bibinfo {pages} {16011} (\bibinfo {year}
  {2024})}\BibitemShut {NoStop}%
\bibitem [{\citenamefont {McClean}\ \emph {et~al.}(2018)\citenamefont
  {McClean}, \citenamefont {Boixo}, \citenamefont {Smelyanskiy}, \citenamefont
  {Babbush},\ and\ \citenamefont {Neven}}]{barren_plateaus}%
  \BibitemOpen
  \bibfield  {author} {\bibinfo {author} {\bibfnamefont {J.~R.}\ \bibnamefont
  {McClean}}, \bibinfo {author} {\bibfnamefont {S.}~\bibnamefont {Boixo}},
  \bibinfo {author} {\bibfnamefont {V.~N.}\ \bibnamefont {Smelyanskiy}},
  \bibinfo {author} {\bibfnamefont {R.}~\bibnamefont {Babbush}},\ and\ \bibinfo
  {author} {\bibfnamefont {H.}~\bibnamefont {Neven}},\ }\bibfield  {journal}
  {\bibinfo  {journal} {Nature Communications}\ }\textbf {\bibinfo {volume}
  {9}},\ \href {https://doi.org/10.1038/s41467-018-07090-4}
  {10.1038/s41467-018-07090-4} (\bibinfo {year} {2018})\BibitemShut {NoStop}%
\bibitem [{\citenamefont {Zhu}\ \emph {et~al.}(2022)\citenamefont {Zhu},
  \citenamefont {Tang}, \citenamefont {Barron}, \citenamefont
  {Calderon-Vargas}, \citenamefont {Mayhall}, \citenamefont {Barnes},\ and\
  \citenamefont {Economou}}]{zhu2022adaptive}%
  \BibitemOpen
  \bibfield  {author} {\bibinfo {author} {\bibfnamefont {L.}~\bibnamefont
  {Zhu}}, \bibinfo {author} {\bibfnamefont {H.~L.}\ \bibnamefont {Tang}},
  \bibinfo {author} {\bibfnamefont {G.~S.}\ \bibnamefont {Barron}}, \bibinfo
  {author} {\bibfnamefont {F.~A.}\ \bibnamefont {Calderon-Vargas}}, \bibinfo
  {author} {\bibfnamefont {N.~J.}\ \bibnamefont {Mayhall}}, \bibinfo {author}
  {\bibfnamefont {E.}~\bibnamefont {Barnes}},\ and\ \bibinfo {author}
  {\bibfnamefont {S.~E.}\ \bibnamefont {Economou}},\ }\href
  {https://doi.org/10.1103/PhysRevResearch.4.033029} {\bibfield  {journal}
  {\bibinfo  {journal} {Phys. Rev. Res.}\ }\textbf {\bibinfo {volume} {4}},\
  \bibinfo {pages} {033029} (\bibinfo {year} {2022})}\BibitemShut {NoStop}%
\bibitem [{\citenamefont {Hadfield}\ \emph {et~al.}(2019)\citenamefont
  {Hadfield}, \citenamefont {Wang}, \citenamefont {O’gorman}, \citenamefont
  {Rieffel}, \citenamefont {Venturelli},\ and\ \citenamefont
  {Biswas}}]{hadfield2019quantum}%
  \BibitemOpen
  \bibfield  {author} {\bibinfo {author} {\bibfnamefont {S.}~\bibnamefont
  {Hadfield}}, \bibinfo {author} {\bibfnamefont {Z.}~\bibnamefont {Wang}},
  \bibinfo {author} {\bibfnamefont {B.}~\bibnamefont {O’gorman}}, \bibinfo
  {author} {\bibfnamefont {E.~G.}\ \bibnamefont {Rieffel}}, \bibinfo {author}
  {\bibfnamefont {D.}~\bibnamefont {Venturelli}},\ and\ \bibinfo {author}
  {\bibfnamefont {R.}~\bibnamefont {Biswas}},\ }\href
  {https://doi.org/https://doi.org/10.3390/a12020034} {\bibfield  {journal}
  {\bibinfo  {journal} {Algorithms}\ }\textbf {\bibinfo {volume} {12}},\
  \bibinfo {pages} {34} (\bibinfo {year} {2019})}\BibitemShut {NoStop}%
\bibitem [{\citenamefont {Shaydulin}\ and\ \citenamefont
  {Wild}(2021)}]{shaydulin2021exploiting}%
  \BibitemOpen
  \bibfield  {author} {\bibinfo {author} {\bibfnamefont {R.}~\bibnamefont
  {Shaydulin}}\ and\ \bibinfo {author} {\bibfnamefont {S.~M.}\ \bibnamefont
  {Wild}},\ }\href {https://doi.org/https://doi.org/10.1109/TQE.2021.3066275}
  {\bibfield  {journal} {\bibinfo  {journal} {IEEE Transactions on Quantum
  Engineering}\ }\textbf {\bibinfo {volume} {2}},\ \bibinfo {pages} {1}
  (\bibinfo {year} {2021})}\BibitemShut {NoStop}%
\bibitem [{\citenamefont {Shaydulin}\ \emph {et~al.}(2021)\citenamefont
  {Shaydulin}, \citenamefont {Hadfield}, \citenamefont {Hogg},\ and\
  \citenamefont {Safro}}]{Shaydulin2021classical}%
  \BibitemOpen
  \bibfield  {author} {\bibinfo {author} {\bibfnamefont {R.}~\bibnamefont
  {Shaydulin}}, \bibinfo {author} {\bibfnamefont {S.}~\bibnamefont {Hadfield}},
  \bibinfo {author} {\bibfnamefont {T.}~\bibnamefont {Hogg}},\ and\ \bibinfo
  {author} {\bibfnamefont {I.}~\bibnamefont {Safro}},\ }\bibfield  {journal}
  {\bibinfo  {journal} {Quantum Information Processing}\ }\textbf {\bibinfo
  {volume} {20}},\ \href {https://doi.org/10.1007/s11128-021-03298-4}
  {10.1007/s11128-021-03298-4} (\bibinfo {year} {2021})\BibitemShut {NoStop}%
\bibitem [{\citenamefont {Tsvelikhovskiy}\ \emph {et~al.}(2024)\citenamefont
  {Tsvelikhovskiy}, \citenamefont {Safro},\ and\ \citenamefont
  {Alexeev}}]{tsvelikhovskiy2024symmetries}%
  \BibitemOpen
  \bibfield  {author} {\bibinfo {author} {\bibfnamefont {B.}~\bibnamefont
  {Tsvelikhovskiy}}, \bibinfo {author} {\bibfnamefont {I.}~\bibnamefont
  {Safro}},\ and\ \bibinfo {author} {\bibfnamefont {Y.}~\bibnamefont
  {Alexeev}},\ }\href {https://arxiv.org/abs/2309.13787} {\bibinfo {title}
  {Symmetries and dimension reduction in quantum approximate optimization
  algorithm}} (\bibinfo {year} {2024}),\ \Eprint
  {https://arxiv.org/abs/2309.13787} {arXiv:2309.13787 [quant-ph]} \BibitemShut
  {NoStop}%
\bibitem [{\citenamefont {Kazi}\ \emph {et~al.}(2024)\citenamefont {Kazi},
  \citenamefont {Larocca}, \citenamefont {Farinati}, \citenamefont {Coles},
  \citenamefont {Cerezo},\ and\ \citenamefont {Zeier}}]{kazi2024analyzing}%
  \BibitemOpen
  \bibfield  {author} {\bibinfo {author} {\bibfnamefont {S.}~\bibnamefont
  {Kazi}}, \bibinfo {author} {\bibfnamefont {M.}~\bibnamefont {Larocca}},
  \bibinfo {author} {\bibfnamefont {M.}~\bibnamefont {Farinati}}, \bibinfo
  {author} {\bibfnamefont {P.~J.}\ \bibnamefont {Coles}}, \bibinfo {author}
  {\bibfnamefont {M.}~\bibnamefont {Cerezo}},\ and\ \bibinfo {author}
  {\bibfnamefont {R.}~\bibnamefont {Zeier}},\ }\href
  {https://arxiv.org/abs/2410.05187} {\bibinfo {title} {Analyzing the quantum
  approximate optimization algorithm: ans\"atze, symmetries, and lie algebras}}
  (\bibinfo {year} {2024}),\ \Eprint {https://arxiv.org/abs/2410.05187}
  {arXiv:2410.05187 [quant-ph]} \BibitemShut {NoStop}%
\bibitem [{\citenamefont {Guerreschi}\ and\ \citenamefont
  {Smelyanskiy}(2017)}]{guerreschi2017compare}%
  \BibitemOpen
  \bibfield  {author} {\bibinfo {author} {\bibfnamefont {G.~G.}\ \bibnamefont
  {Guerreschi}}\ and\ \bibinfo {author} {\bibfnamefont {M.}~\bibnamefont
  {Smelyanskiy}},\ }\href {https://arxiv.org/abs/1701.01450} {\bibinfo {title}
  {Practical optimization for hybrid quantum-classical algorithms}} (\bibinfo
  {year} {2017}),\ \Eprint {https://arxiv.org/abs/1701.01450} {arXiv:1701.01450
  [quant-ph]} \BibitemShut {NoStop}%
\bibitem [{\citenamefont {Nannicini}(2019)}]{nannicini2019compare}%
  \BibitemOpen
  \bibfield  {author} {\bibinfo {author} {\bibfnamefont {G.}~\bibnamefont
  {Nannicini}},\ }\href {https://doi.org/10.1103/PhysRevE.99.013304} {\bibfield
   {journal} {\bibinfo  {journal} {Phys. Rev. E}\ }\textbf {\bibinfo {volume}
  {99}},\ \bibinfo {pages} {013304} (\bibinfo {year} {2019})}\BibitemShut
  {NoStop}%
\bibitem [{\citenamefont {Shaydulin}\ \emph {et~al.}(2019)\citenamefont
  {Shaydulin}, \citenamefont {Safro},\ and\ \citenamefont
  {Larson}}]{shaydulin2019multistart}%
  \BibitemOpen
  \bibfield  {author} {\bibinfo {author} {\bibfnamefont {R.}~\bibnamefont
  {Shaydulin}}, \bibinfo {author} {\bibfnamefont {I.}~\bibnamefont {Safro}},\
  and\ \bibinfo {author} {\bibfnamefont {J.}~\bibnamefont {Larson}},\ }in\
  \href {https://doi.org/10.1109/HPEC.2019.8916288} {\emph {\bibinfo
  {booktitle} {2019 IEEE High Performance Extreme Computing Conference
  (HPEC)}}}\ (\bibinfo {year} {2019})\ pp.\ \bibinfo {pages} {1--8}\BibitemShut
  {NoStop}%
\bibitem [{\citenamefont {Bonet-Monroig}\ \emph {et~al.}(2023)\citenamefont
  {Bonet-Monroig}, \citenamefont {Wang}, \citenamefont {Vermetten},
  \citenamefont {Senjean}, \citenamefont {Moussa}, \citenamefont {B\"ack},
  \citenamefont {Dunjko},\ and\ \citenamefont {O'Brien}}]{bonet2023compare}%
  \BibitemOpen
  \bibfield  {author} {\bibinfo {author} {\bibfnamefont {X.}~\bibnamefont
  {Bonet-Monroig}}, \bibinfo {author} {\bibfnamefont {H.}~\bibnamefont {Wang}},
  \bibinfo {author} {\bibfnamefont {D.}~\bibnamefont {Vermetten}}, \bibinfo
  {author} {\bibfnamefont {B.}~\bibnamefont {Senjean}}, \bibinfo {author}
  {\bibfnamefont {C.}~\bibnamefont {Moussa}}, \bibinfo {author} {\bibfnamefont
  {T.}~\bibnamefont {B\"ack}}, \bibinfo {author} {\bibfnamefont
  {V.}~\bibnamefont {Dunjko}},\ and\ \bibinfo {author} {\bibfnamefont {T.~E.}\
  \bibnamefont {O'Brien}},\ }\href
  {https://doi.org/10.1103/PhysRevA.107.032407} {\bibfield  {journal} {\bibinfo
   {journal} {Phys. Rev. A}\ }\textbf {\bibinfo {volume} {107}},\ \bibinfo
  {pages} {032407} (\bibinfo {year} {2023})}\BibitemShut {NoStop}%
\bibitem [{\citenamefont {Hao}\ \emph {et~al.}(2024)\citenamefont {Hao},
  \citenamefont {He}, \citenamefont {Shaydulin}, \citenamefont {Larson},\ and\
  \citenamefont {Pistoia}}]{hao2024fewshots}%
  \BibitemOpen
  \bibfield  {author} {\bibinfo {author} {\bibfnamefont {T.}~\bibnamefont
  {Hao}}, \bibinfo {author} {\bibfnamefont {Z.}~\bibnamefont {He}}, \bibinfo
  {author} {\bibfnamefont {R.}~\bibnamefont {Shaydulin}}, \bibinfo {author}
  {\bibfnamefont {J.}~\bibnamefont {Larson}},\ and\ \bibinfo {author}
  {\bibfnamefont {M.}~\bibnamefont {Pistoia}},\ }\href
  {https://arxiv.org/abs/2408.00557} {\bibinfo {title} {End-to-end protocol for
  high-quality qaoa parameters with few shots}} (\bibinfo {year} {2024}),\
  \Eprint {https://arxiv.org/abs/2408.00557} {arXiv:2408.00557 [quant-ph]}
  \BibitemShut {NoStop}%
\bibitem [{\citenamefont {Alam}\ \emph {et~al.}(2020)\citenamefont {Alam},
  \citenamefont {Ash-Saki},\ and\ \citenamefont
  {Ghosh}}]{alam2020accelerating}%
  \BibitemOpen
  \bibfield  {author} {\bibinfo {author} {\bibfnamefont {M.}~\bibnamefont
  {Alam}}, \bibinfo {author} {\bibfnamefont {A.}~\bibnamefont {Ash-Saki}},\
  and\ \bibinfo {author} {\bibfnamefont {S.}~\bibnamefont {Ghosh}},\ }in\ \href
  {https://doi.org/https://doi.org/10.48550/arXiv.2002.01089} {\emph {\bibinfo
  {booktitle} {2020 Design, Automation \& Test in Europe Conference \&
  Exhibition (DATE)}}}\ (\bibinfo {organization} {IEEE},\ \bibinfo {year}
  {2020})\ pp.\ \bibinfo {pages} {686--689}\BibitemShut {NoStop}%
\bibitem [{\citenamefont {Khairy}\ \emph {et~al.}(2020)\citenamefont {Khairy},
  \citenamefont {Shaydulin}, \citenamefont {Cincio}, \citenamefont {Alexeev},\
  and\ \citenamefont {Balaprakash}}]{khairy2020learning}%
  \BibitemOpen
  \bibfield  {author} {\bibinfo {author} {\bibfnamefont {S.}~\bibnamefont
  {Khairy}}, \bibinfo {author} {\bibfnamefont {R.}~\bibnamefont {Shaydulin}},
  \bibinfo {author} {\bibfnamefont {L.}~\bibnamefont {Cincio}}, \bibinfo
  {author} {\bibfnamefont {Y.}~\bibnamefont {Alexeev}},\ and\ \bibinfo {author}
  {\bibfnamefont {P.}~\bibnamefont {Balaprakash}},\ }in\ \href
  {https://doi.org/https://doi.org/10.1609/aaai.v34i03.5616} {\emph {\bibinfo
  {booktitle} {Proceedings of the AAAI conference on artificial
  intelligence}}},\ Vol.~\bibinfo {volume} {34}\ (\bibinfo {year} {2020})\ pp.\
  \bibinfo {pages} {2367--2375}\BibitemShut {NoStop}%
\bibitem [{\citenamefont {Yao}\ \emph {et~al.}(2020)\citenamefont {Yao},
  \citenamefont {Bukov},\ and\ \citenamefont {Lin}}]{yao2020policy}%
  \BibitemOpen
  \bibfield  {author} {\bibinfo {author} {\bibfnamefont {J.}~\bibnamefont
  {Yao}}, \bibinfo {author} {\bibfnamefont {M.}~\bibnamefont {Bukov}},\ and\
  \bibinfo {author} {\bibfnamefont {L.}~\bibnamefont {Lin}},\ }in\ \href
  {http://proceedings.mlr.press/v107/yao20a/yao20a.pdf} {\emph {\bibinfo
  {booktitle} {Mathematical and scientific machine learning}}}\ (\bibinfo
  {organization} {PMLR},\ \bibinfo {year} {2020})\ pp.\ \bibinfo {pages}
  {605--634}\BibitemShut {NoStop}%
\bibitem [{\citenamefont {Moussa}\ \emph {et~al.}(2022)\citenamefont {Moussa},
  \citenamefont {Wang}, \citenamefont {B{\"a}ck},\ and\ \citenamefont
  {Dunjko}}]{moussa2022unsupervised}%
  \BibitemOpen
  \bibfield  {author} {\bibinfo {author} {\bibfnamefont {C.}~\bibnamefont
  {Moussa}}, \bibinfo {author} {\bibfnamefont {H.}~\bibnamefont {Wang}},
  \bibinfo {author} {\bibfnamefont {T.}~\bibnamefont {B{\"a}ck}},\ and\
  \bibinfo {author} {\bibfnamefont {V.}~\bibnamefont {Dunjko}},\ }\href
  {https://doi.org/https://doi.org/10.1140/epjqt/s40507-022-00131-4} {\bibfield
   {journal} {\bibinfo  {journal} {EPJ Quantum Technology}\ }\textbf {\bibinfo
  {volume} {9}},\ \bibinfo {pages} {11} (\bibinfo {year} {2022})}\BibitemShut
  {NoStop}%
\bibitem [{\citenamefont {Wauters}\ \emph {et~al.}(2020)\citenamefont
  {Wauters}, \citenamefont {Panizon}, \citenamefont {Mbeng},\ and\
  \citenamefont {Santoro}}]{wauters2020rl}%
  \BibitemOpen
  \bibfield  {author} {\bibinfo {author} {\bibfnamefont {M.~M.}\ \bibnamefont
  {Wauters}}, \bibinfo {author} {\bibfnamefont {E.}~\bibnamefont {Panizon}},
  \bibinfo {author} {\bibfnamefont {G.~B.}\ \bibnamefont {Mbeng}},\ and\
  \bibinfo {author} {\bibfnamefont {G.~E.}\ \bibnamefont {Santoro}},\ }\href
  {https://doi.org/10.1103/PhysRevResearch.2.033446} {\bibfield  {journal}
  {\bibinfo  {journal} {Phys. Rev. Res.}\ }\textbf {\bibinfo {volume} {2}},\
  \bibinfo {pages} {033446} (\bibinfo {year} {2020})}\BibitemShut {NoStop}%
\bibitem [{\citenamefont {Xie}\ \emph {et~al.}(2023)\citenamefont {Xie},
  \citenamefont {Lee}, \citenamefont {Cai}, \citenamefont {Saito},\ and\
  \citenamefont {Asai}}]{xie2023quantum}%
  \BibitemOpen
  \bibfield  {author} {\bibinfo {author} {\bibfnamefont {N.}~\bibnamefont
  {Xie}}, \bibinfo {author} {\bibfnamefont {X.}~\bibnamefont {Lee}}, \bibinfo
  {author} {\bibfnamefont {D.}~\bibnamefont {Cai}}, \bibinfo {author}
  {\bibfnamefont {Y.}~\bibnamefont {Saito}},\ and\ \bibinfo {author}
  {\bibfnamefont {N.}~\bibnamefont {Asai}},\ }in\ \href
  {https://doi.org/10.1088/1742-6596/2595/1/012001} {\emph {\bibinfo
  {booktitle} {Journal of Physics: Conference Series}}},\ Vol.\ \bibinfo
  {volume} {2595}\ (\bibinfo {organization} {IOP Publishing},\ \bibinfo {year}
  {2023})\ p.\ \bibinfo {pages} {012001}\BibitemShut {NoStop}%
\bibitem [{\citenamefont {Patel}\ \emph {et~al.}(2024)\citenamefont {Patel},
  \citenamefont {Jerbi}, \citenamefont {B{\"a}ck},\ and\ \citenamefont
  {Dunjko}}]{patel2024reinforcement}%
  \BibitemOpen
  \bibfield  {author} {\bibinfo {author} {\bibfnamefont {Y.~J.}\ \bibnamefont
  {Patel}}, \bibinfo {author} {\bibfnamefont {S.}~\bibnamefont {Jerbi}},
  \bibinfo {author} {\bibfnamefont {T.}~\bibnamefont {B{\"a}ck}},\ and\
  \bibinfo {author} {\bibfnamefont {V.}~\bibnamefont {Dunjko}},\ }\href
  {https://doi.org/https://doi.org/10.1140/epjqt/s40507-023-00214-w} {\bibfield
   {journal} {\bibinfo  {journal} {EPJ Quantum Technology}\ }\textbf {\bibinfo
  {volume} {11}},\ \bibinfo {pages} {6} (\bibinfo {year} {2024})}\BibitemShut
  {NoStop}%
\bibitem [{\citenamefont {Brandao}\ \emph {et~al.}(2018)\citenamefont
  {Brandao}, \citenamefont {Broughton}, \citenamefont {Farhi}, \citenamefont
  {Gutmann},\ and\ \citenamefont {Neven}}]{brandao2018fixedcontrol}%
  \BibitemOpen
  \bibfield  {author} {\bibinfo {author} {\bibfnamefont {F.~G. S.~L.}\
  \bibnamefont {Brandao}}, \bibinfo {author} {\bibfnamefont {M.}~\bibnamefont
  {Broughton}}, \bibinfo {author} {\bibfnamefont {E.}~\bibnamefont {Farhi}},
  \bibinfo {author} {\bibfnamefont {S.}~\bibnamefont {Gutmann}},\ and\ \bibinfo
  {author} {\bibfnamefont {H.}~\bibnamefont {Neven}},\ }\href
  {https://arxiv.org/abs/1812.04170} {\bibinfo {title} {For fixed control
  parameters the quantum approximate optimization algorithm's objective
  function value concentrates for typical instances}} (\bibinfo {year}
  {2018}),\ \Eprint {https://arxiv.org/abs/1812.04170} {arXiv:1812.04170
  [quant-ph]} \BibitemShut {NoStop}%
\bibitem [{\citenamefont {Basso}\ \emph {et~al.}(2022)\citenamefont {Basso},
  \citenamefont {Farhi}, \citenamefont {Marwaha}, \citenamefont {Villalonga},\
  and\ \citenamefont {Zhou}}]{basso2022concen}%
  \BibitemOpen
  \bibfield  {author} {\bibinfo {author} {\bibfnamefont {J.}~\bibnamefont
  {Basso}}, \bibinfo {author} {\bibfnamefont {E.}~\bibnamefont {Farhi}},
  \bibinfo {author} {\bibfnamefont {K.}~\bibnamefont {Marwaha}}, \bibinfo
  {author} {\bibfnamefont {B.}~\bibnamefont {Villalonga}},\ and\ \bibinfo
  {author} {\bibfnamefont {L.}~\bibnamefont {Zhou}},\ }in\ \href
  {https://doi.org/10.4230/LIPIcs.TQC.2022.7} {\emph {\bibinfo {booktitle}
  {17th Conference on the Theory of Quantum Computation, Communication and
  Cryptography (TQC 2022)}}},\ \bibinfo {series} {Leibniz International
  Proceedings in Informatics (LIPIcs)}, Vol.\ \bibinfo {volume} {232},\
  \bibinfo {editor} {edited by\ \bibinfo {editor} {\bibfnamefont
  {F.}~\bibnamefont {Le~Gall}}\ and\ \bibinfo {editor} {\bibfnamefont
  {T.}~\bibnamefont {Morimae}}}\ (\bibinfo  {publisher} {Schloss Dagstuhl --
  Leibniz-Zentrum f{\"u}r Informatik},\ \bibinfo {address} {Dagstuhl,
  Germany},\ \bibinfo {year} {2022})\ pp.\ \bibinfo {pages}
  {7:1--7:21}\BibitemShut {NoStop}%
\bibitem [{\citenamefont {Farhi}\ \emph {et~al.}(2022)\citenamefont {Farhi},
  \citenamefont {Goldstone}, \citenamefont {Gutmann},\ and\ \citenamefont
  {Zhou}}]{farhi2022quantum}%
  \BibitemOpen
  \bibfield  {author} {\bibinfo {author} {\bibfnamefont {E.}~\bibnamefont
  {Farhi}}, \bibinfo {author} {\bibfnamefont {J.}~\bibnamefont {Goldstone}},
  \bibinfo {author} {\bibfnamefont {S.}~\bibnamefont {Gutmann}},\ and\ \bibinfo
  {author} {\bibfnamefont {L.}~\bibnamefont {Zhou}},\ }\href
  {https://doi.org/https://doi.org/10.22331/q-2022-07-07-759} {\bibfield
  {journal} {\bibinfo  {journal} {Quantum}\ }\textbf {\bibinfo {volume} {6}},\
  \bibinfo {pages} {759} (\bibinfo {year} {2022})}\BibitemShut {NoStop}%
\bibitem [{\citenamefont {Sureshbabu}\ \emph {et~al.}(2024)\citenamefont
  {Sureshbabu}, \citenamefont {Herman}, \citenamefont {Shaydulin},
  \citenamefont {Basso}, \citenamefont {Chakrabarti}, \citenamefont {Sun},\
  and\ \citenamefont {Pistoia}}]{sureshbabu2024parameter}%
  \BibitemOpen
  \bibfield  {author} {\bibinfo {author} {\bibfnamefont {S.~H.}\ \bibnamefont
  {Sureshbabu}}, \bibinfo {author} {\bibfnamefont {D.}~\bibnamefont {Herman}},
  \bibinfo {author} {\bibfnamefont {R.}~\bibnamefont {Shaydulin}}, \bibinfo
  {author} {\bibfnamefont {J.}~\bibnamefont {Basso}}, \bibinfo {author}
  {\bibfnamefont {S.}~\bibnamefont {Chakrabarti}}, \bibinfo {author}
  {\bibfnamefont {Y.}~\bibnamefont {Sun}},\ and\ \bibinfo {author}
  {\bibfnamefont {M.}~\bibnamefont {Pistoia}},\ }\href
  {https://doi.org/https://doi.org/10.22331/q-2024-01-18-1231} {\bibfield
  {journal} {\bibinfo  {journal} {Quantum}\ }\textbf {\bibinfo {volume} {8}},\
  \bibinfo {pages} {1231} (\bibinfo {year} {2024})}\BibitemShut {NoStop}%
\bibitem [{\citenamefont {Vijendran}\ \emph {et~al.}(2025)\citenamefont
  {Vijendran}, \citenamefont {Koh}, \citenamefont {Bae}, \citenamefont {Kwon},
  \citenamefont {Lam},\ and\ \citenamefont
  {Assad}}]{vijendran2025nearoptimalparametertuninglevel1}%
  \BibitemOpen
  \bibfield  {author} {\bibinfo {author} {\bibfnamefont {V.}~\bibnamefont
  {Vijendran}}, \bibinfo {author} {\bibfnamefont {D.~E.}\ \bibnamefont {Koh}},
  \bibinfo {author} {\bibfnamefont {E.}~\bibnamefont {Bae}}, \bibinfo {author}
  {\bibfnamefont {H.}~\bibnamefont {Kwon}}, \bibinfo {author} {\bibfnamefont
  {P.~K.}\ \bibnamefont {Lam}},\ and\ \bibinfo {author} {\bibfnamefont {S.~M.}\
  \bibnamefont {Assad}},\ }\href {https://arxiv.org/abs/2501.16419} {\bibinfo
  {title} {Near-optimal parameter tuning of level-1 qaoa for ising models}}
  (\bibinfo {year} {2025}),\ \Eprint {https://arxiv.org/abs/2501.16419}
  {arXiv:2501.16419 [quant-ph]} \BibitemShut {NoStop}%
\bibitem [{\citenamefont {Galda}\ \emph {et~al.}(2021)\citenamefont {Galda},
  \citenamefont {Liu}, \citenamefont {Lykov}, \citenamefont {Alexeev},\ and\
  \citenamefont {Safro}}]{galda2021transfer}%
  \BibitemOpen
  \bibfield  {author} {\bibinfo {author} {\bibfnamefont {A.}~\bibnamefont
  {Galda}}, \bibinfo {author} {\bibfnamefont {X.}~\bibnamefont {Liu}}, \bibinfo
  {author} {\bibfnamefont {D.}~\bibnamefont {Lykov}}, \bibinfo {author}
  {\bibfnamefont {Y.}~\bibnamefont {Alexeev}},\ and\ \bibinfo {author}
  {\bibfnamefont {I.}~\bibnamefont {Safro}},\ }in\ \href
  {https://doi.org/10.1109/QCE52317.2021.00034} {\emph {\bibinfo {booktitle}
  {2021 IEEE International Conference on Quantum Computing and Engineering
  (QCE)}}}\ (\bibinfo {year} {2021})\ pp.\ \bibinfo {pages}
  {171--180}\BibitemShut {NoStop}%
\bibitem [{\citenamefont {Shaydulin}\ \emph {et~al.}(2023)\citenamefont
  {Shaydulin}, \citenamefont {Lotshaw}, \citenamefont {Larson}, \citenamefont
  {Ostrowski},\ and\ \citenamefont {Humble}}]{shaydulin2023parameter}%
  \BibitemOpen
  \bibfield  {author} {\bibinfo {author} {\bibfnamefont {R.}~\bibnamefont
  {Shaydulin}}, \bibinfo {author} {\bibfnamefont {P.~C.}\ \bibnamefont
  {Lotshaw}}, \bibinfo {author} {\bibfnamefont {J.}~\bibnamefont {Larson}},
  \bibinfo {author} {\bibfnamefont {J.}~\bibnamefont {Ostrowski}},\ and\
  \bibinfo {author} {\bibfnamefont {T.~S.}\ \bibnamefont {Humble}},\ }\href
  {https://doi.org/https://doi.org/10.1145/3584706} {\bibfield  {journal}
  {\bibinfo  {journal} {ACM Transactions on Quantum Computing}\ }\textbf
  {\bibinfo {volume} {4}},\ \bibinfo {pages} {1} (\bibinfo {year}
  {2023})}\BibitemShut {NoStop}%
\bibitem [{\citenamefont {Montanez-Barrera}\ \emph {et~al.}(2024)\citenamefont
  {Montanez-Barrera}, \citenamefont {Willsch},\ and\ \citenamefont
  {Michielsen}}]{montanez2024transfer}%
  \BibitemOpen
  \bibfield  {author} {\bibinfo {author} {\bibfnamefont {J.~A.}\ \bibnamefont
  {Montanez-Barrera}}, \bibinfo {author} {\bibfnamefont {D.}~\bibnamefont
  {Willsch}},\ and\ \bibinfo {author} {\bibfnamefont {K.}~\bibnamefont
  {Michielsen}},\ }\href {https://arxiv.org/abs/2402.05549} {\bibinfo {title}
  {Transfer learning of optimal qaoa parameters in combinatorial optimization}}
  (\bibinfo {year} {2024}),\ \Eprint {https://arxiv.org/abs/2402.05549}
  {arXiv:2402.05549 [quant-ph]} \BibitemShut {NoStop}%
\bibitem [{\citenamefont {Sakai}\ \emph {et~al.}(2024)\citenamefont {Sakai},
  \citenamefont {Matsuyama}, \citenamefont {Tam}, \citenamefont {Yamashiro},\
  and\ \citenamefont {Fujii}}]{sakai2024linearly}%
  \BibitemOpen
  \bibfield  {author} {\bibinfo {author} {\bibfnamefont {R.}~\bibnamefont
  {Sakai}}, \bibinfo {author} {\bibfnamefont {H.}~\bibnamefont {Matsuyama}},
  \bibinfo {author} {\bibfnamefont {W.-H.}\ \bibnamefont {Tam}}, \bibinfo
  {author} {\bibfnamefont {Y.}~\bibnamefont {Yamashiro}},\ and\ \bibinfo
  {author} {\bibfnamefont {K.}~\bibnamefont {Fujii}},\ }\href
  {https://arxiv.org/abs/2405.00655} {\bibinfo {title} {Linearly simplified
  qaoa parameters and transferability}} (\bibinfo {year} {2024}),\ \Eprint
  {https://arxiv.org/abs/2405.00655} {arXiv:2405.00655 [quant-ph]} \BibitemShut
  {NoStop}%
\bibitem [{\citenamefont {Sud}\ \emph {et~al.}(2024)\citenamefont {Sud},
  \citenamefont {Hadfield}, \citenamefont {Rieffel}, \citenamefont {Tubman},\
  and\ \citenamefont {Hogg}}]{sud2024qaoa}%
  \BibitemOpen
  \bibfield  {author} {\bibinfo {author} {\bibfnamefont {J.}~\bibnamefont
  {Sud}}, \bibinfo {author} {\bibfnamefont {S.}~\bibnamefont {Hadfield}},
  \bibinfo {author} {\bibfnamefont {E.}~\bibnamefont {Rieffel}}, \bibinfo
  {author} {\bibfnamefont {N.}~\bibnamefont {Tubman}},\ and\ \bibinfo {author}
  {\bibfnamefont {T.}~\bibnamefont {Hogg}},\ }\href
  {https://doi.org/10.1103/PhysRevResearch.6.023171} {\bibfield  {journal}
  {\bibinfo  {journal} {Phys. Rev. Res.}\ }\textbf {\bibinfo {volume} {6}},\
  \bibinfo {pages} {023171} (\bibinfo {year} {2024})}\BibitemShut {NoStop}%
\bibitem [{\citenamefont {Dehn}\ \emph {et~al.}(2025)\citenamefont {Dehn},
  \citenamefont {Zaefferer}, \citenamefont {Hellstern}, \citenamefont
  {Reiter},\ and\ \citenamefont {Wellens}}]{dehn2025linearramp}%
  \BibitemOpen
  \bibfield  {author} {\bibinfo {author} {\bibfnamefont {V.}~\bibnamefont
  {Dehn}}, \bibinfo {author} {\bibfnamefont {M.}~\bibnamefont {Zaefferer}},
  \bibinfo {author} {\bibfnamefont {G.}~\bibnamefont {Hellstern}}, \bibinfo
  {author} {\bibfnamefont {F.}~\bibnamefont {Reiter}},\ and\ \bibinfo {author}
  {\bibfnamefont {T.}~\bibnamefont {Wellens}},\ }\href
  {https://arxiv.org/abs/2504.08577} {\bibinfo {title} {Extrapolation method to
  optimize linear-ramp qaoa parameters: Evaluation of qaoa runtime scaling}}
  (\bibinfo {year} {2025}),\ \Eprint {https://arxiv.org/abs/2504.08577}
  {arXiv:2504.08577 [quant-ph]} \BibitemShut {NoStop}%
\bibitem [{\citenamefont {Sack}\ and\ \citenamefont
  {Serbyn}(2021)}]{sack2021quantum}%
  \BibitemOpen
  \bibfield  {author} {\bibinfo {author} {\bibfnamefont {S.~H.}\ \bibnamefont
  {Sack}}\ and\ \bibinfo {author} {\bibfnamefont {M.}~\bibnamefont {Serbyn}},\
  }\href {https://doi.org/https://doi.org/10.22331/q-2021-07-01-491} {\bibfield
   {journal} {\bibinfo  {journal} {quantum}\ }\textbf {\bibinfo {volume} {5}},\
  \bibinfo {pages} {491} (\bibinfo {year} {2021})}\BibitemShut {NoStop}%
\bibitem [{\citenamefont {Kremenetski}\ \emph {et~al.}(2021)\citenamefont
  {Kremenetski}, \citenamefont {Hogg}, \citenamefont {Hadfield}, \citenamefont
  {Cotton},\ and\ \citenamefont {Tubman}}]{kremenetski2021linearramp}%
  \BibitemOpen
  \bibfield  {author} {\bibinfo {author} {\bibfnamefont {V.}~\bibnamefont
  {Kremenetski}}, \bibinfo {author} {\bibfnamefont {T.}~\bibnamefont {Hogg}},
  \bibinfo {author} {\bibfnamefont {S.}~\bibnamefont {Hadfield}}, \bibinfo
  {author} {\bibfnamefont {S.~J.}\ \bibnamefont {Cotton}},\ and\ \bibinfo
  {author} {\bibfnamefont {N.~M.}\ \bibnamefont {Tubman}},\ }\href
  {https://arxiv.org/abs/2108.13056} {\bibinfo {title} {Quantum alternating
  operator ansatz (qaoa) phase diagrams and applications for quantum
  chemistry}} (\bibinfo {year} {2021}),\ \Eprint
  {https://arxiv.org/abs/2108.13056} {arXiv:2108.13056 [quant-ph]} \BibitemShut
  {NoStop}%
\bibitem [{\citenamefont {Kremenetski}\ \emph {et~al.}(2023)\citenamefont
  {Kremenetski}, \citenamefont {Apte}, \citenamefont {Hogg}, \citenamefont
  {Hadfield},\ and\ \citenamefont {Tubman}}]{kremenetski2023linramp}%
  \BibitemOpen
  \bibfield  {author} {\bibinfo {author} {\bibfnamefont {V.}~\bibnamefont
  {Kremenetski}}, \bibinfo {author} {\bibfnamefont {A.}~\bibnamefont {Apte}},
  \bibinfo {author} {\bibfnamefont {T.}~\bibnamefont {Hogg}}, \bibinfo {author}
  {\bibfnamefont {S.}~\bibnamefont {Hadfield}},\ and\ \bibinfo {author}
  {\bibfnamefont {N.~M.}\ \bibnamefont {Tubman}},\ }\href
  {https://arxiv.org/abs/2305.04455} {\bibinfo {title} {Quantum alternating
  operator ansatz (qaoa) beyond low depth with gradually changing unitaries}}
  (\bibinfo {year} {2023}),\ \Eprint {https://arxiv.org/abs/2305.04455}
  {arXiv:2305.04455 [quant-ph]} \BibitemShut {NoStop}%
\bibitem [{\citenamefont {Montanez-Barrera}\ and\ \citenamefont
  {Michielsen}(2024{\natexlab{b}})}]{montanezbarrera2024qaoa}%
  \BibitemOpen
  \bibfield  {author} {\bibinfo {author} {\bibfnamefont {J.~A.}\ \bibnamefont
  {Montanez-Barrera}}\ and\ \bibinfo {author} {\bibfnamefont {K.}~\bibnamefont
  {Michielsen}},\ }\href
  {https://doi.org/https://doi.org/10.48550/arXiv.2405.09169} {\bibinfo {title}
  {Towards a universal qaoa protocol: Evidence of a scaling advantage in
  solving some combinatorial optimization problems}} (\bibinfo {year}
  {2024}{\natexlab{b}}),\ \Eprint {https://arxiv.org/abs/2405.09169}
  {arXiv:2405.09169 [quant-ph]} \BibitemShut {NoStop}%
\bibitem [{\citenamefont {Apte}\ \emph {et~al.}(2025)\citenamefont {Apte},
  \citenamefont {Sureshbabu}, \citenamefont {Shaydulin}, \citenamefont
  {Boulebnane}, \citenamefont {He}, \citenamefont {Herman}, \citenamefont
  {Sud},\ and\ \citenamefont {Pistoia}}]{apte2025iterativeinterp}%
  \BibitemOpen
  \bibfield  {author} {\bibinfo {author} {\bibfnamefont {A.}~\bibnamefont
  {Apte}}, \bibinfo {author} {\bibfnamefont {S.~H.}\ \bibnamefont
  {Sureshbabu}}, \bibinfo {author} {\bibfnamefont {R.}~\bibnamefont
  {Shaydulin}}, \bibinfo {author} {\bibfnamefont {S.}~\bibnamefont
  {Boulebnane}}, \bibinfo {author} {\bibfnamefont {Z.}~\bibnamefont {He}},
  \bibinfo {author} {\bibfnamefont {D.}~\bibnamefont {Herman}}, \bibinfo
  {author} {\bibfnamefont {J.}~\bibnamefont {Sud}},\ and\ \bibinfo {author}
  {\bibfnamefont {M.}~\bibnamefont {Pistoia}},\ }\href
  {https://doi.org/https://doi.org/10.48550/arXiv.2504.01694} {\bibinfo {title}
  {Iterative interpolation schedules for quantum approximate optimization
  algorithm}} (\bibinfo {year} {2025}),\ \Eprint
  {https://arxiv.org/abs/2504.01694} {arXiv:2504.01694 [quant-ph]} \BibitemShut
  {NoStop}%
\bibitem [{\citenamefont {Yang}\ \emph {et~al.}(2007)\citenamefont {Yang},
  \citenamefont {Li},\ and\ \citenamefont {Cheng}}]{yang2007efficiency}%
  \BibitemOpen
  \bibfield  {author} {\bibinfo {author} {\bibfnamefont {D.}~\bibnamefont
  {Yang}}, \bibinfo {author} {\bibfnamefont {G.}~\bibnamefont {Li}},\ and\
  \bibinfo {author} {\bibfnamefont {G.}~\bibnamefont {Cheng}},\ }\href
  {https://doi.org/https://doi.org/10.1016/j.chaos.2006.04.057} {\bibfield
  {journal} {\bibinfo  {journal} {Chaos, Solitons \& Fractals}\ }\textbf
  {\bibinfo {volume} {34}},\ \bibinfo {pages} {1366} (\bibinfo {year}
  {2007})}\BibitemShut {NoStop}%
\bibitem [{\citenamefont {Luo}\ \emph {et~al.}(2008)\citenamefont {Luo},
  \citenamefont {Tang},\ and\ \citenamefont {Zhou}}]{luo2008chaos}%
  \BibitemOpen
  \bibfield  {author} {\bibinfo {author} {\bibfnamefont {Y.-Z.}\ \bibnamefont
  {Luo}}, \bibinfo {author} {\bibfnamefont {G.-J.}\ \bibnamefont {Tang}},\ and\
  \bibinfo {author} {\bibfnamefont {L.-N.}\ \bibnamefont {Zhou}},\ }\href
  {https://doi.org/https://doi.org/10.1016/j.asoc.2007.05.013} {\bibfield
  {journal} {\bibinfo  {journal} {Applied Soft Computing}\ }\textbf {\bibinfo
  {volume} {8}},\ \bibinfo {pages} {1068} (\bibinfo {year} {2008})}\BibitemShut
  {NoStop}%
\bibitem [{\citenamefont {Zhang}\ \emph {et~al.}(2024)\citenamefont {Zhang},
  \citenamefont {Lu}, \citenamefont {Zhao}, \citenamefont {Li},\ and\
  \citenamefont {Yan}}]{zhang2024chaos}%
  \BibitemOpen
  \bibfield  {author} {\bibinfo {author} {\bibfnamefont {Y.}~\bibnamefont
  {Zhang}}, \bibinfo {author} {\bibfnamefont {J.}~\bibnamefont {Lu}}, \bibinfo
  {author} {\bibfnamefont {C.}~\bibnamefont {Zhao}}, \bibinfo {author}
  {\bibfnamefont {Z.}~\bibnamefont {Li}},\ and\ \bibinfo {author}
  {\bibfnamefont {J.}~\bibnamefont {Yan}},\ }\href
  {https://doi.org/10.1142/S0218127424502055} {\bibfield  {journal} {\bibinfo
  {journal} {International Journal of Bifurcation and Chaos}\ }\textbf
  {\bibinfo {volume} {34}},\ \bibinfo {pages} {2450205} (\bibinfo {year}
  {2024})},\ \Eprint
  {https://arxiv.org/abs/https://doi.org/10.1142/S0218127424502055}
  {https://doi.org/10.1142/S0218127424502055} \BibitemShut {NoStop}%
\bibitem [{\citenamefont {Hansen}\ and\ \citenamefont
  {Jaumard}(1990)}]{hansen1990algorithms}%
  \BibitemOpen
  \bibfield  {author} {\bibinfo {author} {\bibfnamefont {P.}~\bibnamefont
  {Hansen}}\ and\ \bibinfo {author} {\bibfnamefont {B.}~\bibnamefont
  {Jaumard}},\ }\href@noop {} {\bibfield  {journal} {\bibinfo  {journal}
  {Computing}\ }\textbf {\bibinfo {volume} {44}},\ \bibinfo {pages} {279}
  (\bibinfo {year} {1990})}\BibitemShut {NoStop}%
\bibitem [{\citenamefont {Strogatz}(2018)}]{strogatz2018nonlinear}%
  \BibitemOpen
  \bibfield  {author} {\bibinfo {author} {\bibfnamefont {S.~H.}\ \bibnamefont
  {Strogatz}},\ }\href {https://doi.org/https://doi.org/10.1201/9780429492563}
  {\emph {\bibinfo {title} {Nonlinear dynamics and chaos: with applications to
  physics, biology, chemistry, and engineering}}}\ (\bibinfo  {publisher} {CRC
  press},\ \bibinfo {year} {2018})\BibitemShut {NoStop}%
\bibitem [{\citenamefont {Cook}(1971)}]{Cook1971}%
  \BibitemOpen
  \bibfield  {author} {\bibinfo {author} {\bibfnamefont {S.~A.}\ \bibnamefont
  {Cook}},\ }in\ \href {https://doi.org/10.1145/800157.805047} {\emph {\bibinfo
  {booktitle} {Proceedings of the Third Annual ACM Symposium on Theory of
  Computing}}},\ \bibinfo {series and number} {STOC '71}\ (\bibinfo
  {publisher} {Association for Computing Machinery},\ \bibinfo {address} {New
  York, NY, USA},\ \bibinfo {year} {1971})\ p.\ \bibinfo {pages}
  {151–158}\BibitemShut {NoStop}%
\bibitem [{\citenamefont {Garey}\ \emph {et~al.}(1976)\citenamefont {Garey},
  \citenamefont {Johnson},\ and\ \citenamefont {Stockmeyer}}]{GAREY1976237}%
  \BibitemOpen
  \bibfield  {author} {\bibinfo {author} {\bibfnamefont {M.}~\bibnamefont
  {Garey}}, \bibinfo {author} {\bibfnamefont {D.}~\bibnamefont {Johnson}},\
  and\ \bibinfo {author} {\bibfnamefont {L.}~\bibnamefont {Stockmeyer}},\
  }\href {https://doi.org/https://doi.org/10.1016/0304-3975(76)90059-1}
  {\bibfield  {journal} {\bibinfo  {journal} {Theoretical Computer Science}\
  }\textbf {\bibinfo {volume} {1}},\ \bibinfo {pages} {237} (\bibinfo {year}
  {1976})}\BibitemShut {NoStop}%
\bibitem [{\citenamefont {Crescenzi}\ and\ \citenamefont
  {Trevisan}(1999)}]{CRESCENZI199965}%
  \BibitemOpen
  \bibfield  {author} {\bibinfo {author} {\bibfnamefont {P.}~\bibnamefont
  {Crescenzi}}\ and\ \bibinfo {author} {\bibfnamefont {L.}~\bibnamefont
  {Trevisan}},\ }\href
  {https://doi.org/https://doi.org/10.1016/S0304-3975(98)00200-X} {\bibfield
  {journal} {\bibinfo  {journal} {Theoretical Computer Science}\ }\textbf
  {\bibinfo {volume} {225}},\ \bibinfo {pages} {65} (\bibinfo {year}
  {1999})}\BibitemShut {NoStop}%
\bibitem [{\citenamefont {Goerdt}(1996)}]{Goerdt1996}%
  \BibitemOpen
  \bibfield  {author} {\bibinfo {author} {\bibfnamefont {A.}~\bibnamefont
  {Goerdt}},\ }\href {https://doi.org/https://doi.org/10.1006/jcss.1996.0081}
  {\bibfield  {journal} {\bibinfo  {journal} {Journal of Computer and System
  Sciences}\ }\textbf {\bibinfo {volume} {53}},\ \bibinfo {pages} {469}
  (\bibinfo {year} {1996})}\BibitemShut {NoStop}%
\bibitem [{\citenamefont {Akshay}\ \emph {et~al.}(2020)\citenamefont {Akshay},
  \citenamefont {Philathong}, \citenamefont {Morales},\ and\ \citenamefont
  {Biamonte}}]{Akshay_2020}%
  \BibitemOpen
  \bibfield  {author} {\bibinfo {author} {\bibfnamefont {V.}~\bibnamefont
  {Akshay}}, \bibinfo {author} {\bibfnamefont {H.}~\bibnamefont {Philathong}},
  \bibinfo {author} {\bibfnamefont {M.}~\bibnamefont {Morales}},\ and\ \bibinfo
  {author} {\bibfnamefont {J.}~\bibnamefont {Biamonte}},\ }\bibfield  {journal}
  {\bibinfo  {journal} {Physical Review Letters}\ }\textbf {\bibinfo {volume}
  {124}},\ \href {https://doi.org/10.1103/physrevlett.124.090504}
  {10.1103/physrevlett.124.090504} (\bibinfo {year} {2020})\BibitemShut
  {NoStop}%
\bibitem [{\citenamefont {Santra}\ \emph {et~al.}(2014)\citenamefont {Santra},
  \citenamefont {Quiroz}, \citenamefont {Ver~Steeg},\ and\ \citenamefont
  {Lidar}}]{max2sat_108_qubits}%
  \BibitemOpen
  \bibfield  {author} {\bibinfo {author} {\bibfnamefont {S.}~\bibnamefont
  {Santra}}, \bibinfo {author} {\bibfnamefont {G.}~\bibnamefont {Quiroz}},
  \bibinfo {author} {\bibfnamefont {G.}~\bibnamefont {Ver~Steeg}},\ and\
  \bibinfo {author} {\bibfnamefont {D.~A.}\ \bibnamefont {Lidar}},\ }\href
  {https://doi.org/10.1088/1367-2630/16/4/045006} {\bibfield  {journal}
  {\bibinfo  {journal} {New Journal of Physics}\ }\textbf {\bibinfo {volume}
  {16}},\ \bibinfo {pages} {045006} (\bibinfo {year} {2014})}\BibitemShut
  {NoStop}%
\bibitem [{\citenamefont {Crawford}\ and\ \citenamefont
  {Auton}(1996)}]{Crawford1996}%
  \BibitemOpen
  \bibfield  {author} {\bibinfo {author} {\bibfnamefont {J.~M.}\ \bibnamefont
  {Crawford}}\ and\ \bibinfo {author} {\bibfnamefont {L.~D.}\ \bibnamefont
  {Auton}},\ }\href
  {https://doi.org/https://doi.org/10.1016/0004-3702(95)00046-1} {\bibfield
  {journal} {\bibinfo  {journal} {Artificial Intelligence}\ }\textbf {\bibinfo
  {volume} {81}},\ \bibinfo {pages} {31} (\bibinfo {year} {1996})},\ \bibinfo
  {note} {frontiers in Problem Solving: Phase Transitions and
  Complexity}\BibitemShut {NoStop}%
\bibitem [{\citenamefont {Akshay}\ \emph {et~al.}(2022)\citenamefont {Akshay},
  \citenamefont {Philathong}, \citenamefont {Campos}, \citenamefont
  {Rabinovich}, \citenamefont {Zacharov}, \citenamefont {Zhang},\ and\
  \citenamefont {Biamonte}}]{Akshay2022}%
  \BibitemOpen
  \bibfield  {author} {\bibinfo {author} {\bibfnamefont {V.}~\bibnamefont
  {Akshay}}, \bibinfo {author} {\bibfnamefont {H.}~\bibnamefont {Philathong}},
  \bibinfo {author} {\bibfnamefont {E.}~\bibnamefont {Campos}}, \bibinfo
  {author} {\bibfnamefont {D.}~\bibnamefont {Rabinovich}}, \bibinfo {author}
  {\bibfnamefont {I.}~\bibnamefont {Zacharov}}, \bibinfo {author}
  {\bibfnamefont {X.-M.}\ \bibnamefont {Zhang}},\ and\ \bibinfo {author}
  {\bibfnamefont {J.~D.}\ \bibnamefont {Biamonte}},\ }\href
  {https://doi.org/10.1103/PhysRevA.106.042438} {\bibfield  {journal} {\bibinfo
   {journal} {Phys. Rev. A}\ }\textbf {\bibinfo {volume} {106}},\ \bibinfo
  {pages} {042438} (\bibinfo {year} {2022})}\BibitemShut {NoStop}%
\bibitem [{\citenamefont {Whittaker}(1991)}]{Whittaker1991}%
  \BibitemOpen
  \bibfield  {author} {\bibinfo {author} {\bibfnamefont {J.~V.}\ \bibnamefont
  {Whittaker}},\ }\href {http://www.jstor.org/stable/2324869} {\bibfield
  {journal} {\bibinfo  {journal} {The American Mathematical Monthly}\ }\textbf
  {\bibinfo {volume} {98}},\ \bibinfo {pages} {489} (\bibinfo {year}
  {1991})}\BibitemShut {NoStop}%
\bibitem [{\citenamefont {Hirsch}\ \emph {et~al.}(2013)\citenamefont {Hirsch},
  \citenamefont {Smale},\ and\ \citenamefont {Devaney}}]{Hirsch2013}%
  \BibitemOpen
  \bibfield  {author} {\bibinfo {author} {\bibfnamefont {M.~W.}\ \bibnamefont
  {Hirsch}}, \bibinfo {author} {\bibfnamefont {S.}~\bibnamefont {Smale}},\ and\
  \bibinfo {author} {\bibfnamefont {R.~L.}\ \bibnamefont {Devaney}},\ }in\
  \href {https://doi.org/https://doi.org/10.1016/B978-0-12-382010-5.00015-4}
  {\emph {\bibinfo {booktitle} {Differential Equations, Dynamical Systems, and
  an Introduction to Chaos (Third Edition)}}},\ \bibinfo {editor} {edited by\
  \bibinfo {editor} {\bibfnamefont {M.~W.}\ \bibnamefont {Hirsch}}, \bibinfo
  {editor} {\bibfnamefont {S.}~\bibnamefont {Smale}},\ and\ \bibinfo {editor}
  {\bibfnamefont {R.~L.}\ \bibnamefont {Devaney}}}\ (\bibinfo  {publisher}
  {Academic Press},\ \bibinfo {address} {Boston},\ \bibinfo {year} {2013})\
  \bibinfo {edition} {third edition}\ ed.,\ pp.\ \bibinfo {pages}
  {329--359}\BibitemShut {NoStop}%
\bibitem [{\citenamefont {Spall}(1997)}]{SPALL1997109}%
  \BibitemOpen
  \bibfield  {author} {\bibinfo {author} {\bibfnamefont {J.~C.}\ \bibnamefont
  {Spall}},\ }\href
  {https://doi.org/https://doi.org/10.1016/S0005-1098(96)00149-5} {\bibfield
  {journal} {\bibinfo  {journal} {Automatica}\ }\textbf {\bibinfo {volume}
  {33}},\ \bibinfo {pages} {109} (\bibinfo {year} {1997})}\BibitemShut
  {NoStop}%
\bibitem [{\citenamefont {Spall}(1998)}]{spall_spsa_implementation}%
  \BibitemOpen
  \bibfield  {author} {\bibinfo {author} {\bibfnamefont {J.}~\bibnamefont
  {Spall}},\ }\href {https://doi.org/10.1109/7.705889} {\bibfield  {journal}
  {\bibinfo  {journal} {IEEE Transactions on Aerospace and Electronic Systems}\
  }\textbf {\bibinfo {volume} {34}},\ \bibinfo {pages} {817} (\bibinfo {year}
  {1998})}\BibitemShut {NoStop}%
\bibitem [{\citenamefont {Sadegh}(1997)}]{SADEGH1997281}%
  \BibitemOpen
  \bibfield  {author} {\bibinfo {author} {\bibfnamefont {P.}~\bibnamefont
  {Sadegh}},\ }\href
  {https://doi.org/https://doi.org/10.1016/S1474-6670(17)42860-6} {\bibfield
  {journal} {\bibinfo  {journal} {IFAC Proceedings Volumes}\ }\textbf {\bibinfo
  {volume} {30}},\ \bibinfo {pages} {281} (\bibinfo {year} {1997})},\ \bibinfo
  {note} {iFAC Symposium on System Identification (SYSID'97), Kitakyushu,
  Fukuoka, Japan, 8-11 July 1997}\BibitemShut {NoStop}%
\bibitem [{\citenamefont {Kechris}(1995)}]{Kechris1995}%
  \BibitemOpen
  \bibfield  {author} {\bibinfo {author} {\bibfnamefont {A.~S.}\ \bibnamefont
  {Kechris}},\ }\href
  {https://doi.org/https://doi.org/10.1007/978-1-4612-4190-4} {\emph {\bibinfo
  {title} {Classical Descriptive Set Theory}}},\ \bibinfo {edition} {1st}\ ed.\
  (\bibinfo  {publisher} {Springer New York, NY},\ \bibinfo {year}
  {1995})\BibitemShut {NoStop}%
\bibitem [{\citenamefont {Kallenberg}(2021)}]{Kallenberg2021}%
  \BibitemOpen
  \bibfield  {author} {\bibinfo {author} {\bibfnamefont {O.}~\bibnamefont
  {Kallenberg}},\ }\href
  {https://doi.org/https://doi.org/10.1007/978-3-030-61871-1} {\emph {\bibinfo
  {title} {Foundations of Modern Probability}}},\ \bibinfo {edition} {3rd}\
  ed.\ (\bibinfo  {publisher} {Springer Cham},\ \bibinfo {year}
  {2021})\BibitemShut {NoStop}%
\bibitem [{\citenamefont {Einsiedler}(2011)}]{Einsiedler2011}%
  \BibitemOpen
  \bibfield  {author} {\bibinfo {author} {\bibfnamefont {M.}~\bibnamefont
  {Einsiedler}},\ }\href
  {https://doi.org/https://doi.org/10.1007/978-0-85729-021-2} {\emph {\bibinfo
  {title} {Ergodic Theory}}}\ (\bibinfo  {publisher} {Springer London},\
  \bibinfo {year} {2011})\BibitemShut {NoStop}%
\bibitem [{\citenamefont {Birkhoff}(1931)}]{Birkhoff1931}%
  \BibitemOpen
  \bibfield  {author} {\bibinfo {author} {\bibfnamefont {G.~D.}\ \bibnamefont
  {Birkhoff}},\ }\href {https://doi.org/10.1073/pnas.17.2.656} {\bibfield
  {journal} {\bibinfo  {journal} {Proceedings of the National Academy of
  Sciences}\ }\textbf {\bibinfo {volume} {17}},\ \bibinfo {pages} {656}
  (\bibinfo {year} {1931})},\ \Eprint
  {https://arxiv.org/abs/https://www.pnas.org/doi/pdf/10.1073/pnas.17.2.656}
  {https://www.pnas.org/doi/pdf/10.1073/pnas.17.2.656} \BibitemShut {NoStop}%
\bibitem [{\citenamefont {Moore}(2015)}]{Moore2015}%
  \BibitemOpen
  \bibfield  {author} {\bibinfo {author} {\bibfnamefont {C.~C.}\ \bibnamefont
  {Moore}},\ }\href {https://doi.org/10.1073/pnas.1421798112} {\bibfield
  {journal} {\bibinfo  {journal} {Proceedings of the National Academy of
  Sciences}\ }\textbf {\bibinfo {volume} {112}},\ \bibinfo {pages} {1907}
  (\bibinfo {year} {2015})},\ \Eprint
  {https://arxiv.org/abs/https://www.pnas.org/doi/pdf/10.1073/pnas.1421798112}
  {https://www.pnas.org/doi/pdf/10.1073/pnas.1421798112} \BibitemShut {NoStop}%
\bibitem [{\citenamefont {Eckmann}\ and\ \citenamefont
  {Ruelle}(1985)}]{Eckmann1985}%
  \BibitemOpen
  \bibfield  {author} {\bibinfo {author} {\bibfnamefont {J.~P.}\ \bibnamefont
  {Eckmann}}\ and\ \bibinfo {author} {\bibfnamefont {D.}~\bibnamefont
  {Ruelle}},\ }\href {https://doi.org/10.1103/RevModPhys.57.617} {\bibfield
  {journal} {\bibinfo  {journal} {Rev. Mod. Phys.}\ }\textbf {\bibinfo {volume}
  {57}},\ \bibinfo {pages} {617} (\bibinfo {year} {1985})}\BibitemShut
  {NoStop}%
\bibitem [{\citenamefont {Abarbanel}\ \emph {et~al.}(1991)\citenamefont
  {Abarbanel}, \citenamefont {Brown},\ and\ \citenamefont
  {Kennel}}]{Abarbanel1991}%
  \BibitemOpen
  \bibfield  {author} {\bibinfo {author} {\bibfnamefont {H.~D.~I.}\
  \bibnamefont {Abarbanel}}, \bibinfo {author} {\bibfnamefont {R.}~\bibnamefont
  {Brown}},\ and\ \bibinfo {author} {\bibfnamefont {M.~B.}\ \bibnamefont
  {Kennel}},\ }\href {https://doi.org/10.1007/BF01209065} {\bibfield  {journal}
  {\bibinfo  {journal} {Journal of Nonlinear Science}\ }\textbf {\bibinfo
  {volume} {1}},\ \bibinfo {pages} {175} (\bibinfo {year} {1991})}\BibitemShut
  {NoStop}%
\bibitem [{\citenamefont {Eckhardt}\ and\ \citenamefont
  {Yao}(1993)}]{Eckhardt1993}%
  \BibitemOpen
  \bibfield  {author} {\bibinfo {author} {\bibfnamefont {B.}~\bibnamefont
  {Eckhardt}}\ and\ \bibinfo {author} {\bibfnamefont {D.}~\bibnamefont {Yao}},\
  }\href {https://doi.org/https://doi.org/10.1016/0167-2789(93)90007-N}
  {\bibfield  {journal} {\bibinfo  {journal} {Physica D: Nonlinear Phenomena}\
  }\textbf {\bibinfo {volume} {65}},\ \bibinfo {pages} {100} (\bibinfo {year}
  {1993})}\BibitemShut {NoStop}%
\bibitem [{\citenamefont {Maryak}\ and\ \citenamefont
  {Chin}(2001)}]{maryak2001}%
  \BibitemOpen
  \bibfield  {author} {\bibinfo {author} {\bibfnamefont {J.}~\bibnamefont
  {Maryak}}\ and\ \bibinfo {author} {\bibfnamefont {D.}~\bibnamefont {Chin}},\
  }in\ \href {https://doi.org/10.1109/WSC.2001.977290} {\emph {\bibinfo
  {booktitle} {Proceeding of the 2001 Winter Simulation Conference (Cat.
  No.01CH37304)}}},\ Vol.~\bibinfo {volume} {1}\ (\bibinfo {year} {2001})\ pp.\
  \bibinfo {pages} {307--312 vol.1}\BibitemShut {NoStop}%
\bibitem [{\citenamefont {Quiroz}\ \emph {et~al.}(2025)\citenamefont {Quiroz},
  \citenamefont {Titum}, \citenamefont {Lotshaw}, \citenamefont {Lougovski},
  \citenamefont {Schultz}, \citenamefont {Dumitrescu},\ and\ \citenamefont
  {Hen}}]{quiroz2025qaoa}%
  \BibitemOpen
  \bibfield  {author} {\bibinfo {author} {\bibfnamefont {G.}~\bibnamefont
  {Quiroz}}, \bibinfo {author} {\bibfnamefont {P.}~\bibnamefont {Titum}},
  \bibinfo {author} {\bibfnamefont {P.}~\bibnamefont {Lotshaw}}, \bibinfo
  {author} {\bibfnamefont {P.}~\bibnamefont {Lougovski}}, \bibinfo {author}
  {\bibfnamefont {K.}~\bibnamefont {Schultz}}, \bibinfo {author} {\bibfnamefont
  {E.}~\bibnamefont {Dumitrescu}},\ and\ \bibinfo {author} {\bibfnamefont
  {I.}~\bibnamefont {Hen}},\ }\href
  {https://doi.org/10.1103/PhysRevResearch.7.023240} {\bibfield  {journal}
  {\bibinfo  {journal} {Phys. Rev. Res.}\ }\textbf {\bibinfo {volume} {7}},\
  \bibinfo {pages} {023240} (\bibinfo {year} {2025})}\BibitemShut {NoStop}%
\bibitem [{\citenamefont {Falk}(1984)}]{Falk1984}%
  \BibitemOpen
  \bibfield  {author} {\bibinfo {author} {\bibfnamefont {H.}~\bibnamefont
  {Falk}},\ }\href
  {https://doi.org/https://doi.org/10.1016/0375-9601(84)90645-5} {\bibfield
  {journal} {\bibinfo  {journal} {Physics Letters A}\ }\textbf {\bibinfo
  {volume} {105}},\ \bibinfo {pages} {101} (\bibinfo {year}
  {1984})}\BibitemShut {NoStop}%
\end{thebibliography}

%
\end{document}